\documentclass[prd,preprint,superscriptaddress,floatfix,eqsecnum,nofootinbib,12pt]{revtex4}
\usepackage{hyperref,url}
\usepackage{amssymb,amsmath,amsthm,amsfonts}
\usepackage{graphicx}
\usepackage{ulem}
\usepackage{subfig,epsfig}
\usepackage[labelfont=bf]{caption}
\usepackage[usenames]{color}
\usepackage[letterpaper,left=2.cm,right=2.cm,top=2.5cm,bottom=2.5cm]{geometry}
\usepackage[T1]{fontenc}
\usepackage{float}

\def\di{\mathrm{d}}

\usepackage{notes2bib}
\begin{document}

\title{The no-boundary proposal in biaxial Bianchi IX minisuperspace}

\author{O. Janssen}
\email{opj202@nyu.edu}
\affiliation{Center for Cosmology and Particle Physics, NYU, NY 10003, USA}
\author{J. J. Halliwell}
\email{j.halliwell@imperial.ac.uk}
\affiliation{Blackett Laboratory, Imperial College, London SW7 2BZ, UK}
\author{T. Hertog}
\email{thomas.hertog@fys.kuleuven.be}
\affiliation{Institute for Theoretical Physics, KU Leuven, 3001 Leuven, Belgium \vspace{2cm}}

\begin{abstract}
\noindent
We implement the no-boundary proposal for the wave function of the universe in an exactly solvable Bianchi IX minisuperspace model with two scale factors. We extend our earlier work (Phys. Rev. Lett. \textbf{121}, 081302, 2018) to include the contribution from the $\mathbb{C}\text{P}^2 \setminus B^4$ topology. The resulting wave function yields normalizable probabilities and thus fits into a predictive framework for semiclassical quantum cosmology. We find that the amplitude is low for large anisotropies. In the isotropic limit the usual Hartle-Hawking wave function for the de Sitter minisuperspace model is recovered. Inhomogeneous perturbations in an extended minisuperspace are shown to be initially in their ground state. We also demonstrate that the precise mathematical implementation of the no-boundary proposal as a functional integral in minisuperspace depends on detailed aspects of the model, including the choice of gauge-fixing. This shows in particular that the choice of contour cannot be fundamental, adding weight to the recent proposal that the semiclassical no-boundary wave function should be defined solely in terms of a collection of saddle points. We adopt this approach in most of this paper. Finally we show that the semiclassical tunneling wave function of the universe is essentially equal to the no-boundary state in this particular minisuperspace model, at least in the subset of the classical domain where the former is known.

\end{abstract}

\maketitle

\newpage
\tableofcontents

\newpage
\section{Introduction} \label{introsec}

\subsection{Quantum cosmology and the no-boundary proposal} \label{introsec1}
\noindent Quantum cosmology is concerned with the search for a quantum-mechanical theory that describes the origin and evolution of the universe. A central element in this is a wave function $\Psi$ for a closed universe that is a functional of the three-metric and matter fields on a spacelike three-surface $\Sigma$. A wave function of the universe is held to be a solution to the Wheeler-DeWitt (WDW) equation,
\begin{equation} \label{WDWeq}
H \Psi = 0 \,,
\end{equation}
where $H$ is the Hamiltonian for the gravitational and matter modes of the system. The WDW equation has potentially an infinite number of solutions so boundary conditions or some more general principle are required to limit the possible solutions, or even select a solution uniquely. Of the possible proposals to accomplish this, the most frequently-utilized one is the no-boundary proposal of Hartle and Hawking, which picks out a solution to the WDW equation that is traditionally defined by a functional integral over gravitational and matter fields on compact four-manifolds whose only boundary is the three-surface $\Sigma$ \cite{PhysRevD.28.2960,Hawking1983}.

Since its inception, the no-boundary proposal has been through a progression of development to determine exactly how it is implemented in specific models and to elucidate its physical predictions. Most of these models have been simple minisuperspace models, essentially quantum-mechanical models in which the gravitational and matter modes are artificially constrained to have a finite number of degrees of freedom. In such simple models it has been argued that the no-boundary proposal successfully predicts important features of our observed universe such as the existence of classical histories \cite{PhysRevD.28.2960,Hawking1983,HHH2008}, an early period of inflation \cite{HHH2008,Hartle:2007gi} and a nearly-Gaussian spectrum of primordial density fluctuations \cite{Halliwell:1984eu,Hartle:2010vi,Hartle:2010dq,Hertog:2013mra}. 

The need to refine the definition and implementation of the no-boundary proposal has acquired some urgency in light of recent criticisms questioning the solidity of these successful predictions \cite{FLT2,FLT3}. Motivated by this we have recently shown \cite{DiazDorronsoro:2018wro} in a minisuperspace model that there exists a precise mathematical implementation of the no-boundary proposal, expressed in terms of a gravitational path integral, that yields a well-defined state in which large universes behave classically and large perturbations are damped. This lends supports to the viability of the no-boundary wave function (NBWF) as the state of our observed universe and it refutes the recent claim that the NBWF is ill-defined due to problems with large perturbations \cite{FLT2,FLT3}.

The biaxial Bianchi IX (BB9) minisuperspace we considered in Ref. \cite{DiazDorronsoro:2018wro} is a homogeneous but anisotropic minisuperspace approximation to gravity coupled to a positive cosmological constant and no matter. The classical histories in this minisuperspace are known as BB9 cosmologies. The configuration space on which the wave function is defined in this model consists of squashed three-spheres specified by two scale factors; one specifying the size of the two-sphere and the other specifying the size of the circle, when the three-sphere is viewed as a fibration of a circle over a two-sphere. We evaluated the NBWF as a functional integral in this model on the four-disk in Ref. \cite{DiazDorronsoro:2018wro} and found that non-zero squashings, i.e. anisotropies, are suppressed. In the present paper we refine and extend our analysis of the BB9 minisuperspace model in a number of ways.

\subsection{Implementation of the no-boundary proposal} \label{introsec2}
\noindent Despite the clear geometric and intuitive appeal of the no-boundary proposal, its implementation in specific minisuperspace models requires a certain amount of additional input.
First, it must be specified what exactly is meant by no-boundary initial conditions. Classically, regularity of the no-boundary saddle points implies constraints on the metric and its first derivatives. These enter as variables and conjugate momenta in the quantum theory. In BB9 minisuperspace we showed that a proper implementation of the no-boundary idea as a functional integral over geometries on the four-disk requires the two-sphere scale factor to be zero initially together with a carefully chosen regularity condition on the {\it momentum} conjugate to the scale factor of the circle \cite{DiazDorronsoro:2018wro}.

Second, the contour of integration must be specified, at least in the functional integral formulation of the NBWF. It was clearly stated from the outset in the 1980s that the no-boundary path integral must be carried out over a suitable complex contour for physical reasons. The general requirements that such a contour must satisfy in order to yield a physically viable wave function were thoroughly investigated \cite{HarHal1990,Halliwell:1990qr} and numerous models were worked out explicitly (see e.g. \cite{Halliwell1988,Halliwell1989,Halliwell1990,Garay1990,PhysRevD.40.4011}). In the BB9 minisuperspace model this concerns the choice of contour for the lapse integral. In \cite{DiazDorronsoro:2018wro} we took this to be a closed contour encircling the origin $N = 0$, which yields a well-defined state and predictions that agree with observation.\footnote{Closed contours in the context of the NBWF have been considered before (see e.g. \cite{Halliwell1988,Halliwell1990,HarHal1990,PhysRevD.40.4011}). Together with infinite contours they provide the only evident ways of generating wave functions constructed as path integrals.}

\subsection{A Lorentzian path integral approach} \label{introsec3}
\noindent By contrast Feldbrugge et al. in their recent papers \cite{FLT1,FLT2,FLT3} revived \cite{Brown1990} an alternative approach to path integral quantum cosmology, based on a purely Lorentzian path integral construction, which comes with a contour for the lapse that runs over the positive real axis only. This proposal bears some resemblance to the tunneling proposal \cite{Vilenkin1982,Vilenkin1984,Vilenkin:2018dch,Vilenkin:2018oja} but differs in some key respects. The positive real line choice of contour does not yield a solution of the WDW equation but rather a Green's function. More significantly, in the semiclassical limit it selects a saddle point that is different from that specifying the NBWF in the BB9 model and which fails to provide a reasonable physical basis for a predictive framework for cosmology. Feldbrugge et al. advance their Lorentzian approach on the grounds that it encodes a primitive notion of causality. However this is not the case. Histories of geometry are curves in the superspace of three-geometries. There is no physical notion of one three-geometry being `before' or `after' another. Furthermore, the lapse integration is not directly related to the observed arrows of time such as those defined by the increase in entropy, the retardation of radiation, and the growth of fluctuations. As shown in \cite{Halliwell:1984eu} and as much subsequent work confirmed \cite{Hawking:1993tu,HH2011}, these physical arrows arise because the NBWF predicts that fluctuations are small when the universe was small. We also note that all physical predictions in any quantum-mechanical system are derived most directly from a wave function, for which a well-defined formalism exists for the computation of probabilities, and not from a Green's function. Hence any claims made on the basis of the properties of Green's functions must include a specification of the way in which they are used to compute probabilities.

\subsection{The no-boundary proposal as a collection of saddle points} \label{introsec4}
\noindent Having said this, the debate over the correct choice of contour, and over the focus on solutions to the WDW equation versus Green's functions, is significantly neutralized in the semiclassical approximation to the wave function, since the latter is logically independent of any integral representation. Given that the minisuperspace functional integral implementation is only meaningful in the semiclassical approximation, it is in many ways appealing and certainly simpler to specify the semiclassical NBWF directly in terms of a collection of saddle points that satisfy a minimal set of criteria that encapsulate its physical principles, without relying on a functional integral of any kind. This approach was recently advanced in \cite{Halliwell:2018ejl} where the NBWF was given as a collection of specific saddle points of the dynamical theory, and this is the approach we adopt, for the large part, in the main text of this paper (although we will address some aspects of path integral representations too).

\subsection{This paper} \label{introsec5}
\noindent In detail, the analysis of this paper will cover the following issues. First, we compute a second topological contribution to the NBWF in the BB9 model, coming from the $\mathbb{C}\text{P}^2 \setminus B^4$ topology. No-boundary initial conditions in this case amount to setting the scale factor of the circle to zero and fixing the momentum conjugate to the two-sphere scale factor. We compare this contribution with the one coming from the four-disk topology we considered in Ref. \cite{DiazDorronsoro:2018wro}. We will also show that both of these contributions are semiclassical approximations to exact solutions to the WDW equation which are normalizable, in the sense that they have normalizable flux across surfaces of interest. This therefore means that the theory delivers well-defined probabilities and so fits into a predictive framework of quantum cosmology. We recover the prediction of the original Hartle-Hawking state that the amplitude of large anisotropies is suppressed \cite{HAWKING198483,PhysRevD.31.3073}. In particular we find there is no contribution from ``wrong sign'' saddle points or any other source that would favor anisotropic configurations. We also extend the probability distribution to arbitrarily large anisotropies by appropriately adjusting the prefactor of the wave function -- which the semiclassical analysis leaves largely undetermined. 

Second, we analyze the behavior of the wave function in the isotropic limit. Clearly one does not expect the wave function of the two-dimensional BB9 model to agree exactly with the wave function of a one-dimensional de Sitter (dS) minisuperspace model in this limit. However one does expect agreement between both theories at the classical level and hence for the exponential behavior of the wave functions to coincide, which we show is the case indeed.

Third, we consider an extension of BB9 minisuperspace that includes inhomogeneous massless scalar fluctuations. We compute the wave function of such fluctuations and show we recover quantum field theory in curved spacetime, with inhomogeneities initially in their ground state. 

Fourth, although we have focused on the definition of the NBWF as a collection of saddle points, we explore various aspects of its minisuperspace functional integral representation through some elementary examples. These examples reinforce the argument that features such as the choice of lapse contour and the choice of initial conditions at the south pole of the saddle points can both depend on the model and even on its specific parameterization (or gauge-fixing). Thus they should not be regarded as universal or fundamental features of the NBWF functional integral. General requirements on those facets of functional integrals cannot therefore be advanced to falsify the NBWF.

The outline of the remainder of this paper is as follows. We start in Section \ref{B9sec} with a description of the biaxial Bianchi IX model and its metric. We derive its WDW equation and exhibit the general exact solution. We also describe the classical solutions for the model. We then describe, in Section \ref{NBWFsec}, the construction of no-boundary wave functions in the BB9 model for the two different topologies of interest.
A detailed description of the saddle points for the $\overline{B^4}$ topology is given in Section \ref{NUTsec} and for the  $\mathbb{C}\text{P}^2 \setminus B^4$ topology is given in Section \ref{Boltsec}. In Section \ref{normsec} we explain the sense in which the probabilities constructed from the wave function are normalizable. We then in Section \ref{PsiHHsec} construct the wave function arising from contributions from saddle points with both topologies. We discuss the isotropic limit in Section \ref{isotropicsec} and in Section \ref{inhomogeneoussec} we discuss inhomogeneous perturbations about the two types of saddle points discussed above. In Section \ref{TWFsec} we compare the NBWF with the tunneling wave function in the BB9 model, showing that the two coincide in certain regimes where the latter is known and making a conjecture about their coincidence in a larger portion of the minisuperspace. We conclude in Section \ref{conclusionsec}.

Some of the details of our work are relegated to a set of appendices.  Appendix \ref{A1} gives a more detailed discussed of the $\mathbb{C}\text{P}^2 \setminus B^4$  saddle points and the phase transitions between the various saddle points is described in Appendix \ref{A2}. In Appendices \ref{nonlinearsec} and \ref{nonlinearsec2} we describe how the BB9 model may be viewed as a non-linear extension of the dS minisuperspace model perturbed by a single mode of either a tensor field or massless minimally coupled scalar. In Appendix \ref{offshellsec} we discuss the off-shell structure of minisuperspace path integrals and show, as promised earlier on in this Introduction, that they depend sensitively on the details of the model and its parameterization. Building on this we respond to the criticisms of Feldbrugge et al. \cite{FLT2,FLT3,FLT4} in Appendix \ref{FLTsec}.

\clearpage
\newpage
\section{Biaxial Bianchi IX minisuperspace} \label{B9sec}

\subsection{Metric}
\noindent In the BB9 minisuperspace model, the wave function of the universe, $\Psi$, is a function on the superspace of single-squashed three-sphere ($S^3$) geometries. It specifies the amplitude that a spatially closed universe, which for simplicity we will assume has the topology of $\mathbb{R}_\text{time} \times S^3$, contains a spacelike section which is a squashed $S^3$. Such geometries may be parametrized by two coordinates $p$ and $q$ living in the quadrant $\{ p,q \geq 0 \}$, which appear in the metric on the $S^3$ as
\begin{equation} \label{S3metric}
	\di \ell^2 = \frac{p}{4} \left( \sigma_1^2 + \sigma_2^2 \right) + \frac{q}{4} \sigma_3^2 \,.
\end{equation}
Thus $\Psi = \Psi(p,q)$. In Eq. \eqref{S3metric} $\sigma_{1,2,3}$ are the left-invariant one-forms of SU(2) given by
\begin{equation}\label{eqn:left1forms}
 \sigma_1 = -\sin\psi \, \di\theta + \cos\psi \sin\theta \, \di\phi \,, ~~
 \sigma_2 = \cos\psi \, \di\theta + \sin\psi \sin\theta \, \di\phi \,, ~~
 \sigma_3 = \di\psi + \cos\theta \, \di\phi \,,
\end{equation}
with $0 \leq \theta \leq \pi$, $0 \leq \phi < 2\pi$ and $0 \leq \psi < 4\pi, \psi \cong \psi + 4 \pi$ the Euler angles on the $S^3$. In these coordinates the $S^3$ is represented as the fibration of an $S^1$ over an $S^2$, and the values $p$ and $q$ determine the sizes of the $S^2$ base and the $S^1$ fibers respectively. When $p = q$ the metric is proportional to $\delta^{ij}\sigma_i \sigma_j/4$, which is the round metric on the unit $S^3$. When $p \neq q$ the metric represents a deformed or ``squashed'' $S^3$ and the space is anisotropic. The degree of squashing is conveniently expressed through the quantity
\begin{equation}
	\alpha \equiv \frac{p}{q} - 1 \,,
\end{equation}
with $\alpha = 0$ corresponding to the round $S^3$, $\alpha > 0$ corresponding to a ``prolate symmetric top'' or cigar and $\alpha < 0$ corresponding to an ``oblate symmetric top'' or pancake (here $p \sim I_1 = I_2, q \sim I_3$ are interpreted as the moments of inertia of the spheroid about its principal axes of rotation). The wave function may alternatively be viewed as a function of the ``size'' $p$ of the $S^3$ and the squashing $\alpha \in (-1,\infty)$, $\Psi = \Psi(p,\alpha)$, or any other combination for that matter. Since $\Psi$ should assign the same amplitude to the same configuration, it transfoms as a scalar under such coordinate transformations. (We neglect a possible phase factor.)

The wave function of the universe for this model satisfies the WDW equation, Eq. \eqref{WDWeq}. In the BB9 minisuperspace this is a second order (linear) partial differential equation (PDE) which requires boundary conditions on a line in superspace to define a unique solution. Its form (given in Eq. \eqref{Hpqposition} below) depends on the underlying dynamical theory, which we take to be Einstein gravity with a positive cosmological constant. Later on in \S\ref{inhomogeneoussec} we will extend this minisuperspace to contain a small, massless and minimally coupled scalar field.

As we detail in Appendix \ref{nonlinearsec}, the BB9 minisuperspace can be viewed as a non-linear extension of the dS minisuperspace model (that is, the model with $p = q$) perturbed by a particular (the ``$n=2$'') transverse and traceless tensor mode, or as a non-linear extension of the dS minisuperspace model containing a specific mode of a massless minimally coupled scalar (Appendix \ref{nonlinearsec2}). The BB9 model is also a restricted version of the mixmaster universe \cite{PhysRevLett.22.1071} where two out of the three scale factors are set equal. For earlier and related work on this model in quantum cosmology we refer the reader to Refs. \cite{PhysRevD.31.3073,Jensen1991,Daughton:1998aa,DiazDorronsoro:2018wro,HAWKING198483,delCampo:1989hy,WRIGHT1985115}. (Ref. \cite{PhysRevD.31.3073} contains errors but arrives at a correct qualitative conclusion.) Ref. \cite{Jensen1991} in particular discusses the same object as the one we are mainly interested in here, namely the NBWF in the BB9 model with a positive cosmological constant. While our results are consistent with the ones presented in that work, we provide more details and also extend the known results by giving accurate analytic approximations in a large subset of superspace to various quantities of interest for any value of the squashing of the $S^3$ in the argument of the wave function.

\subsection{Wheeler-DeWitt equation and solution} \label{WDWsec}
\noindent To derive the WDW equation for the BB9 minisuperspace, we consider homogeneous four-metrics on the spacetime $\mathbb{R}_\text{time} \times S^3$ of the form
\begin{equation} \label{qpansatz}
	2 \pi^2 \, \di s^2 = -\frac{N(t)^2}{q(t)} \di t^2 + \frac{p(t)}{4} \left( \sigma_1^2 + \sigma_2^2 \right) + \frac{q(t)}{4} \sigma_3^2 \,.
\end{equation}
Here $p(t)$ and $q(t)$ are real, positive functions. $N(t)$ is the lapse function, which is arbitrary and represents our freedom to perform time reparametrizations. (We have chosen the particular form of the $00$ component of the metric because it simplifies the analysis.) By convention we will study transition amplitudes between three-geometries at the times $t = 0$ and $t = 1$. The bulk part of the Einstein-Hilbert action evaluated on the metric \eqref{qpansatz}, setting $M_\text{Pl} = 1$ and including a vacuum energy density $2 \pi^2 \Lambda$, reads
\begin{equation} \label{Sqp}
	S[p,q;N] = \int_0^1 \di t \, N \left[ \frac{1}{2N^2} \left( - \frac{q}{2 p} \dot{p}^2 - \dot{p} \dot{q} \right) - \left( \frac{q}{p} + \Lambda p - 4 \right) \right] \,,
\end{equation}
where we have chosen the branch $\sqrt{N^2} = +N$.\footnote{This choice of sign is inconsequential for this section, i.e. for the WDW equation and classical paths. However since the action and canonical momenta (see Eq. \eqref{canmomenta}) are sensitive to the choice of sign, the discussion of the NBWF in Sections \ref{NUTsec} and \ref{Boltsec} is sensitive to the choice $\sqrt{N^2} = \pm N$. If one were to choose $\sqrt{N^2} = -N$, one should replace $N$ with $-N$ everywhere in those sections.} Note that we can absorb $\Lambda$ into the other variables by the redefinitions $p \rightarrow \Lambda p, q \rightarrow \Lambda q, N \rightarrow \Lambda N, S \rightarrow \Lambda S$. We will do this for now and reinstate $\Lambda$ at a later stage. The action \eqref{Sqp} can then be abbrieviated as
\begin{equation} \label{QCactionconvention}
	S[x;N] = \int_0^1 \di t \, N \left( \frac{1}{2 N^2} f_{\alpha \beta}(x) \dot{x}^\alpha \dot{x}^\beta - U(x) \right) \,,
\end{equation}
where $x \equiv (p,q)$ and we have defined the DeWitt metric on minisuperspace
\begin{equation} \label{fMSS}
   f =
  \frac{-1}{2} \left( {\begin{array}{cc}
   q/p & 1 \\
   1 & 0 \\
  \end{array} } \right) \,,
\end{equation}
with components ordered as $(p,q)$, and the potential
\begin{equation}
	U(p,q) = \frac{q}{p} + p - 4 \,.
\end{equation}
This is the action for particle $x$ moving on a curved background with metric $f$ under the influence of a potential $U$. The momenta conjugate to $p$ and $q$ are
\begin{equation} \label{canmomenta}
	\Pi_\alpha = \frac{1}{N} f_{\alpha \beta} \dot{x}^\beta \,,
\end{equation}
or
\begin{align}
	\Pi_p &= -\frac{1}{2N} \left( \frac{q \dot{p}}{p} + \dot{q} \right) \,, \label{pi-p} \\
	\Pi_q &= -\frac{1}{2N} \, \dot{p} \,. \label{pi-q}
\end{align}
With this Eq. \eqref{QCactionconvention} can be rewritten as
\begin{equation} \label{pqactionH}
	S[x,\Pi;N] = \int_0^1 \di t \left( \Pi_\alpha \dot{x}^\alpha - N H \right) \,,
\end{equation}
where
\begin{equation} \label{pqhamiltonian}
	H = \Pi_q \frac{q}{p} \Pi_q - 2 \Pi_q \Pi_p + \frac{q}{p} + p - 4 \,.
\end{equation}
In Eq. \eqref{pqhamiltonian} the operator ordering is chosen which, upon making the canonical quantization replacements $\Pi_p \rightarrow -i \hbar \partial_p, \Pi_q \rightarrow - i \hbar \partial_q$, gives rise to a Laplacian ordering of the derivatives. That is, in position space,
\begin{align}
	\hat{H} &= - \frac{\hbar^2}{2} \nabla^2 + U \notag \\
	&= - \hbar^2 \left( \frac{q}{p} \partial_q^2 + \frac{1}{p} \partial_q - 2 \partial_p \partial_q \right) + \frac{q}{p} + p - 4 \,, \label{Hpqposition}
\end{align}
where $\nabla^2$ is the scalar Laplacian with respect to $f$. This is the WDW operator for the BB9 minisuperspace. The particular factor ordering in Eq. \eqref{Hpqposition} ensures that any potential wave function of the universe transforms as a scalar under redefinitions of the minisuperspace coordinates $p$ and $q$ in Eq. \eqref{qpansatz}, as we have mentioned it should above. Finally, in two dimensions, the scalar Laplacian is conformal, so that $\Psi$ transforms in a simple way under redefinitions of the lapse \cite{HalliwellWdW1988,Moss:1988wk}.\footnote{This means that if we were to send $N(t) \rightarrow \omega(x(t))^2 N(t)$ for arbitrary $\omega$, $\Psi$ would transform as $\Psi \rightarrow \omega^{D/2-1} \Psi$ where $D > 1$ is the dimension of the minisuperspace.} In fact in two-dimensional minisuperspaces like the BB9 model, $\Psi$ is invariant under redefinitions of the lapse function.

A salient feature of the BB9 minisuperspace model is that as a two-dimensional quantum system, it is essentially classical \cite{DiazDorronsoro:2018wro}. This is because two out of the four phase space coordiantes, namely $q$ and $\Pi_p$, appear linearly in the Hamiltonian \eqref{pqhamiltonian}. Thus sums-over-histories that fix $p$ and $\Pi_q$ at the boundary of the time interval will only contain a single path -- the classical one corresponding to the boundary data. In other words, the semiclassical ``approximation'' to transition amplitudes is exact.

Closely related is that one can solve the WDW equation \eqref{WDWeq} for the BB9 model in closed form, for arbitrary boundary conditions. To achieve this one can go to the representation in which $\Psi$ is a function of $p$ and $\Pi_q$, in terms of which the WDW equation is a first order PDE which can be solved by the method of characteristics. One then Fourier transforms the result, and obtains
\begin{align} \label{WDWgeneralsol}
	\Psi_\text{WDW}(p,q) &= \notag \\ \int_\mathbb{R} \di \Pi_q \, \exp &\left( \frac{i \Pi_q q}{\hbar} \right) \exp \left[ \frac{i p \Pi_q}{3 \hbar(1+\Pi_q^2)^2} \left( p (3+\Pi_q^2) - 12(1+\Pi_q^2) \right) \right] \frac{1}{\sqrt{1+\Pi_q^2}} \, f_2 \left( \frac{p}{1+\Pi_q^2} \right) \,,
\end{align}
with $f_2$ a function to be fixed by the boundary conditions. We will not use this form in what follows, but it could be a useful starting point for discussing other proposals for the wave function in the BB9 model. We also note that the quantity $p/(1 + \Pi_q^2)$ is conserved under classical and quantum evolution.

Finally we recall the equations that an approximate WKB solution to Eq. \eqref{WDWeq}, $\Psi \approx \mathcal{A} \, \exp \left( i \bar{S}_0 / \hbar \right)$, must satisfy in the semiclassical limit $\hbar \rightarrow 0$. These are the Hamilton-Jacobi equation corresponding to zero energy,
\begin{equation}
	\frac{1}{2} \left( \nabla \bar{S}_0 \right)^2 + U = 0 \,, \label{h0eq} \\
\end{equation}
and a continuity equation for the prefactor,
\begin{equation}
	\nabla \cdot \left( \mathcal{A}^2 \nabla \bar{S}_0 \right) = 0 \,. \label{h1eq}
\end{equation}
They are the leading and next-to-leading order in $\hbar$ components of the WDW equation respectively. Inner products in Eqns. \eqref{h0eq}-\eqref{h1eq} are taken with respect to the metric \eqref{fMSS}.

\subsection{Classical paths} \label{saddlessec}
\noindent From here on we will choose a gauge in which the lapse function $N(t)$ is a constant, $N$. Setting variations of \eqref{Sqp} with respect to $q$ and $p$ to zero yields the second order classical EOM
\begin{align}
	p \ddot{p} - \frac{\dot{p}^2}{2} - 2N^2 &= 0 \,, \label{p-EOM} \\
	p \left( p \ddot{q} + q \ddot{p} \right) + \dot{p} \left( p \dot{q} - \frac{q \dot{p}}{2} \right) + 2N^2 \left( q - p^2 \right) &= 0 \,, \label{q-EOM}
\end{align}
while setting variations with respect to the lapse to zero yields the first order Hamiltonian constraint
\begin{equation} \label{Hamiltonianconstraint}
	-\frac{1}{2N^2} \left( \frac{q}{2p} \dot{p}^2 + \dot{p}\dot{q} \right) + \frac{q}{p} + p - 4 = 0 \,.
\end{equation}
These equations are equivalent to the full Einstein equations for the metric Ansatz \eqref{qpansatz}. Note that the third derivative of $p$ and the fifth derivative of $p q$ are identically zero for a classical path. For each $N$ there are two solutions to the second order EOM which interpolate between the configuration $(p_0,q_0)$ at $t = 0$ and $(p_1,q_1)$ at $t = 1$, where one is given by
\begin{align}
	\bar{p}(t;N) &= p_0 + 2 \left( \sqrt{p_0 p_1 - N^2} - p_0 \right) t + \left( p_0 + p_1 - 2 \sqrt{p_0 p_1 - N^2} \right) t^2 \,, \label{p-sol-general} \\
	\bar{q}(t;N) &= \frac{p_0 q_0 + c_1 t + c_2 t^2 + c_3 t^3 + c_4 t^4}{\bar{p}(t;N)} \label{q-sol-general} \,,
\end{align}
with
\begin{align}
	c_1 &= \frac{p_0}{3 \sqrt{p_0 p_1 - N^2}} \left[ 3 p_1 (q_0 + q_1) - N^2 p_1 - 2 \left( 3 q_0 + N^2 \right) \sqrt{p_0 p_1 - N^2} \right] \,, \\
	c_2 &= \frac{1}{3 \sqrt{p_0 p_1 - N^2}} \left\{ 2 N^4 + p_1 \left( 3 q_1 - N^2 \right) \sqrt{p_0 p_1 - N^2} \, + \right. \\
	&\left. \hspace{3.5cm} p_0 \left[ \left( 3 q_0 + 5 N^2 \right) \sqrt{p_0 p_1 - N^2} - p_1 \left( N^2 + 3 (q_0 + q_1) \right) \right] \right\} \,, \notag \\
	c_3 &= \frac{4 N^2}{3} \left( \sqrt{p_0 p_1 - N^2} - p_0 \right) \,, \\
	c_4 &= \frac{N^2}{3} \left( p_0 + p_1 - 2 \sqrt{p_0 p_1 - N^2} \right) \,, \label{c4}
\end{align}
where some choice of branch cut for the square root is made, and the other solution is obtained from this one for each $N$ by replacing every square root by its negative. The solution to the second order EOM with other boundary conditions, for example those that fix $p$ and $\Pi_q$ at $t = 0$, can readily be obtained from the solution given above.

At this stage there remains a single undetermined parameter, $N$. Its value is determined by the Hamiltonian constraint \eqref{Hamiltonianconstraint}. This is consistent since the left-hand side (LHS) of that equation is constant in $t$ when evaluated on Eqns. \eqref{p-sol-general}-\eqref{q-sol-general}, which yields an algebraic equation for $N$ that depends on the boundary data. We will denote solutions to this equation by $N_s$.

\section{No-boundary wave function in biaxial Bianchi IX \\ minisuperspace} \label{NBWFsec}

\subsection{Topological contributions} \label{topsec}
\noindent In the semiclassical limit, the NBWF for the BB9 minisuperspace model is determined by regular solutions to the complexified Einstein equations which live on a compact four-manifold with only boundary an $S^3$. (In an abuse of terminology we will sometimes call these solutions ``instantons''.) There are infinitely many such solutions and their general classification is unknown. We will deal with this situation in the usual simplistic manner, which is to consider only a handful of highly symmetric solutions (of which one hopes that they give the dominant contributions to the wave function in the semiclassical limit \cite{PhysRevD.42.2458,doi:10.1063/1.526571}). The particular ones we will study live on the four-manifolds $\overline{B^4}$ (\S\ref{NUTsec}, see also \cite{DiazDorronsoro2017,Jensen1991}) and $\mathbb{C}\text{P}^2 \setminus B^4$ (\S\ref{Boltsec}, see also \cite{Jensen1991}), respectively the closed four-ball or four-disk and the two-dimensional complex projective plane with an open four-ball cut out, and both can be written in the form
\begin{equation} \label{qpansatz2}
	2 \pi^2 \, \di s^2 = -\frac{N^2}{q(\tau)} \di \tau^2 + \frac{p(\tau)}{4} \left( \sigma_1^2 + \sigma_2^2 \right) + \frac{q(\tau)}{4} \sigma_3^2 \,.
\end{equation}
At least one other no-boundary solution of the form \eqref{qpansatz2} is known \cite{Daughton:1998aa} -- it lives on the manifold $\mathbb{R}\text{P}^4 \setminus B^4$ -- but we do not discuss this contribution here.

In contrast to our discussion in \S\ref{B9sec}, and although the notation is similar, we emphasize that \eqref{qpansatz2} is \textit{not} a metric on the Lorentzian spacetime with topology $\mathbb{R}_\text{time} \times S^3$. Instead here $\tau$ is a (real) radial coordinate on either $\overline{B^4}$ or $\mathbb{C}\text{P}^2 \setminus B^4$ and the metric \eqref{qpansatz2} is defined on these manifolds. Additionally, the functions $p$ and $q$ in Eq. \eqref{qpansatz2} will generally be complex if they contribute to the semiclassical NBWF. Thus with the NBWF we are in general dealing with complex metrics on real manifolds.

\subsection{Path integral} \label{PIsec}
\noindent To construct the NBWF in the BB9 model, in our previous work \cite{DiazDorronsoro:2018wro} we followed the general minisuperspace functional integral approach detailed e.g. in Ref. \cite{Halliwell1990} and briefly reviewed in Appendix \ref{offshellsec}. That is, we first computed the propagator $K(x_1,N;x_0,0)$, which is a solution to the Schr\"odinger equation
\begin{equation}
	\hat{H} K = i \hbar \, \partial_N K \,,
\end{equation}
and represents the amplitude for the geometry to evolve from a state with shape and size $(p_0,q_0)$ to one specified by $(p_1,q_1)$ in a ``time'' $N$ \bibnote{We stress again that the lapse is not the time identified by an observer who lives in a member of the ensemble of classical, Lorentzian histories predicted by the wave function. Time is not a fundamental input in quantum cosmology, as the WDW equation shows. Instead it emerges as a strong correlation between the superspace coordinates and momenta, predicted by the wave function in a certain subset of the parameter space. Statements which connect the lapse, which is a gauge parameter, to the flow of time in the classical histories that are predicted are misguided. We merely use the term ``time'' here because it appears in the same position as does the time in the non-relativistic quantum system described by the classical action \eqref{Sqp} with $t \rightarrow \tau$. (Indeed $\tau$ is a radial coordinate!) Similar comments appear in \textsc{J. B. Hartle} and \textsc{S. W. Hawking}, ``Path integral derivation of black hole radiance'', \underline{Phys. Rev.} \textbf{D13} (1976) 2188-2203.}. To construct a solution relevant to the no-boundary proposal, we performed a generalized Laplace transformation on the coordinates $(p_0,q_0)$, thereby transferring to a mixed representation $(p_0,\Pi_{q,0})$ on the $\tau = 0$ radial slice. Boundary conditions $\mathcal{B}$ on $p_0$ and $\Pi_{q,0}$ were then carefully chosen such that 1) they correspond to the behavior of a regular solution to the Einstein equations near $\tau = 0$, where by convention the spatial volume of the Ansatz \eqref{qpansatz} shrinks to zero, and 2) they ultimately lead to a normalizable wave function. Finally $K$ was integrated over a certain contour $\mathcal{C}$ in the complex lapse-plane (the closed contour around the origin in this case) to yield a specific solution to the WDW equation -- the NBWF $\Psi_\text{HH}$.

The algorithm described above gives only a particular contribution to the NBWF, one coming from a particular compact four-manifold. As we mentioned above, in the no-boundary sum over geometries (i.e. metrics \textit{and} manifolds) one should include contributions from other compact manifolds $\mathcal{M}$. The general programme outlined above can then be expressed schematically as (cf. Eq. (3.1) in \cite{DiazDorronsoro:2018wro})
\begin{equation} \label{NBWFminisuperspace}
\Psi_\text{HH}(y) = \sum_\mathcal{M} \int_\mathcal{C}\di N \hspace{-3mm} \overset{\hspace{3mm}x(1) = y}{\int_\mathcal{B}} \hspace{-3mm} \mathcal{D}x^\alpha \, \mathcal{D}\Pi_\alpha ~ e^{i S[x,\Pi;N]/\hbar} ~.
\end{equation}

The choice of lapse contour $\mathcal{C}$ and of boundary conditions $\mathcal{B}$ in the minisuperspace functional integral definition \eqref{NBWFminisuperspace} depend not only on the minisuperspace model but even on the parameterization (:= gauge choice for the lapse) of a specific model. We clarify this statement and illustrate it via two elementary examples in Appendix \ref{offshellsec}. Moreover, we do not expect the off-shell structure of the lapse function integrand in minisuperspace (such as the precise flow of the steepest descent contours) to contain factual information about quantum gravity at all -- it is clearly an artefact of the minisuperspace truncation. In particular, in a semiclassical evaluation of the path integral in Eq. \eqref{NBWFminisuperspace}, the prefactors $\mathcal{A}(y;I)$ (written in Eq. \eqref{NBWFsaddledef} below), depend on the fluctuations of all fields in the full theory and cannot be calculated in minisuperspace.\footnote{However to interpret the wave function as giving a probability distribution over classical histories in a particular minisuperspace model, we must associate to it a conserved current which does rely on an appropriate prefactor. We will return to this issue in detail in \S\ref{normsec}, and see that the constraint of current conservation leaves considerable freedom for the choice of prefactor.} The ingredients $(\mathcal{C},\mathcal{B})$ specifying the minisuperspace functional integral form \eqref{NBWFminisuperspace} of the NBWF -- and of any other wave function -- should therefore not be regarded as fundamental.

\subsection{NBWF as a collection of saddle points}
\noindent For these reasons we have recently argued that it is in many ways simpler and cleaner to specify the semiclassical NBWF of the universe without relying on a functional integral of any kind \cite{Halliwell:2018ejl}. After all the semiclassical approximation to the wave function is logically independent of any integral representation. The wave function would then be given as a sum of specific saddle points of the dynamical theory that satisfy conditions of regularity on geometry and field and which together yield a time neutral state that is normalizable in an appropriate inner product. This specifies a predictive framework of semiclassical quantum cosmology that is adequate to make probabilistic predictions.

\noindent In this approach the wave function can be written as
\begin{equation} \label{NBWFsaddledef}
	\Psi_\text{HH}(y) = \sum_{I \in \{(\mathcal{M},i)\}} \Psi_\text{HH}(y;I) = \sum_I \mathcal{A}(y;I) \, \exp \left( \frac{i}{\hbar} \bar{S}_0(y;I) \right) \left[ 1 + \mathcal{O}(\hbar) \right] \,,
\end{equation}
where the index $I$ runs over the semiclassical contributions from each compact four-manifold $\mathcal{M}$ which fills in the $S^3$, $\bar{S}_0(y;I)$ is the action of a regular (generally complex) solution to the Einstein equations on $\mathcal{M}$ which induces $y$ on $S^3$ and $\mathcal{A}$ is a prefactor. This definition is of course inherently semiclassical, and thus restricted, but since one does not expect the minisuperspace approximation to contain information about quantum gravity beyond the semiclassical limit this definition is essentially equivalent to the minisuperspace functional integral approach in its regime of applicability.\footnote{In \cite{DiazDorronsoro2017,DiazDorronsoro:2018wro} we went beyond the semiclassical approximation in minisuperspace to show that the NBWF has a fully consistent definition in terms of a minisuperspace path integral, addressing claims of the contrary made in \cite{FLT2,FLT3}.}

The choice of which saddle points to include as contributions to the semiclassical NBWF in this restricted definition replaces the contour choice. One general statement we can make is that the saddle points in Eq. \eqref{NBWFsaddledef} will appear in pairs \cite{Halliwell1988,HarHal1990}. These pairs have the same imaginary part of $\bar{S}_0$ but opposite real part, thus the exponential factors are each others complex conjugate. We will assume the same applies to the prefactors of the pairs, so that the semiclassical NBWF is real, in accordance with \cite{PhysRevD.28.2960}.\footnote{One might be inclined to say that saddles will appear in pairs of pairs, where in each pair the imaginary part of $\bar{S}_0$ is the same and the real part opposite, and between pairs the imaginary part of $\bar{S}_0$ is opposite \cite{Halliwell1988,HarHal1990,DiazDorronsoro2017,FLT1}. However one of these pairs can be eliminated by an appropriate choice of boundary conditions $\mathcal{B}$ on the instantons \cite{DiazDorronsoro:2018wro}.} Another general statement is that one expects only a few saddle points to be relevant in the semiclassical definition \eqref{NBWFsaddledef}.

\section{No-boundary saddle points on $\overline{B^4}$} \label{NUTsec}
\noindent In this section we discuss the contributions to the semiclassical NBWF arising from instantons that live on $\overline{B^4}$, a.k.a. the Taub-NUT-dS solutions \cite{carter1968}. (These were also discussed in detail in Ref. \cite{Daughton:1998aa}.) In this case both scale factors tend to zero at the origin of the spherical coordinate system, $\tau = 0$, i.e. $p_0 = 0 = q_0$. By definition we take the slice at $\tau = 1$ to represent the closed three-surface on which the argument of the wave function lives (here, an $S^3$). The solutions to the Einstein equations in the Ansatz \eqref{qpansatz2} are of the form \eqref{p-sol-general}-\eqref{q-sol-general} with $t \rightarrow \tau$. Judiciously choosing a sign for the square roots appearing in those solutions \cite{DiazDorronsoro:2018wro}\footnote{The choice of sign can be expressed by the choice $\Pi_{q,0} = (p_0 - \sqrt{p_0 p - N^2})/N \rightarrow -i$.}, the Hamiltonian constraint can be expressed as
\begin{equation} \label{saddlepointeq}
	\frac{2 i}{3} N_s^3 + \left( \frac{p}{3} - 4 \right) N_s^2 - p q = 0 \,,
\end{equation}
where $p$ and $q$ are the arguments of the wave function (we have dropped the index 1) and we have included an index $s$ on the lapse to indicate that $N_s$ is the particular, generally complex value of the lapse that causes the Einstein equations to be satisfied. By combining Eqns. \eqref{p-sol-general}-\eqref{q-sol-general}-\eqref{saddlepointeq} one can verify that the solution $(\bar{p}(\tau;N_s),\bar{q}(\tau;N_s))$ can be written in the form
\begin{align}
	\bar{p}(\rho;L) &= 4 \left( \rho^2 - L^2 \right) \,, \label{p-sol-literature} \\
	\bar{q}(\rho;L) &= \frac{16 L^2 \Delta(\rho)}{\rho^2 - L^2} \,, \label{q-sol-literature}
\end{align}
where
\begin{align}
	\Delta(\rho) &= (\rho-L)^2 - \frac{1}{3} \left( \rho + 3L \right) (\rho-L)^3 \,, \\
	\rho &= \frac{i N_s}{4L} \tau + L \,, \label{rhotaurelation} \\
	L &= \frac{N_s}{2 \sqrt{2iN_s - p}} \,, 
\end{align}
which often appears in the literature (e.g. \cite{Akbar:2003ed}). The no-boundary instantons on $\overline{B^4}$ in this minisuperspace are thus Taub-NUT-dS solutions with complex NUT parameter $L$.

\subsection{Saddles $N_s$ and on-shell action} \label{B4saddlessec}
\noindent Defining $x \equiv i N_s$, Eq. \eqref{saddlepointeq} is recast into the form
\begin{equation} \label{f(x)=0}
	f_3(x) \equiv x^3 + a x^2 + b = 0 \,,
\end{equation}
where
\begin{align}
	a &\equiv \frac{p}{2} - 6 \,, \\
	b &\equiv \frac{3 p q}{2} \,.
\end{align}
There is always a real solution $x$, and since $f_3(0) = b > 0$ it is always negative, corresponding to a solution $N_s$ lying on the positive imaginary axis. Further one can show that there are genuinely complex solutions -- a necessary requirement to predict classical spacetime (e.g. \cite{HarHal1990} and references therein) -- if and only if
\begin{equation} \label{largevolumecondition1}
	4 a^3 + 27 b > 0 ~,
\end{equation}
or, in terms of $p$ and $q$,
\begin{align} \label{largevolumecondition2}
q > \frac{p^2}{81}\left(\frac{12}{p}-1\right)^3 \,.
\end{align}
(If this condition is not satisfied, there are three purely imaginary solutions $N_s$, one on the positive imaginary axis and the two others on the negative imaginary axis.) In the regime \eqref{largevolumecondition2} the three solutions to Eq. \eqref{saddlepointeq} are given by
\begin{align}
	N_\text{IM} &= \frac{i}{3} \left( a + \frac{2^{1/3} a^2}{\Delta} + \frac{\Delta}{2^{1/3}}  \right) \,, \label{NIM} \\
	N_+ &= \frac{1}{2^{2/3} \sqrt{3}} \left( \frac{a^2}{\Delta} - \frac{\Delta}{2^{2/3}} \right) + \frac{i}{3} \left[ a - \frac{1}{2^{2/3}} \left( \frac{a^2}{\Delta} + \frac{\Delta}{2^{2/3}} \right) \right] \,, \label{Nplus} \\
	N_- &= - N_+^* \,, \label{Nminus}
\end{align}
where
\begin{equation}
	\Delta \equiv \left( 2 a^3 + 27 b - \sqrt{27} \sqrt{4 a^3 b + 27 b^2} \right)^{1/3} \,,
\end{equation}
and in this last formula we mean the positive real square and cubic roots, which are well-defined in the regime of interest \eqref{largevolumecondition1}. Note $N_+$ always has positive real part, and $N_-$ has a negative real part.

Using the saddle point equation \eqref{saddlepointeq}, one can show that the on-shell action\footnote{\label{bconditionsftn}That is, the action \eqref{Sqp} evaluated on the complex Taub-NUT-dS solutions \eqref{p-sol-literature}-\eqref{q-sol-literature}. No additional boundary terms are needed to make the variational problem (fixing $(p_1,q_1,p_0,q_0)$ or $(p_1,q_1,p_0,\Pi_{q,0})$) well-defined. We use notation consistent with our previous works \cite{DiazDorronsoro2017,DiazDorronsoro:2018wro}: $S_0(N)$ is the action evaluated on a solution to the second order EOM, Eqns. \eqref{p-EOM}-\eqref{q-EOM} here, while $\bar{S}_0$ is the on-shell action, $S_0(N_s)$.} is equal to
\begin{equation} \label{on-shellS}
	\bar{S}_0 \equiv S_0(N_s) = - i N_s^2 + 2 \left( 4 - \frac{p}{3} \right) N_s + i q \,.
\end{equation}
Since the saddle $N_\text{IM}$ in Eq. \eqref{NIM} does not give rise to classical spacetimes when it is used as a saddle point contribution to a wave function of the universe, we do not include it as a contribution to the NBWF. We do include the other two saddles $N_\pm$. Thus we have (cf. Eq. \eqref{NBWFsaddledef})
\begin{equation}
	\Psi_\text{HH}(p,q;\overline{B^4}) \approx \mathcal{A}_+(p,q;\overline{B^4}) \, e^{i \bar{S}_0^+ / \hbar} + \mathcal{A}_-(p,q;\overline{B^4}) \, e^{i \bar{S}_0^- / \hbar} \,,
\end{equation}
where $\bar{S}_0^\pm \equiv S_0(N_\pm)$. Note that indeed $\bar{S}_0^+ = S_0(-N_-^*) = - S_0(N_-)^* = - (\bar{S}_0^-)^*$, as we have discussed in \S\ref{NBWFsec}. Also, as discussed in that section, we will assume $\mathcal{A}_- = \mathcal{A}_+^*$, so
\begin{equation} \label{PsiHHB4}
	\Psi_\text{HH}(p,q;\overline{B^4}) \approx \mathcal{A}_+(p,q;\overline{B^4}) \, e^{i \bar{S}_0^+ / \hbar} + \text{(complex conjugate)} \,.
\end{equation}

\subsection{Classical histories} \label{B4classicalsec}
\begin{figure}[h!]
\centering
\includegraphics[width=0.5\textwidth]{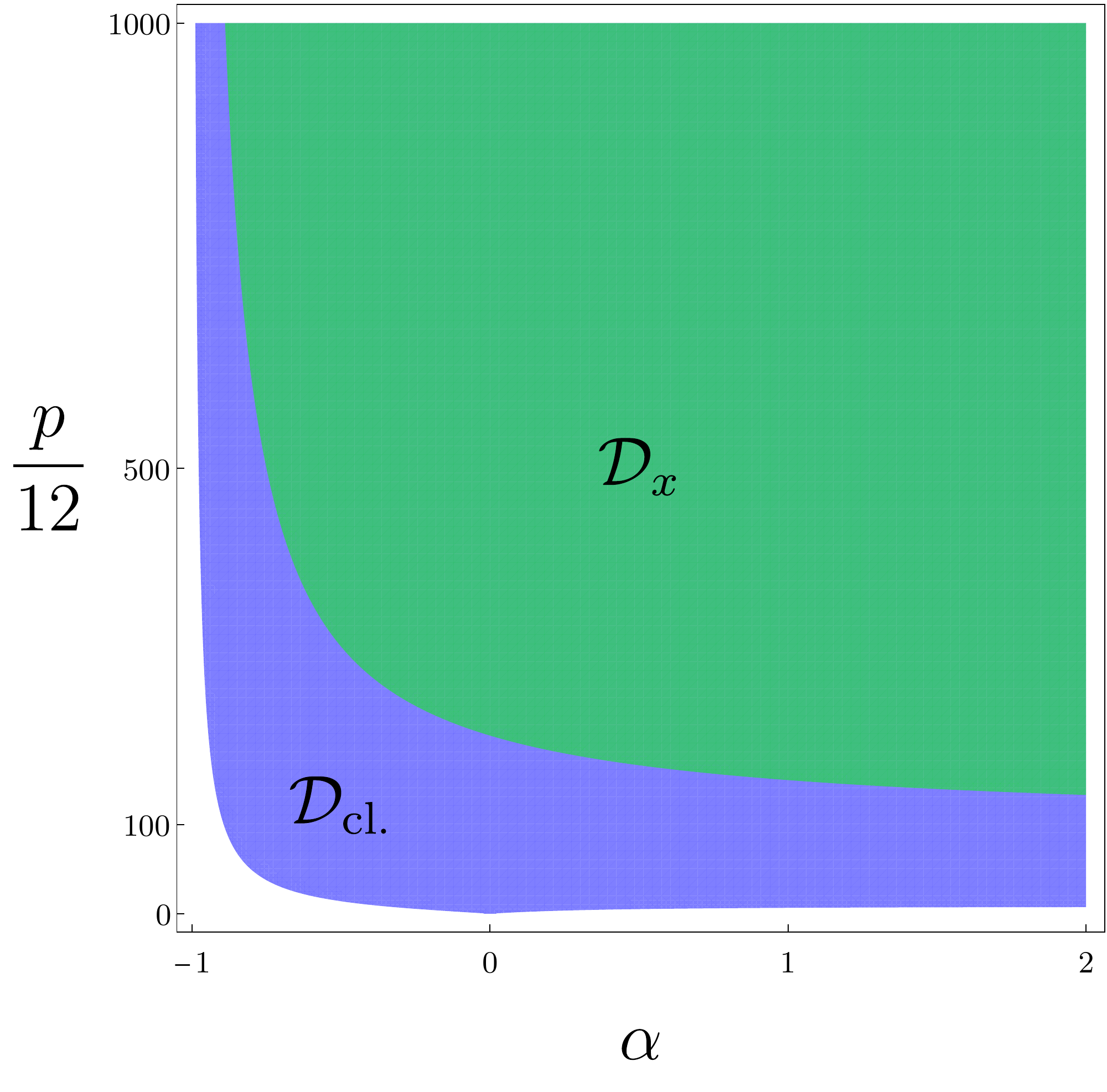}
\caption{\small The regime $\mathcal{D}_\text{cl.}$, where the classicality conditions \eqref{classicalityconditions} hold, is colored blue, and the regime $\mathcal{D}_x$, defined in Eq. \eqref{xdomain} and in which simple approximations to the quantities of interest in BB9 quantum cosmology exist, is colored green. We have $\mathcal{D}_x \subset \mathcal{D}_\text{cl.}$. Numerically we have taken $\gg 1 \times (\dots)$ to mean $> 100 \times (\dots)$.}
\label{DclDxfigure}
\end{figure}

\noindent In the regime where the wave function can be expressed as a rapidly varying phase times a slowly varying amplitude, it predicts strong correlations between configurations that lie along the same integral curve of the phase's argument \cite{Lapchinsky:1979fd,Banks:1984cw,Hartle1987,Halliwell:1987eu,Singh:1989ct}. In the form \eqref{PsiHHB4}, we require the classicality condition (e.g. \cite{HHH2008})
\begin{equation} \label{classicalityconditions}
	\left| \nabla \text{Re} \left( i \bar{S}_0^+ \right) \right| \ll \left| \nabla \text{Im} \left( i \bar{S}_0^+ \right) \right| ~~~ (\mathcal{D}_\text{cl.})
\end{equation}
to hold, where we have dubbed this regime $\mathcal{D}_\text{cl.}$.\footnote{The effects of the prefactor in this expression are usually neglected since they are subleading in $\hbar$.} (By $|v|$ we mean $\sqrt{|f^{\alpha \beta} v_\alpha v_\beta|}$, with $f^{\alpha \beta} = (f^{-1})_{\alpha \beta}$.) With equations \eqref{Nplus} and \eqref{on-shellS} we can evaluate this condition numerically, and the result shown in Figure \ref{DclDxfigure}.

Since the exact expressions are cumbersome, we will use a simplifying approximation to gain insight. The approximation is inspired by the necessary condition \eqref{largevolumecondition2} to have complex saddle points, which is surely satisfied if $p > 12$. In the approximation we will assume $p \gg 12$. This is not precise enough, however, since for any $p$ there is a non-classical regime for $\alpha$ close enough to $-1$. A convenient perturbation parameter that takes this into account is
\begin{equation} \label{xexpansionparameter}
	x \equiv \frac{12 q}{p^2} = \frac{12}{p (1+\alpha)} = \frac{12}{q (1+\alpha)^2} \,,
\end{equation}
where we recall the definition of the squashing parameter
\begin{equation}
	\alpha \equiv \frac{p}{q} - 1 \,.
\end{equation}
Trading $(p,q)$ for $(\alpha,x)$, and expanding in $x$, we find
\begin{equation} \label{classicalityratio}
\frac{\left| \nabla \text{Re} \left( i \bar{S}_0^+ \right) \right|}{\left| \nabla \text{Im} \left( i \bar{S}_0^+ \right) \right|} = \frac{\alpha x^{3/2}}{2} \left[ 1 + \frac{12\alpha - 15}{8} x + \mathcal{O}(x^2) \right] ~~ \text{as } x \rightarrow 0 \,.
\end{equation}
First, in order for the correction factor to be close to $1$, we require $x \ll 1$ when $\alpha$ is small and $\alpha x \ll 1$ when $\alpha$ is large.\footnote{The higher order corrections in Eq. \eqref{classicalityratio} display the same qualitative behavior as the first order correction term -- at small $\alpha$ the correction behaves as $x^n$ and at large $\alpha$ the correction behaves as $(\alpha x)^n$.} These conditions can be summarized in $(p,\alpha)$-coordinates by
\begin{equation} \label{xdomain}
	p \cdot \frac{\alpha + 1}{\alpha + 2} \gg 12 ~~~ (\mathcal{D}_x) \,,
\end{equation}
and we will call this regime $\mathcal{D}_x$. Then, in $\mathcal{D}_x$, the leading factor in Eq. \eqref{classicalityratio} is indeed small. This conclusion is also visualized in Figure \ref{DclDxfigure}.

\begin{figure}[t]
\centering
\includegraphics[width=0.75\textwidth]{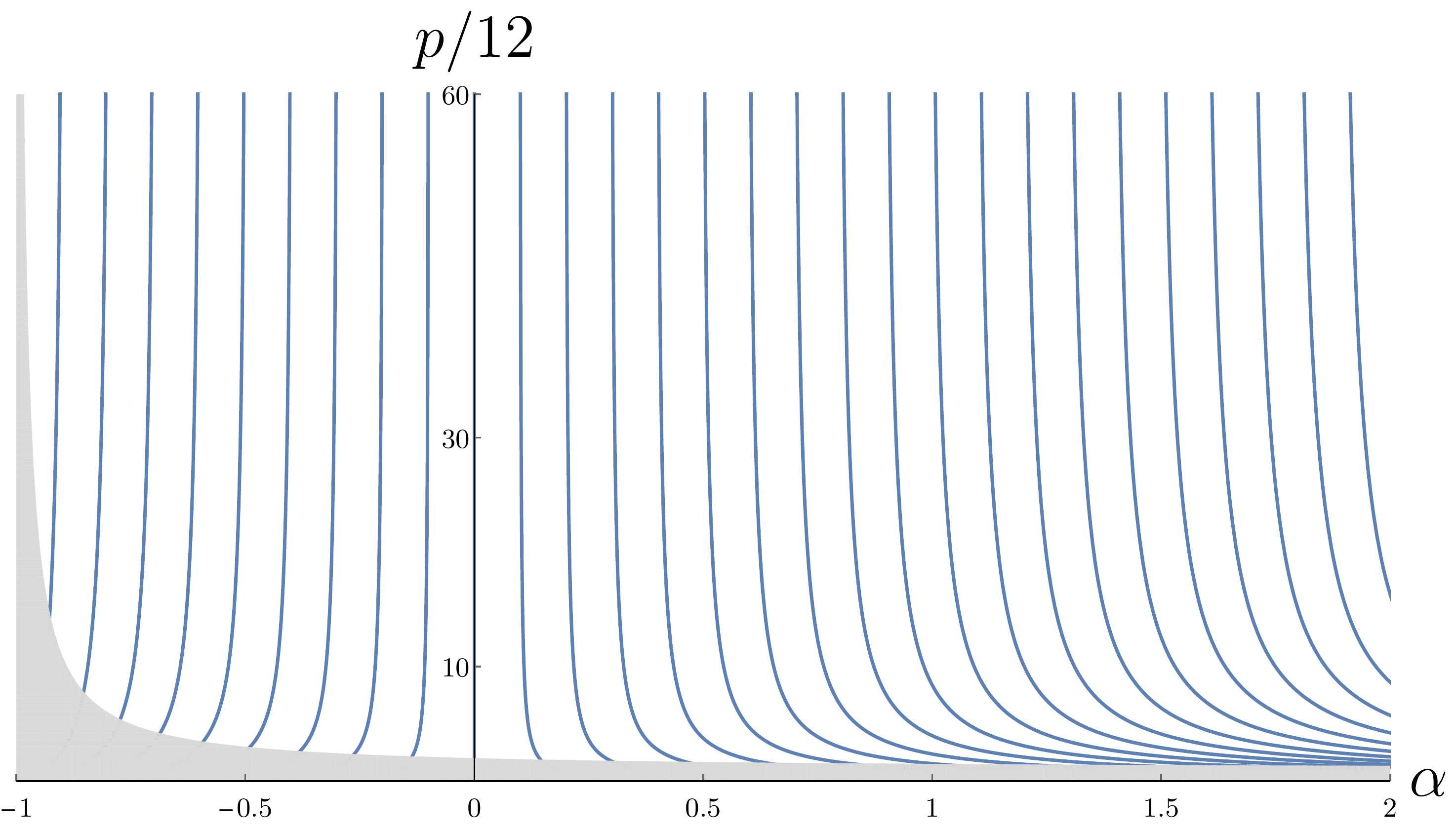}
\caption{\small The ensemble of classical, Lorentzian histories on $\mathbb{R}_\text{time} \times S^3$ predicted by the NBWF in the BB9 minisuperspace model, with four-metric given in Eq. \eqref{classicalmetric}. Each trajectory can be labeled by its asymptotic squashing parameter $\displaystyle\lim_{p \rightarrow \infty} \alpha(p)$. The grey region is where the classicality conditions do not hold. In this plot we took this region to be $p / 12 < (\alpha+2)/(\alpha+1)$, which is a good approximation to the complement of the actual regime $\mathcal{D}_\text{cl.}$ defined in Eq. \eqref{classicalityconditions}.}
\label{classicaltrajectories}
\end{figure}

\noindent The classical histories predicted by the $\overline{B^4}$ contributions to the NBWF are the integral curves of $\text{Re}(\bar{S}_0^+)$, i.e. they satisfy\footnote{Again the effects of the prefactor are usually neglected here. Strictly speaking a non-constant phase of the prefactor would add a subleading term in $\hbar$ to the right-hand side (RHS) of Eq. \eqref{classicalhistODE}, $\hbar \, \partial_\alpha \arg(\mathcal{A}_+)$, providing a quantum correction to the classical trajectories reminiscent of the de Broglie-Bohm approach to quantum mechanics.}
\begin{equation} \label{classicalhistODE}
	\Pi_\alpha \left( \frac{\di x^\beta}{\di \lambda} \right) = \partial_\alpha \text{Re}(\bar{S}_0^+) \,,
\end{equation}
where on the LHS the momenta are expressed in terms of the first derivatives of the minisuperspace coordinates according to Eqns. \eqref{pi-p}-\eqref{pi-q}. Because $\bar{S}_0^+$ satisfies the Hamilton-Jacobi equation \eqref{h0eq} and the classicality condition \eqref{classicalityconditions} holds in $\mathcal{D}_\text{cl.}$, $\text{Re}(\bar{S}_0^+)$ approximately satisfies the Hamilton-Jacobi equation in $\mathcal{D}_\text{cl.}$. This in turn implies that the solutions $\{ x(\lambda) \}$ to Eq. \eqref{classicalhistODE} approximately satisfy the Einstein equations in $\mathcal{D}_\text{cl.}$. These classical histories, which have the metric
\begin{equation} \label{classicalmetric}
	2 \pi^2 \, \di s^2 = -\frac{1 + \alpha(\lambda)}{p(\lambda)} \di \lambda^2 + \frac{p(\lambda)}{4} \left( \sigma_1^2 + \sigma_2^2 + \frac{1}{1 + \alpha(\lambda)} \sigma_3^2 \right) \,,
\end{equation}
are shown in the $(p,\alpha)$-plane in Figure \ref{classicaltrajectories}. For clarity we note again that these histories should not be confused with the no-boundary instantons \eqref{p-sol-literature}-\eqref{q-sol-literature}, which are complex and live on $\overline{B^4}$. The solution \eqref{classicalmetric} instead is a real, Lorentzian signature metric on the manifold $\mathbb{R}_\text{time} \times S^3$.

In $\mathcal{D}_x$ one can show that approximate solutions to Eq. \eqref{classicalhistODE} are given by the rays $p/q$ = constant, which is suggested in Figure \ref{classicaltrajectories}. This makes $\displaystyle\lim_{p \rightarrow \infty} \alpha(p)$ a good label for the classical trajectories. This observation also follows from the Hamilton-Jacobi equation since $\text{Im}(\bar{S}_0^+)$ is constant along the integral curves of $\text{Re}(\bar{S}_0^+)$, and it only depends on $\alpha$ to leading order in $\mathcal{D}_x$ (see Eq. \eqref{Splus} below).

\begin{figure}[b]
\includegraphics[width=0.60\textwidth]{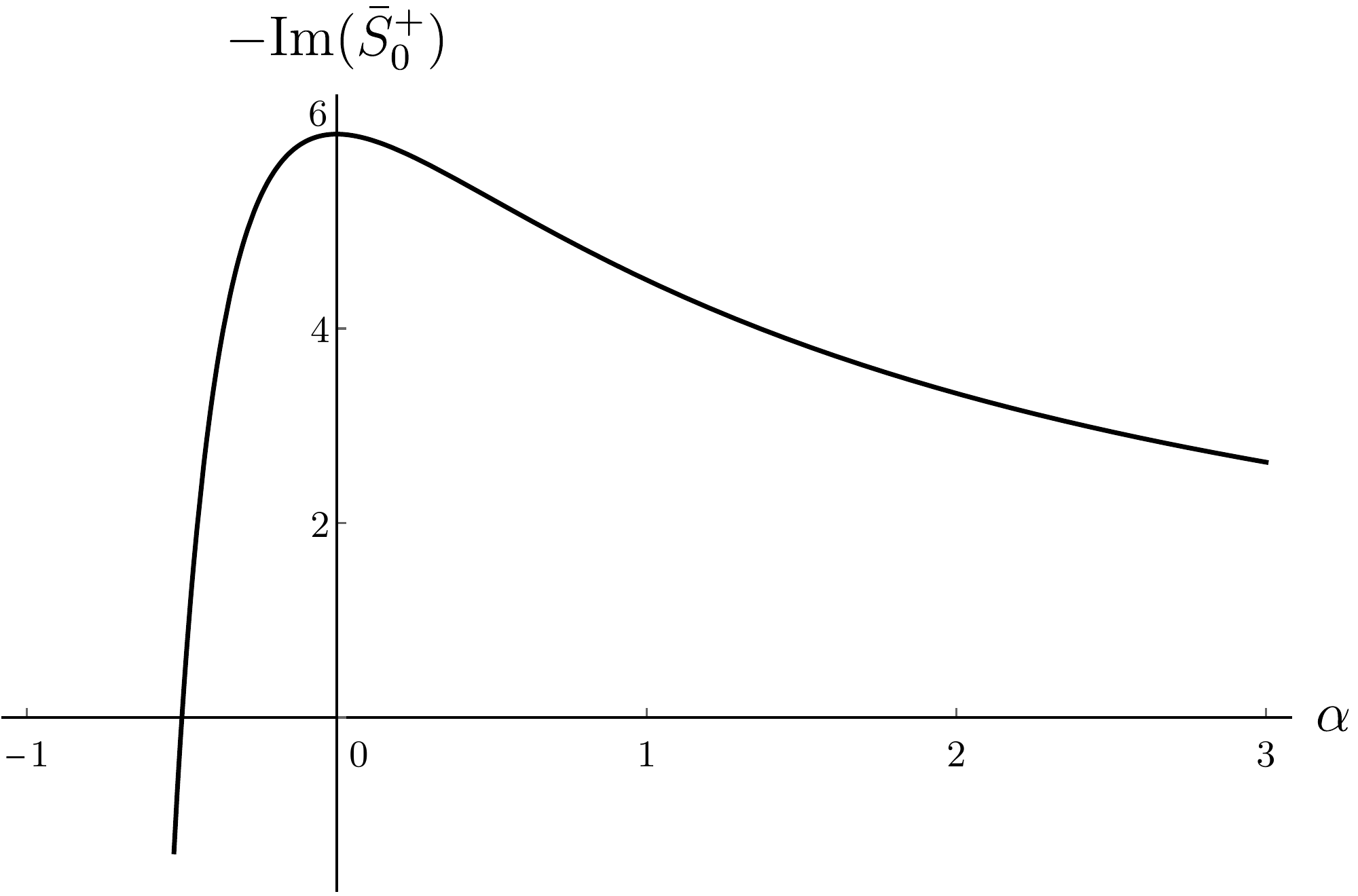}
\caption{\small The imaginary part of the on-shell action (approximately in $\mathcal{D}_x$ or at large volume) for the Taub-NUT-dS contribution to the NBWF as a function of the squashing parameter $\alpha \equiv p/q - 1$ of the $S^3$. To leading order in $\hbar$ and in $\mathcal{D}_x$ the magnitude of the wave function is determined by this function via $|\Psi_\text{HH}(p,\alpha)| \sim e^{-\text{Im}(\bar{S}_0^+)/\hbar}$.}
\label{NUTaction}
\end{figure}

Using the notation of the previous section, we have
\begin{align}
	N_\pm &= \pm \sqrt{3q} \left[ 1 + \frac{4\alpha - 1}{8}x + \mathcal{O}\left( x^2 \right) \right] - \frac{3 i}{1+\alpha} \left[ 1 + 2\alpha x + \mathcal{O}\left( x^2 \right) \right] \,, \label{Nplusx} \\
	\bar{S}_0^\pm &= \mp \frac{2 \, p^{3/2}}{\sqrt{3(1+\alpha)}} \left[ 1 - \frac{4\alpha + 3}{8}x + \mathcal{O}\left( x^2 \right) \right] - \frac{6(1+2\alpha)i}{(1+\alpha)^2} \left[ 1 + \frac{2\alpha}{2\alpha + 1} \alpha x + \mathcal{O}\left( x^2 \right) \right] \label{Splus} \,.
\end{align}
So, from Eq. \eqref{PsiHHB4},
\begin{equation} \label{PsiHHB4approx}
	\Psi_\text{HH}(p,\alpha;\overline{B^4}) \approx |\mathcal{A}_+| \, \exp\left( \frac{6}{\hbar \Lambda} \frac{(1+2\alpha)}{(1+\alpha)^2} \right) \, \cos\left( \frac{2}{\hbar \Lambda} \frac{\left( \Lambda p \right)^{3/2}}{\sqrt{3(1+\alpha)}} + \arg\left( \mathcal{A}_+ \right) \right) \,,
\end{equation}
where we have reinstated $\Lambda$ and neglected an overall factor of $2$, and the approximation is valid in $\mathcal{D}_x$. The argument of the exponential function in Eq. \eqref{PsiHHB4approx}, $-\text{Im}(\bar{S}_0^+)$, is plotted to leading order in $\mathcal{D}_x$ in Figure \ref{NUTaction}. We discuss the prefactor $\mathcal{A}_+$, whose properties at large $\alpha$ in $\mathcal{D}_x$ determine whether the contribution is normalizable, in \S\ref{normsec}.

\newpage

\clearpage
\newpage
\section{No-boundary saddle points on $\mathbb{C}\text{P}^2 \setminus B^4$} \label{Boltsec}
\noindent In this section we discuss the contributions to the semiclassical NBWF arising from instantons that live on the compact four-manifold $\mathbb{C}\text{P}^2 \setminus B^4$, a.k.a. the Taub-Bolt-dS solutions \cite{Eguchi:1978xp}. In this case the $S^1$ fiber shrinks to zero size at $\tau = 0$ while the $S^2$ remains at a finite size there. A regular metric satisfies $q_0 = 0, \Pi_{p,0} = -i$.\footnote{Recall that the NUT instantons were obtained by the boundary conditions $p_0 = 0,\Pi_{q,0} = -i$ \bibnote{In the minisuperspace path integral approach of Eq. \eqref{NBWFminisuperspace}, the boundary conditions $\mathcal{B} = \{ p_0 = 0, q_0 = 0 \}$ are more subtle to deal with due to an ambiguity in the propagator of the system \cite{DiazDorronsoro:2018wro}. This ambiguity is resolved by a choice of sign for the initial momentum of one of the scale factors. The all-momentum boundary conditions $\mathcal{B} = \{ \Pi_{p,0} = -2i, \Pi_{q,0} = -i \}$ proposed in \cite{Louko1988} can also be considered, but an off-shell minisuperspace analysis \`a la \cite{Halliwell1990} is not possible in that case because the second order EOM generally have no solutions for these boundary conditions (the exceptions being at the on-shell points $N = N_s$). This curiosity can be traced back to the existence of a conserved quantity in this system, $p / (1+\Pi_q^2)$.}.} In the general solutions \eqref{p-sol-general}-\eqref{q-sol-general} to the second order EOM, the regularity condition on the momentum conjugate to $p$ at $\tau = 0$ determines the (off-shell) size of the two-sphere at $\tau = 0$ in terms of the lapse parameter:
\begin{equation} \label{p0Bolt}
	p_0 = \frac{4N^6 + 24iN^5 + (p^2-36)N^4 - 6p^2 q N^2 + 9 p^2 q^2}{4 N^2 (N + 3i)^2 p} \,,
\end{equation}
where again $p$ and $q$ are the arguments of the wave function. This time the Hamiltonian constraint reads
\begin{equation} \label{BoltHamiltonianconstraint}
	\frac{3 (N_s - i q) (N_s - 3 i q)p}{N_s^3 (N_s + 3 i)} - \frac{9 i (N_s - i q)^2 p}{N_s^3 (N_s + 3 i)^2} - \frac{4 N_s^2}{p} + \frac{6 p q}{N_s^2} - \frac{8 i N_s}{p} - p + 16 = 0 \,,
\end{equation}
which generally has seven solutions.\footnote{Note that the on-shell size of the two-sphere at $\tau = 0$, $p_0(N_s)$, is a non-constant function of the argument of the wave function $(p,q)$. This implies it is impossible \cite{Halliwell1990} to obtain the Taub-Bolt-dS contribution to the NBWF by imposing Dirichlet boundary conditions on both minisuperspace coordinates at $\tau = 0$.} As in the NUT case these solutions have the property that if $N_s$ is a solution, so too is $-N_s^*$, i.e. the sign of the real part is flipped.

As we did for the no-boundary NUT-type solutions, we can also write the no-boundary Bolt-type solutions in a more conventional form. For this we can copy Eqns. \eqref{p-sol-literature} and \eqref{q-sol-literature}, but now
\begin{align}
	\Delta(\rho) &= (1+2L^2)(\rho^2-\rho_\text{B}^2) - (\rho^4-\rho_\text{B}^4)/3 - 2 M (\rho-\rho_\text{B}) \,, \\
	2 M \rho_\text{B} &= (1+2L^2) \rho_\text{B}^2 - \rho_\text{B}^4/3 + L^2(1+L^2) \,, \\
	\rho_\text{B} &= \frac{-1 \pm \sqrt{1+16L^2(1+L^2)}}{4L} \,, \\
	\rho &= \frac{i N_s}{4L} \tau + \rho_\text{B} \,,
\end{align}
and $L$ is related to $N_s$ (and thus to the arguments of the wave function $p$ and $q$) through the equation
\begin{equation}
	4(\rho_\text{B}^2 - L^2) = p_0(N_s) \,,
\end{equation}
where $p_0(N_s)$ is given in Eq. \eqref{p0Bolt}. The no-boundary instantons on $\mathbb{C}\text{P}^2 \setminus B^4$ in this minisuperspace are thus Taub-Bolt-dS solutions with a complex Bolt parameter $L$ that depends on the arguments on the wave function.

\subsection{Saddles $N_s$ and on-shell action} \label{Boltsaddlessec}
\noindent Contrary to the NUT case, an analytic expression for the solutions of Eq. \eqref{BoltHamiltonianconstraint} is not evident to write down and use. We can however, as in \S\ref{B4classicalsec}, restrict ourselves to a certain regime $\tilde{\mathcal{D}}_x$ where convenient approximations exist also in this case. The regime $\tilde{\mathcal{D}}_x$ is defined by
\begin{equation} \label{xdomain2}
	p \cdot \frac{\alpha + 1}{(\alpha + 2)^2} \gg 12 ~~~ (\tilde{\mathcal{D}}_x) \,,
\end{equation}
which is more restrictive than the regime $\mathcal{D}_x$ (defined in Eq. \eqref{xdomain}) at large $\alpha$. The saddle points of interest can be determined by plugging the Ansatz
\begin{align} \label{NBolt}
	N_\pm^\text{Bolt} \approx \pm \sqrt{A q} + i f_4(\alpha)
\end{align}
into Eq. \eqref{BoltHamiltonianconstraint} (where $\approx$ means valid up to small corrections in $\tilde{\mathcal{D}}_x$) and demanding that the equation be satisfied as well as it can, i.e. perturbatively in $x$, by choosing $A$ and $f_4$ appropriately. We find \begin{align}
	A &= 3 \,, \\
	f_4(\alpha) = f_\pm(\alpha) &= \left\{ \begin{array}{ll}
		-1 & \mbox{~~if } \alpha \in (5 - 3\sqrt{3} , 5 + 3\sqrt{3}) \\
        \displaystyle\frac{-1-\alpha\pm\sqrt{\alpha^2-10\alpha-2}}{1+\alpha} & \mbox{~~if } \alpha \in (-1,5-3\sqrt{3}] \, \cup \, [ 5 + 3\sqrt{3} ,\infty)
    \end{array} \right. \label{fpm}
\end{align}
must be taken. The two possibilities for $f_4$, $f_\pm$, exist for both saddles $N_\pm^\text{Bolt}$ independently. (For clarity by $N_\pm^\text{Bolt}$ we mean, as in the NUT discussion in \S\ref{B4saddlessec}, the saddle points with positive/negative real part, which are further distinguished in this case by a binary choice of $f_+$ or $f_-$.) So by this method we have found four of the seven solutions to Eq. \eqref{BoltHamiltonianconstraint} approximately in $\tilde{\mathcal{D}}_x$, and they are all potentially interesting since they are not purely imaginary. One can show that the other three solutions of Eq. \eqref{BoltHamiltonianconstraint} \textit{are} purely imaginary in $\tilde{\mathcal{D}}_x$ and so, as we have mentioned in \S\ref{B4saddlessec}, are not of interest in quantum cosmology since they lead to a purely real contribution to the wave function at large volume which does not predict classical spacetime.

The action of the Taub-Bolt-dS solutions is given by Eq. \eqref{pqactionH}, evaluated on the solutions \eqref{p-sol-general}-\eqref{q-sol-general} with $p_0$ as in Eq. \eqref{p0Bolt}, plus the boundary term $\Pi_{p,0} \, p_0$. This term is required (and does not vanish, in contrast to the analogous term $\Pi_{q,0} q_0 = 0$ in the NUT discussion) to make the variational problem of fixing $\Pi_p$ and $q$ at $\tau = 0$ and $p$ and $q$ at $\tau = 1$ well-defined. In terms of the lapse -- the only remaining parameter at this stage -- one obtains
\begin{align} \label{S0bolt}
	-S_\text{Bolt}(N) &= \frac{3 N \left[ 6 p^2 q + i N \left(4 N^2 + 16 i N p + p^2 \right)\right] + N^3 \left( 4 N^2 + 3 p^2 \right) + \frac{9 p^2 (N - i q)^2}{N+3 i}}{12 N^2 p} \,.
\end{align}
For the on-shell action (i.e. Eq. \eqref{S0bolt} evaluated on \eqref{NBolt}) one obtains
\begin{equation} \label{Boltonshellaction}
	S(N_\pm^\text{Bolt}; f_\pm) \approx \mp \frac{2 \, p^{3/2}}{\sqrt{3(1+\alpha)}}  + i \times \frac{27 + f_\pm(\alpha) \left[ 9(1+4\alpha) + (1+\alpha)^2 f_\pm(\alpha) \left( 3 + f_\pm(\alpha) \right) \right]}{9(1+\alpha)} \,.
\end{equation}
For a detailed discussion of the approximations written above we refer the reader to Appendix \ref{A1}. For a plot of the imaginary part of Eq. \eqref{Boltonshellaction} for both contributions, see Figure \ref{BoltactionsFIG} in Appendix \ref{A2}.

\subsection{Classical histories} \label{Boltclassicalsec}
\noindent To leading order in $\tilde{\mathcal{D}}_x$ (and in fact to first subleading order as well), the real part of the on-shell action for the contributions to the NBWF from the manifold $\mathbb{C}\text{P}^2 \setminus B^4$, Eq. \eqref{Boltonshellaction}, is identical to its analog in the $\overline{B^4}$ case, Eq. \eqref{Splus} (valid in the regime $\mathcal{D}_x$, and $\tilde{\mathcal{D}}_x \subset \mathcal{D}_x$). This implies that the classical histories on which this branch of the wave function has support are, at large volume, the same as those discussed in \S\ref{B4classicalsec}. That is, they are curves of constant $\alpha$. At smaller volume but still in $\mathcal{D}_\text{cl.}$, we expect the classical trajectories in both ensembles to differ \bibnote{We have not discussed exactly what $\mathcal{D}_\text{cl.}$ is for the $\mathbb{C}\text{P}^2 \setminus B^4$ contributions to the NBWF. This is challenging because Eq. \eqref{BoltHamiltonianconstraint} is hard to solve. Since our main concern in this paper is with the properties of the classical spacetimes predicted by the NBWF at large volume, we leave the question of what happens at smaller volume to future work. We do note that the regime $\tilde{\mathcal{D}}_x$, where the approximations we have written are valid, lies completely in $\mathcal{D}_\text{cl.}(\mathbb{C}\text{P}^2 \setminus B^4)$ (the analogous statement in the $\overline{B^4}$ discussion is true too, see \S\ref{B4classicalsec}). One can show this from Eq. \eqref{Boltonshellaction}.}.

\subsection{Choice of saddles} \label{Boltnormalizationsec}
\noindent Above we have found two types instantons, which we distinguished by the choice $f_\pm$. Since near $\alpha = -1$ (and in $\tilde{\mathcal{D}}_x$), $\text{Re}[i S(N_\pm^\text{Bolt}; f_-)]$ diverges to negative infinity and $\text{Re}[i S(N_\pm^\text{Bolt}; f_+)]$ diverges to positive infinity, only the solutions with the $f_-$ choice lead to normalizable semiclassical contributions to the wave function. So only the solutions $(N_\pm^\text{Bolt}; f_-)$ can be included as contributions to the semiclassical no-boundary state near $\alpha = -1$. At large $\alpha$, $\text{Re}[i S(N_\pm^\text{Bolt}; f_-)]$ behaves as $-\alpha$ and thus surely corresponds to a normalizable contribution, while $\text{Re}[i S(N_\pm^\text{Bolt}; f_+)]$ tends to zero as $1/\alpha$. In principle either solution could be chosen to contribute to the semiclassical NBWF. The former would be automatically normalizable, while the normalizability of the latter would depend on the details of the prefactor as does the contribution from the four-disk (see \S\ref{normsec} where we discuss the normalization of the semiclassical NBWF in detail). For definiteness we will assume only the $(N_\pm^\text{Bolt}; f_-)$ contribution is relevant at large $\alpha$. So, approximately in $\tilde{\mathcal{D}}_x$ and in the semiclassical limit,
\begin{equation} \label{PsiHHBoltapprox}
	\Psi_\text{HH}(p,\alpha;\mathbb{C}\text{P}^2 \setminus B^4) \approx |\tilde{\mathcal{A}}_+| \, \exp\left( - \frac{\text{Im}\left[S(N_+^\text{Bolt}; f_-)\right]}{\hbar \Lambda} \right) \, \cos\left( \frac{2}{\hbar \Lambda} \frac{\left( \Lambda p \right)^{3/2}}{\sqrt{3(1+\alpha)}} + \arg\left( \tilde{\mathcal{A}}_+ \right) \right) \,,
\end{equation}
with $S(N_+^\text{Bolt}; f_-)$ given in Eq. \eqref{Boltonshellaction}. The undetermined prefactor $\tilde{\mathcal{A}}_+$ is discussed in \S\ref{normsec}.

Some readers may have noticed that based on the information we have given so far, we in fact do not know if we have the freedom -- even in principle -- to choose one type of saddle point near $\alpha = -1$ and the other type at large $\alpha$ (both for $x \in \tilde{\mathcal{D}}_x$) as we suggested above. Based on what we have discussed the choices could be mutually exclusive because the imaginary parts of the actions written in Eq. \eqref{Boltonshellaction} receive corrections at finite $x$. If the corrections would be such that at any finite $x$, the imaginary parts of the on-shell action of the two types of solutions never intersect each other at any $\alpha$, the saddles could never exchange dominance and the choice of saddle near $\alpha = -1$ would be tied to the choice at large $\alpha$. In other words a phase transition would be impossible.

By taking into account the corrections at finite $x$ we have verified that such phase transitions between the solutions on $\mathbb{C}\text{P}^2 \setminus B^4$ \textit{are} possible. We refer the reader to Appendix \ref{A2} for a detailed discussion. Which choices of phase are made at particular values of $\alpha$, or which saddles are picked up, depends on the contour of integration in a more detailed definition of the NBWF.\footnote{In minisuperspace, the choice of dominant saddles we have made above (i.e. the $(N_\pm^\text{Bolt}; f_-)$ saddles at all $\alpha$) can be realized by integrating the lapse in Eq. \eqref{NBWFminisuperspace}, for the $\mathbb{C}\text{P}^2 \setminus B^4$ manifold, over the contour $\mathcal{C} = \mathbb{R} - (1+\varepsilon)i$ with $0 < \varepsilon < 2$. This can be shown by a steepest-descent analysis of the integral $\int_\mathcal{C} \di N \, e^{iS_\text{Bolt}/\hbar}$ with $S_\text{Bolt}$ given in Eq. \eqref{S0bolt}.}

\clearpage
\newpage
\section{Normalization} \label{normsec}
\noindent We come now to the issue of normalization of the wave functions we have calculated and the closely related issue of the construction of properly normalized probabilities. As indicated earlier, we are primarily interested in the saddle point version of the no-boundary proposal (SPNB proposal), the strength of which is that there are reasonable grounds for believing that the wave functions obtained in this approach are approximations to the wave functions in a full quantum theory of gravity. However, in the absence of such a theory, we are unable to say very much about the form of the wave function beyond the lowest order saddle point approximation, e.g. about the prefactors. This is significant since normalizability of WKB-type wave functions typically depends on the asymptotic fall-off of the prefactors. So from the perspective of the SPNB proposal, it is not possible to say much about normalizability, other than imposing some sensible broad requirements, and in particular asking that the semiclassical wave functions do not grow asymptotically \cite{Halliwell:2018ejl}. This sort of heuristic approach is clearly sufficient to rule out the undesirable saddle points of the type encountered in Ref. \cite{FLT2}. (In Ref. \cite{FLT3} a growing contribution from some off-shell structure in the minisuperspace path integral is identified, but we show in Appendix \ref{FLTsec} that the calculation of Ref. \cite{FLT3} is inconsistent and that such a contribution does not exist.)

However, the investigations in this paper are not completely limited to the SPNB proposal -- we are also interested in exploring some aspects of fully quantized minisuperspace models. These have the feature that their normalization can be thoroughly explored so it is clearly of interest to do this, if only to get some idea of how it might fail. Of course in minisuperspace models, an infinite number of modes are simply set to zero and there is no obvious sense in which such models could be approximations to a full quantum theory of gravity, except in the lowest order semiclassical approximation.  Hence normalizability in the minisuperspace context is unlikely to say anything about normalizability in a full theory. However, it does seem reasonable to assert that the {\it absence} of normalizability for a given wave function in a minisuperspace model indicates that there is no corresponding normalizable wave function in a full theory. This means that minisuperspace normalizability could be used as a criterion for ruling out certain wave functions.

A Hilbert space structure for the solutions to the WDW equation for minisuperspace models can be defined using the induced inner product. (See for example Ref. \cite{PhysRevD.80.124032}.) Loosely, one requires that the usual Schr\"odinger inner product between a pair of eigenstates of the WDW operator exists and is proportional to $\delta(E-E')$, where $E$ and $E'$ denote the eigenvalues. This will already eliminate certain solutions to the WDW equation if this inner product does not exist, e.g. if the wave functions grow exponentially at large arguments. One can then use the states belonging to the Hilbert space to construct interesting probabilities by finding operators which commute with $H$ and correspond to physically relevant questions concerning cosmological histories. 

This general structure has been shown, at length \cite{PhysRevD.80.124032}, to boil down to fluxes across surfaces of co-dimension one in minisuperspace, which is the more commonly-employed heuristic interpretation of minisuperspace wave functions, where the flux $\Phi_{\Sigma}(J)$ across surface $\Sigma$ is defined in terms of the conserved current 
\begin{equation} \label{WKBcurrent}
	J = - \frac{i \hbar}{2} \left( \Psi^* \nabla \Psi - \Psi \, \nabla \Psi^* \right) \,,
\end{equation}
by
\begin{equation}
\Phi_{\Sigma} (J) = \int_{\Sigma} J \cdot \di \Sigma \,,
\end{equation}
where $\di \Sigma$ denotes a normal surface element \cite{Vilenkin:1988yd,Halliwell:1990uy}. Because the original wave functions are normalizable in the induced inner product, the flux across a surface remains well-defined even for infinite surfaces, which is an important property for the normalization of the probabilities.

For a real wave function such as the NBWF, which consists of a sum of complex conjugate saddle points, the current \eqref{WKBcurrent} is identically zero. One usually proceeds by taking the (semiclassical) current to be constructed out of ``half'' of the (semiclassical) wave function, i.e. $\Psi \approx \mathcal{A} \, e^{i \bar{S}_0 / \hbar}$ in our case. The argument for doing this is that the coarse-graining involved in computing a flux of interest causes the interference between two different WKB wave functions to average out in the flux, hence it is reasonable to consider the probability for each WKB wave function separately \cite{PhysRevD.80.124032}.

One can also have a more general sum of saddle points, not necessarily complex conjugates of each other, which is the case in the BB9 model. In a sufficiently small regime of configuration space usually only a single kind of saddle point exponentially dominates the behavior of the wave function, however, so to good approximation we may restrict our attention to this contribution only and construct the current from it alone.\footnote{Saddle points may exchange dominance as one explores the superspace, i.e. a phase transition may occur. We return to this interesting phenomenon in \S\ref{PsiHHsec} (see also Appendix \ref{A2}).} For $\Psi \approx \mathcal{A} \, e^{i \bar{S}_0 / \hbar}$ one obtains
\begin{equation} \label{sccurrent1}
	J^{\text{cl.}}(\Psi) \approx |\mathcal{A}|^2 \, e^{-2 \text{Im}(\bar{S}_0)/\hbar} \, \nabla \left[ \text{Re}(\bar{S}_0) + \hbar \arg\left( \mathcal{A} \right) \right] \,,
\end{equation}
which is conserved to next-to-next-to-leading order in $\hbar$ due to Eqns. \eqref{h0eq}-\eqref{h1eq}.\footnote{We remark a last time that the second term in Eq. \eqref{sccurrent1} is usually neglected. The first term is exactly conserved.} \textit{Relative} probabilities are then defined by ratios of flux of $J^{\text{cl.}}$ across certain surfaces, according to
\begin{equation} \label{relativeprobs}
	\text{Prob}(\sigma \, | \, \Sigma) = \frac{\Phi_\sigma(J^{\text{cl.}})}{\Phi_\Sigma(J^{\text{cl.}})} \,,
\end{equation}
where $\sigma \subset \Sigma$ and $\Phi_\mathcal{S}(J^{\text{cl.}})$ is the flux of the current $J^{\text{cl.}}$ across the (co-dimension one) surface $\mathcal{S}$ in minisuperspace. This ratio is interpreted as the probability that a history in the classical ensemble passes through $\sigma$ given that it passes through $\Sigma$. This is (approximately) well-defined for any finite $\Sigma$ since $J^{\text{cl.}}$ is (approximately) conserved.

An \textit{absolute} probability that a classical history passes through a surface $\sigma$ can be defined via Eq. \eqref{relativeprobs} if $\Phi_\Sigma(J^{\text{cl.}})$ is finite for a surface $\Sigma$ that slices through \textit{all} the classical trajectories. As indicated above we expect it to be finite on general grounds but it is useful to see how this can be made to work for a WKB wave function $\Psi \approx \mathcal{A} \, e^{i \bar{S}_0 / \hbar}$. For such states the total flux will probably be finite if $\text{Im}(\bar{S}_0)$ tends to $+\infty$ in all directions on $\Sigma$, and will probably be infinite if $\text{Im}(\bar{S}_0)$ tends to $-\infty$ in any direction on $\Sigma$. With the semiclassical contribution to the NBWF in the BB9 model from the $\mathbb{C}\text{P}^2 \setminus B^4$ topology, Eq. \eqref{PsiHHBoltapprox}, we are in the former scenario and thus this contribution is normalizable over the set of all classical trajectories essentially independently of the properties of the prefactor $\tilde{\mathcal{A}}_+$. The case in between is when $\text{Im}(\bar{S}_0)$ tends to zero along some directions on $\Sigma$ and to $+\infty$ in all others. This is the case we are in with the semiclassical contribution from $\overline{B^4}$, \eqref{PsiHHB4approx}: $\text{Im}(\bar{S}_0^+) \rightarrow +\infty$ as $\alpha \rightarrow -1$ and $\text{Im}(\bar{S}_0^+) \rightarrow 0$ as $\alpha \rightarrow +\infty$. In such cases the normalization of $\Psi$ across all classical histories may depend crucially on the asymptotic behavior of the prefactor, $\mathcal{A}_+$ here. 

We now determine the most general prefactor $\mathcal{A}_+$, approximately in $\mathcal{D}_x$, that would allow one to define an absolute probability distribution over all classical histories via the semiclassical current $J^{\text{cl.}}(\Psi_\text{HH}(\overline{B^4}))$. The conservation equation \eqref{h1eq} for $\mathcal{A}_+$ reads
\begin{equation}
	\left( \frac{2}{\sqrt{x}} + i \alpha x \right) \partial_p \mathcal{A}_+ + \frac{1}{1+\alpha} \left( \frac{2}{\sqrt{x}} - 2 i \alpha x \right) \partial_q \mathcal{A}_+ + \frac{1+\alpha}{8} \left[ \sqrt{x} + i (1-\alpha) x^2 \right] \mathcal{A}_+ = 0 \,,
\end{equation}
where we have kept only the leading terms in $\mathcal{D}_x$ in the coefficient functions. From this equation it follows that, to leading order in $\mathcal{D}_x$,
\begin{equation}
	p \, \partial_p |\mathcal{A}_+| + q \, \partial_q |\mathcal{A}_+| + \frac{3}{4} |\mathcal{A}_+| = 0 \,,
\end{equation}
which has the general solution
\begin{equation} \label{generalprefactor}
	|\mathcal{A}_+| = q^{-3/4} \, f_5(\alpha) \,,
\end{equation}
with $f_5$ an arbitrary real-valued function. It is then possible to show that the flux of $J^{\text{cl.}}(\Psi_\text{HH}(\overline{B^4}))$ across an infinite surface that intersects all classical histories is finite if
\begin{equation} \label{finiteflux}
	\int_{-1}^\infty \di \alpha \, f_5(\alpha)^2 < \infty \,.
\end{equation}

An alternative way of deriving the result \eqref{generalprefactor} is via the general solution \eqref{WDWgeneralsol} to the WDW equation. One can evaluate the integral in the semiclassical limit, assuming $p \gg 1$ and $|\Pi^s_q| \gg 1$, where the superscript $s$ denotes saddle point values. This last assumption turns out to be self-consistent in $\mathcal{D}_x$, and the stationary phase approximation  shows that the leading order behavior for the prefactor in $\mathcal{D}_x$ is as in Eq. \eqref{generalprefactor}.

With the prefactor as in Eqns. \eqref{generalprefactor}-\eqref{finiteflux}, the contribution \eqref{PsiHHB4approx} to the semiclassical NBWF arising from geometries on the four-disk is normalizable in the sense described above (even though as $\alpha \rightarrow \infty$, at constant $p$, the exponential part tends to a non-zero constant). The contribution from $\mathbb{C}\text{P}^2 \setminus B^4$ is normalizable independently of the behavior of the prefactor.

At this stage a comment on our previous work \cite{DiazDorronsoro:2018wro} is in order, in which we wrote down an exact solution to the WDW equation in the BB9 model based on a minisuperspace path integral over geometries on the four-disk. This solution has $f_5(\alpha) \equiv 1$ in the semiclassical limit and so is not normalizable over all classical trajectories of the model in the sense explained in this section. This does not invalidate the main point of \cite{DiazDorronsoro:2018wro}, however, which was to illustrate that no sources of enhanced perturbations of the type purportedly found in \cite{FLT3} appear when the NBWF is carefully defined in terms of a minisuperspace path integral.

Furthermore, we have stressed in this paper that the prefactor for a NBWF cannot be fully fixed by a minisuperspace analysis. This means that we have the freedom to adjust $f_5(\alpha)$ to produce a NBWF which yields properly normalized probabilities. So for example, we may take $f_5(\alpha)$ to be approximately $1$ for small $\alpha$ but which decays sufficiently fast for large $\alpha$ so that Eq. \eqref{finiteflux} is satisfied. In this way we confirm explicitly the expected normalization properties stated both here and in Ref. \cite{DiazDorronsoro:2018wro}.

\clearpage
\newpage
\section{Background no-boundary wave function} \label{PsiHHsec}
\noindent To summarize 
\begin{equation} \label{PsiFinal}
	\Psi_\text{HH}(p,\alpha) = \Psi_\text{HH}(p,\alpha; \overline{B^4}) + \Psi_\text{HH}(p,\alpha;\mathbb{C}\text{P}^2 \setminus B^4) \,,
\end{equation}
where expressions for the two terms on the RHS can be found in Eqns. \eqref{PsiHHB4approx} and \eqref{PsiHHBoltapprox} and the one-loop factors were discussed in \S\ref{normsec}. As mentioned in \S\ref{topsec} these are just two of the contributions to the NBWF in the BB9 model -- there may be others, but these two already suffice to illustrate the behavior of the NBWF when multiple types of instantons are included. In this section we will denote the (complex, Lorentzian) actions of the no-boundary Taub-NUT-dS and Taub-Bolt-dS solutions by $S_\text{NUT}$ and $S_\text{Bolt}$ respectively. (These each contribute to the wave function as $\Psi \sim e^{iS/\hbar}$.)

\begin{figure}[t]
\centering
\includegraphics[width=0.72\textwidth]{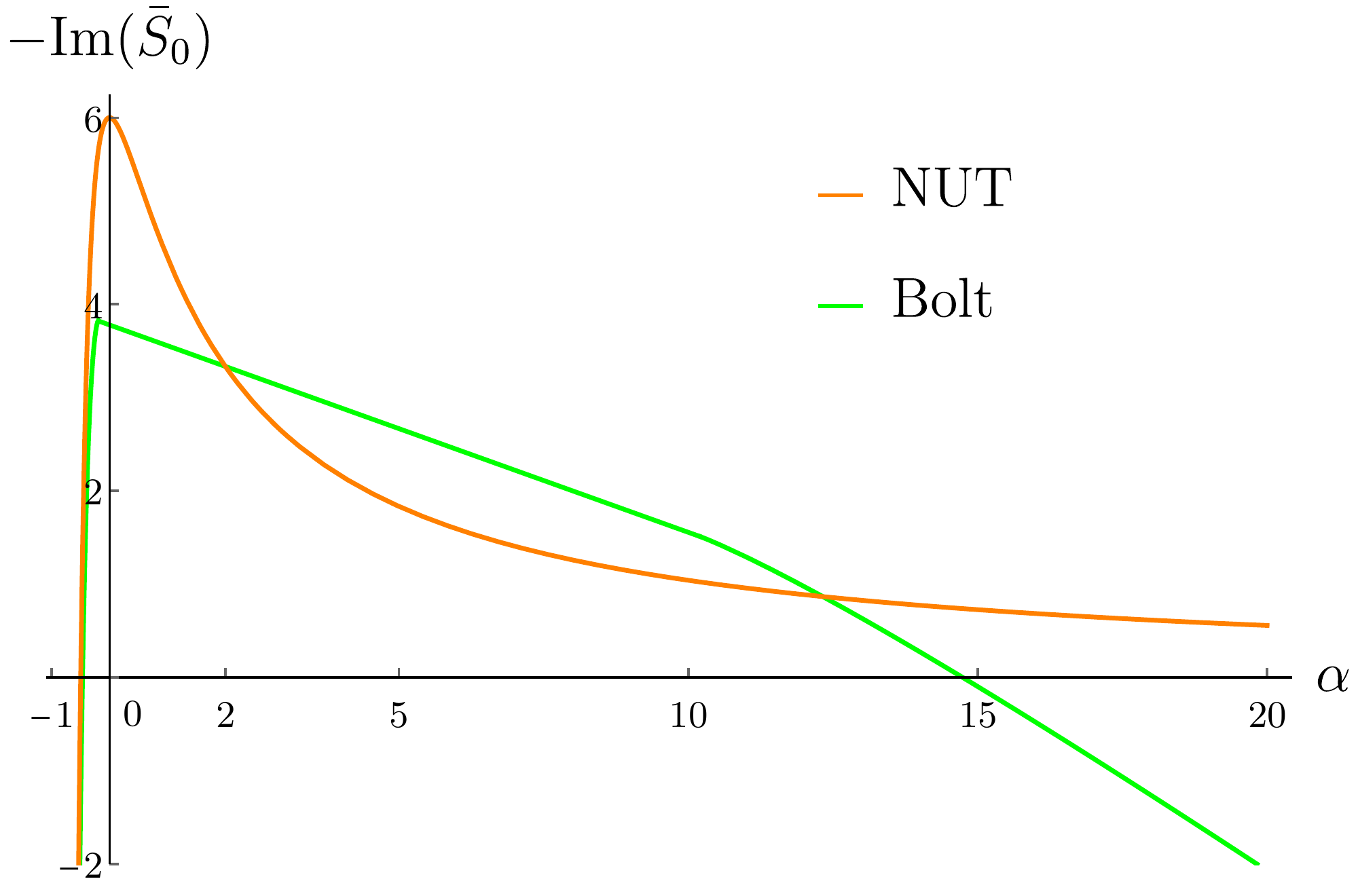}
\caption{\small The imaginary parts of the on-shell actions corresponding to two no-boundary instantons in the BB9 minisuperspace model. The actions are plotted at large volume (more precisely in the regime $\tilde{\mathcal{D}}_x$ defined in Eq. \eqref{xdomain2}), where $\text{Im}(\bar{S}_0)$ is only a function of $\alpha$, the squashing of the $S^3$ in the argument of the NBWF. The contribution labeled by ``NUT'' represents an instanton on the four-disk, while the contribution labeled by ``Bolt'' lives on $\mathbb{C}\text{P}^2 \setminus B^4$.}
\label{allactions}
\end{figure}

The expressions \eqref{PsiHHB4approx} and \eqref{PsiHHBoltapprox} are valid in the parameter regime $\tilde{\mathcal{D}}_x$, defined in Eq. \eqref{xdomain2}, which is a subset of the minisuperspace where the wave function \eqref{PsiFinal} predicts classical correlations between its arguments. (The classical histories are shown in Figure \ref{classicaltrajectories}.) In this regime the functions $\text{Re}(i S_\text{NUT})$ and $\text{Re}(i S_\text{Bolt})$ are approximately only functions of $\alpha$, which labels the classical histories, and both reach a maximum somewhere in this regime. The Taub-Bolt-dS contribution reaches a maximum at a point very close to (but smaller than) $\alpha = 5-3\sqrt{3} \approx -0.2$. The semiclassical contribution to the NBWF from the Bolt topology is thus peaked about an anisotropic classical history. We evaluated the action at this point (or better, \textit{line}) and find
\begin{equation}
	\text{max} \left\{ \text{Re}(i S_\text{Bolt}) \, | \, (\alpha,x) \in \tilde{\mathcal{D}}_x \right\} \approx 3.82 \,.
\end{equation}
However, the relative weight of this configuration compared to the NUT contribution around the isotropic $S^3$ is negligible; from Eq. \eqref{Splus} we obtain
\begin{equation}
	\text{max} \left\{ \text{Re}(i S_\text{NUT}) \, | \, (\alpha,x) \in \mathcal{D}_x \right\} = 6 \,.
\end{equation}
Since the NUT contribution is peaked around $\alpha = 0$ and the Bolt contribution is irrelevant compared to it in a neighborhood of $\alpha = 0$, the NBWF gives the highest probability to the isotropic classical history. This conclusion is visualized in Figure \ref{allactions}.

\begin{figure}[b]
\centering
\includegraphics[width=0.72\textwidth]{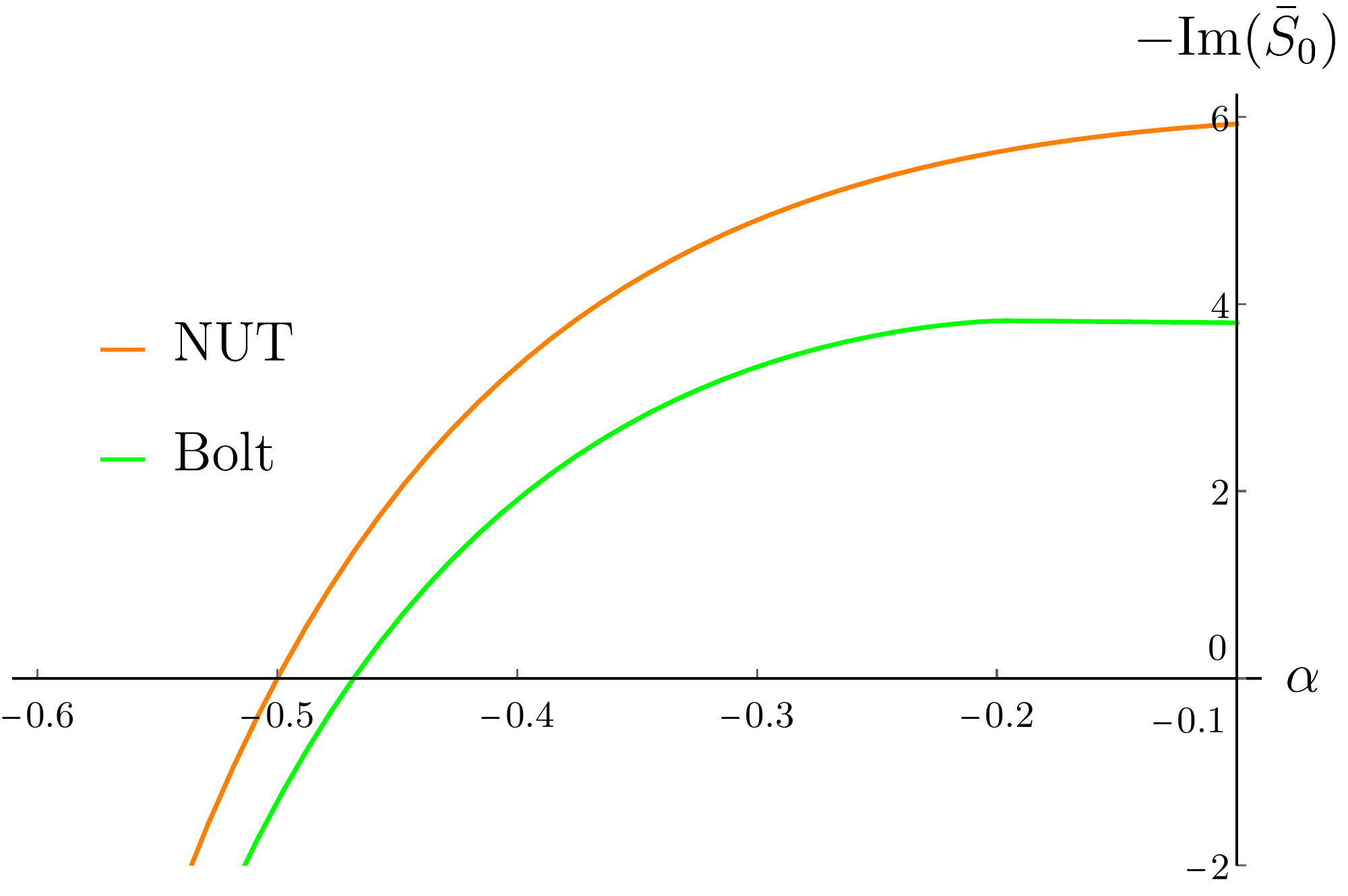}
\caption{\small The same quantities as in Figure \ref{allactions} are shown, focussing on negative values of $\alpha$.}
\label{allactions2}
\end{figure}

As we move to positive $\alpha$, we encounter two phase transitions: the first at $\alpha = 2$ and the second at $\alpha = 2(3+\sqrt{10}) \approx 12.32$. There is no phase transition for negative $\alpha$; the NUT contribution is dominant there. This is illustrated in Figure \ref{allactions2}. This completes our discussion of the NBWF, or at least two of its contributions, in the unperturbed BB9 minisuperspace model.

\newpage

\clearpage
\newpage
\section{Isotropic limit} \label{isotropicsec}
\noindent In this section we discuss the isotropic limit of the NBWF $\Psi_\text{HH}(p,q)$, i.e. its behavior on the isotropic slice $p = q$ (or $\alpha = 0$), about which concern was raised recently in \cite{FLT4}. We argued in \S\ref{PsiHHsec} that only the contribution from $\overline{B^4}$ is relevant at $\alpha = 0$. At this point, the no-boundary Taub-NUT-dS solution discussed in \S\ref{NUTsec} is identical to the homogeneous and isotropic solution originally considered by Hartle and Hawking in \cite{PhysRevD.28.2960} (and reviewed in many articles including our recent work \cite{DiazDorronsoro2017}). Thus the semiclassical wave functions are identical to leading order in $\hbar$, i.e. the exponential parts of the semiclassical expressions are equal. (The possible one-loop factors that follow from the one-dimensional and two-dimensional WDW equations are not equal, but this is not to be expected.) This completes the relevant part of the discussion of the isotropic limit.

The criticism in \cite{FLT4} is directed towards the different implementation of the NBWF as a minisuperspace functional integral in the one-dimensional dS minisuperspace model we used in \cite{DiazDorronsoro2017} vs. its implementation in the two-dimensional BB9 model on the four-disk we used in \cite{DiazDorronsoro:2018wro}. More precisely the claim is that it is inconsistent to choose a different contour $\mathcal{C}$ for the lapse integral in these models (see Appendix \ref{offshellsec} for a brief review of the minisuperspace path integral formalism), because, in the isotropic limit, the two models should coincide. No such consistency condition exists, however. In fact our analysis shows that there cannot be such a consistency condition, precisely because the wave functions we constructed coincide in the isotropic limit even though a different lapse contour $\mathcal{C}$ was chosen in their respective constructions. The more general reason that there is no inconsistency is that the off-shell analysis in minisuperspace models depends sensitively on the details of the path integral including the choice of lapse gauge-fixing and the choice of boundary conditions at the south pole of the geometry. We illustrate this point with simple additional examples in Appendix \ref{offshellsec}.

\clearpage
\newpage
\section{No-boundary wave function of inhomogeneous scalar \\ fluctuations} \label{inhomogeneoussec}
\noindent In this section we consider inhomogeneous, massless and minimally coupled scalar fluctuations around the anisotropic background solutions we discussed in Sections \ref{NUTsec} and \ref{Boltsec}, i.e. the no-boundary Taub-NUT-dS and Taub-Bolt-dS solutions. We emphasize that we are holding the background \textit{fixed} -- the metric is as in Eq. \eqref{qpansatz} with $p(\tau), q(\tau)$ and $N$ taking on definite values. This is different from what the authors of \cite{FLT2,FLT3} have attempted to do, which is to consider a dS plus massless scalar minisuperspace (reviewed in Appendix \ref{nonlinearsec2}) where the background is allowed to fluctuate in response to the scalar and thus the two fields are treated at the same level. We comment on their calculation, which is inconsistent, in Appendix \ref{FLTsec}.

\subsection{Action} \label{fluctactionsec}
\noindent Here instead we are doing the quantum-cosmological analogue of quantum field theory in a (fixed) curved background spacetime \cite{HarHal1990}. In our case the background is complex and lives on a compact four-manifold. The (bulk, Lorentzian) action for a massless minimally coupled scalar $\phi(\tau,\Omega)$ on an anisotropic background specified by $(p(\tau),q(\tau),N_s)$ reads
\begin{align}
	S_\phi &= - \frac{1}{2} \int \di^4 x \, \sqrt{-g} \left( \partial \phi \right)^2 \notag \\
	&= \frac{1}{2 \pi^2} \int_0^1 \di \tau N_s \, \frac{p^{3/2}}{\sqrt{q}} \int_{S^3} \di \Omega \sqrt{g_\Omega} \left( \frac{q}{2 N_s^2} \dot{\phi}^2 + \frac{1}{2p} \, \phi \nabla^2 \phi  \right) \notag \\
	&= \frac{1}{2 \pi^2} \int_0^1 \di \tau N_s \int_{S^3} \di \Omega \sqrt{g_\Omega (\alpha \equiv 0)} \left( \frac{p q}{2 N_s^2} \dot{\phi}^2 + \frac{1}{2} \, \phi \nabla^2 \phi  \right) \,. \label{scalaractionz}
\end{align}
Here $\Omega$ stands for the Euler angles $\theta \in [0,\pi], \phi \in [0,2\pi), \psi \in [0,4\pi)$ on $S^3$ (i.e. the coordinates used in Eq. \eqref{S3metric} but not in Eq. \eqref{tensorminisuperspace})\footnote{We apologize for using $\phi$ twice.} and $(g_\Omega)_{ij}$ is the rescaled spatial part of the metric \eqref{S3metric},
\begin{equation} \label{prolate}
	4 \, (g_\Omega)_{ij} \di \Omega^i \di \Omega^j = \sigma_1^2 + \sigma_2^2 + \frac{1}{1+\alpha(\tau)} \sigma_3^2 \,,
\end{equation}
where $\alpha(\tau) \equiv p(\tau)/q(\tau) - 1$ and $(p(\tau),q(\tau),N_s)$ is one of the complex, no-boundary background solutions discussed in Sections \ref{NUTsec} and \ref{Boltsec}. For clarity, the Laplacian in Eq. \eqref{scalaractionz} is with respect to the $\tau$-dependent metric $g_\Omega$ given in Eq. \eqref{prolate}. We have
\begin{equation}
	\sqrt{g_\Omega} \equiv \sqrt{\det \left[ (g_\Omega)_{ij} \right]} = \frac{\sin \theta}{8 \sqrt{1+\alpha}} = \frac{\sin \theta}{8} \sqrt{\frac{q}{p}} \,.
\end{equation}

\subsection{Fluctuation wave function} \label{fluctwfsec}
\noindent Thus we are considering the NBWF on the extended BB9 minisuperspace spanned by the $S^3$ scale factors $p,q \geq 0$ and the value of the (small) massless scalar field $\varphi(\Omega) = \phi(1,\Omega)$ on the (squashed) $S^3$. The total wave function can be written as a sum of products of the background wave functions and corresponding fluctuation wave functions (e.g. \cite{HarHal1990} and references therein),
\begin{equation}
	\Psi_\text{HH}[p,q,\varphi(\Omega)] = \Psi_\text{HH}(p,q;\overline{B^4}) \times \Psi_\text{fluct.}[\varphi(\Omega) \, | \, \overline{B^4}] + \Psi_\text{HH}(p,q;\mathbb{C}\text{P}^2 \setminus B^4) \times \Psi_\text{fluct.}[\varphi(\Omega) \, | \, \mathbb{C}\text{P}^2 \setminus B^4] \,,
\end{equation}
where the background wave functions are given in Eqns. \eqref{PsiHHB4approx} and \eqref{PsiHHBoltapprox}. The fluctuation wave functions for the no-boundary proposal are determined by a path integral of the form
\begin{equation} \label{fluctPI}
	\Psi_\text{fluct.}[\varphi(\Omega) \, | \, \text{background}] = \int_{\mathcal{B}_\phi}^{\phi(1) = \varphi} \mathcal{D}\phi ~ e^{i S_\phi(\text{background})/\hbar} \,,
\end{equation}
where $\mathcal{B}_\phi$ are boundary conditions on the scalar at the south pole of the background geometry which correspond to the behavior of a regular solution to the scalar field EOM on the background in this regime. Here we will take these to be of the Dirichlet type $\mathcal{B}_\phi = \{ \phi(0,\Omega) \equiv 0 \}$ for both background geometries, following \cite{HarHal1990}, although other options can be explored (e.g. \cite{DiTucci:2019dji}). Since the scalar action \eqref{scalaractionz} -- which is the one appropriate to the Dirichlet boundary conditions we consider -- is quadratic, the evaluation of the path integral \eqref{fluctPI} reduces to finding the solution to the EOM that satisfies the appropriate boundary conditions.

\subsection{Numerical strategy} \label{numericssec}
\noindent Our strategy is to expand $\phi$ and $\varphi$ into harmonics on the (single-)squashed $S^3$ with metric \eqref{prolate}, which are labeled by three numbers $J \in \{0,1/2,1,3/2,\dots \}$ and $K,M \in \{ -J,-J+1,\dots,J \}$. We will denote these quantum numbers collectively by $L$. The harmonics are given explicitly by \cite{Reiche1926,PhysRev.29.262,PhysRev.34.243,WINTER1954274,PhysRevD.8.1048}
\begin{align}
	Y_L(\Omega) &= C_L \, e^{i M \phi} \, e^{i K \psi} \, d_L(\theta) \,, \label{explicitYL} \\
	d_L(\theta) &= \left( \frac{1 - \cos \theta}{2} \right)^{\lambda_1} \left( \frac{1 + \cos \theta}{2} \right)^{\lambda_2} \, {}_2 F_1 \left( \alpha, \beta ; \gamma ; \frac{1-\cos \theta}{2} \right) \,, \label{dL} \\
	\lambda_1 &= \frac{|K-M|}{2} \,, ~~~ \lambda_2 = \frac{|K+M|}{2} \,, ~~~ \alpha = \lambda_1 + \lambda_2 + J + 1 \,, ~~~ \beta = \lambda_1 + \lambda_2 - J \,, ~~~ \gamma = 2\lambda_1 + 1 \,, \notag \\
	C_L &= \frac{1}{\Gamma(\gamma)} \sqrt{\frac{(2J+1) \, \Gamma(\alpha) \, \Gamma(\gamma-\beta)}{\Gamma(1-\beta) \, \Gamma(1+\alpha-\gamma)}} \,. \notag
\end{align}
We stress the $S^3$ coordinate $\phi$ is periodic with $4\pi$, but that to cover the $S^3$ once it runs from $0$ to $2 \pi$. Note that the hypergeometric series in Eq. \eqref{dL} simplifies to a (Jacobi) polynomial in $(1-\cos \theta)/2$ since $\beta = \max \{ |K|,|M| \} - J$ is always a negative integer. These functions satisfy\footnote{Note that the scalar harmonics $Y_L$ do not depend on $\tau$. (The eigenvalues given in Eq. \eqref{eigenvalueYL} \textit{do}.) This feature allows us to proceed analytically for perturbations on the single-squashed $S^3$. For the double-squashed $S^3$ (i.e. $\di s^2 = \sigma_1^2 + (1+\beta)^{-1} \sigma_2^2 + (1+\alpha)^{-1} \sigma_3^2$, $\alpha,\beta \neq 0$) the harmonics depend non-trivially on the squashing parameters and it seems one must proceed numerically \cite{PhysRevD.8.1048,Bobev:2016sap}.}
\begin{align}
	\nabla^2 Y_L &= - 4 \left[ J(J+1) + \alpha(\tau) K^2 \right] Y_L \,, \label{eigenvalueYL} \\
	\int_{S^3} \di \Omega \sqrt{g_\Omega (\alpha \equiv 0)} \, Y_L^* \, Y_{L'} &= 2 \pi^2 \delta_{L L'} \,, \label{orthonormYL} \\
 Y_{(J, K, M)}(\Omega)^* &= Y_{(J, -K, -M)}(\Omega) \,,
\end{align}
and form a basis for the continuous complex functions on $S^3$. We can define a complete set of real harmonics via
\begin{equation} \label{explicitXL}
	X_L(\Omega) = \begin{cases}
               \sqrt{2} \, C_L \sin \left( M \phi + K \psi \right) d_L(\theta) &\text{if } M > 0 \,, \\
               \sqrt{2} \, C_L \sin \left( K \psi \right) d_L(\theta) &\text{if } M = 0, K > 0 \,, \\
               C_L d_L(\theta) &\text{if } M = 0, K = 0 \,, \\
               \sqrt{2} \, C_L \cos \left( K \psi \right) d_L(\theta) &\text{if } M = 0, K < 0 \,, \\
               \sqrt{2} \, C_L \cos \left( M \phi + K \psi \right) d_L(\theta) &\text{if } M < 0 \,.
            \end{cases}
\end{equation}
These functions satisfy the same eigenvalue equation \eqref{eigenvalueYL} as the $Y_L$, form a basis for the real functions on $S^3$ and are orthonormalized in the real sense,
\begin{equation}
	\int_{S^3} \di \Omega \sqrt{g_\Omega (\alpha \equiv 0)} \, X_L X_{L'} = 2 \pi^2 \delta_{L L'} \,. \label{orthonormXL}
\end{equation}
We expand
\begin{align} \label{varphiexpansion}
	\phi(\tau,\Omega) &= \sum_{L} f_L(\tau) \, X_L(\Omega) \,, \\
	\varphi(\Omega) &= \sum_{L} f_{L,1} \, X_L(\Omega) \,,
\end{align}
which gives\footnote{In the isotropic limit $\alpha \equiv 0$ this coincides with previous work: for \cite{DiazDorronsoro2017} identify $J \cong l/2$ and for \cite{HarHal1990} identify $J \cong (n - 1)/2$.}
\begin{equation} \label{fLaction}
	S_\phi = \sum_{L} \int_0^1 \di \tau N_s \left( \frac{p q}{2 N_s^2} \dot{f}_L^2 - 2 \left[ J(J+1) + \alpha K^2 \right] f_L^2 \right) \,.
\end{equation}
All the modes $f_L$ are decoupled from one another. The EOM are, $\forall L$,
\begin{equation} \label{fLEOM}
	\frac{\di}{\di \tau} \left( p q \, \dot{f}_L \right) + 4 N_s^2 \left[ J(J+1) + \alpha K^2 \right] f_L = 0 \,,
\end{equation}
and it is possible to show that the on-shell action takes the form
\begin{equation} \label{fLonshellaction}
	S_\phi^\text{(on-shell)} = \sum_{L} \frac{1}{2N_s} \left[ p q f_L \dot{f}_L \right]^{\tau=1}_{\tau=0} \,.
\end{equation}
Unfortunately we could not find a closed-form solution to Eq. \eqref{fLEOM} for general no-boundary solutions $(p(\tau), q(\tau),N_s)$ with Dirichlet boundary conditions $f_L(0) = 0, f_L(1) = f_{L,1}$. Instead we proceeded numerically. This exercise is greatly simplified by the linearity of Eq. \eqref{fLEOM}: we only need to solve the equation once (for each couple of arguments $(p,q)$ of the background wave function) with $f_L(0) = 0$ and an essentially arbitrary initial value for $\dot{f}_L(0)$.\footnote{In practice one starts the numerical integration at $\tau = \varepsilon \gtrapprox 0$, see below.} We can then simply rescale this solution by a complex number such that $f_L(0) = 0, f_L(1) = 1$. If we call the resulting function $\tilde{f}_L$, the action we are interested in is given by
\begin{equation} \label{fLonshellaction}
	S_\phi^\text{(on-shell)} = \frac{p q}{2N_s} \sum_{L} \dot{\tilde{f}}_L(1) \, f_{L,1}^2 \,,
\end{equation}
where again $p,q$ are the arguments of the background wave function here.

While we cannot analytically solve Eq. \eqref{fLEOM} in general, we can determine the leading behavior of $f_L$ near $\tau = 0$ up to a proportionality factor. For the no-boundary NUT-type backgrounds we have $p(\tau), q(\tau) = 2 i N_s \tau + \mathcal{O}(\tau^2)$ as $\tau \rightarrow 0$. Thus $\alpha(\tau) \rightarrow 0$ as $\tau \rightarrow 0$, and \eqref{fLEOM} becomes
\begin{equation}
	\frac{\di}{\di \tau} \left( \tau^2 \dot{f}_L \right) \approx J(J+1) f_L ~~~ \text{ near } \tau = 0 \text{ for the NUT-type background} \,.
\end{equation}
Trying an Ansatz $f_L(\tau) \propto \tau^A$ yields $A(A+1) = J(J+1)$, so $A = J$ or $A = -1-J$. The latter choice would cause $f_L$ to blow up as $\tau \rightarrow 0$, so we discard this solution since the boundary condition $f_L(0) = 0$ cannot be imposed on it. (Additionally these solutions have infinite action.) Thus for perturbations around the NUT-type background, $f_L(\tau) \propto \tau^J$ near $\tau = 0$.\footnote{\label{homftnote}This holds for $J \neq 0$. For $J = 0$ the $A = J$ solution is identically zero due to the initial boundary condition, and can thus not be made to satisfy the final boundary condition. This holds also for perturbations around the Bolt-type background. We therefore exclude the homogeneous $J = 0$ mode from this analysis, which is better thought of as a background field.} This relation lets us set up the numerical integration problem at a finite value $\tau = \varepsilon \gtrapprox 0$: $f_L(\varepsilon) = A \, \varepsilon^J, \dot{f}_L(\varepsilon) = A J \, \varepsilon^{J-1}$, where $A$ is chosen arbitrarily at first and later rescaled, following the discussion above. For the Bolt-type background one can similarly derive $f_L(\tau) \propto \tau^{|K|}$ for $K \neq 0$ near $\tau = 0$ and set up the numerical integrator accordingly. For $K = 0$ we have $f_L(\tau) \propto \tau^{\pm \sqrt{c}}$ with $c = 2J(J+1)i N_s/p_0$ and the sign is chosen such that $\text{Re}(\pm \sqrt{c}) > 0$.

\subsection{Fluctuations around the NUT solution} \label{NUTinhomogeneoussec}
\begin{figure}[h]
\centering
\includegraphics[width=0.75\textwidth]{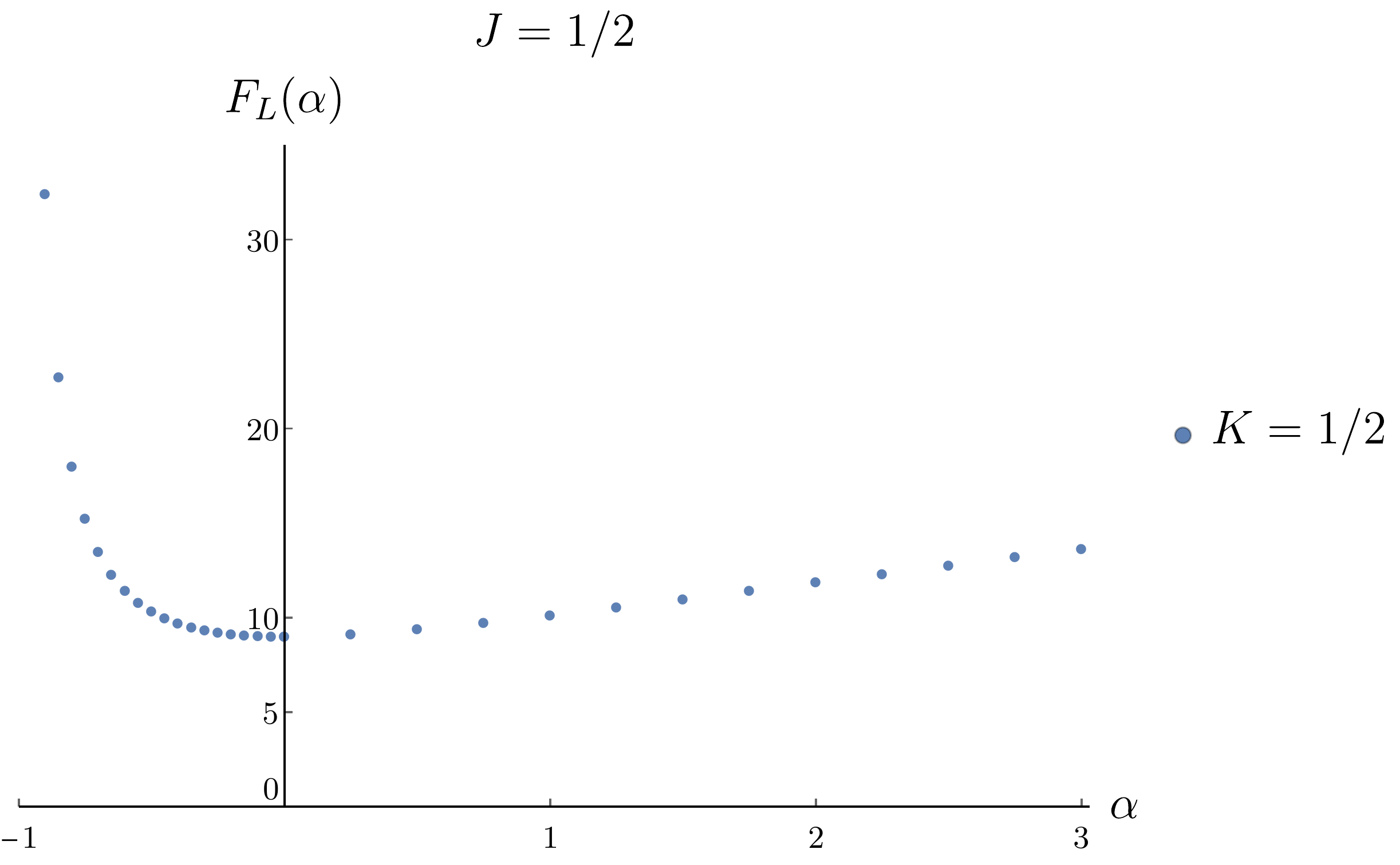}
\caption{\small The coefficient-functions $F_L(\alpha)$, for perturbations labeled by the quantum numbers $L = (J,K,M) \in (1/2, 1/2,M)$, in $\left| \Psi_\text{fluct.}(f_L \, | \, \overline{B^4}) \right| \sim \exp \left[ -F_L(\alpha) f_L^2 / \hbar \Lambda \right]$. Here $| \, \overline{B^4}$ means we are considering massless scalar perturbations around a fixed, anisotropic, no-boundary Taub-NUT-dS instanton at large volume and with squashing $\alpha$.}
\label{J=1/2}
\end{figure}

\begin{figure}[t]
\centering
\includegraphics[width=0.75\textwidth]{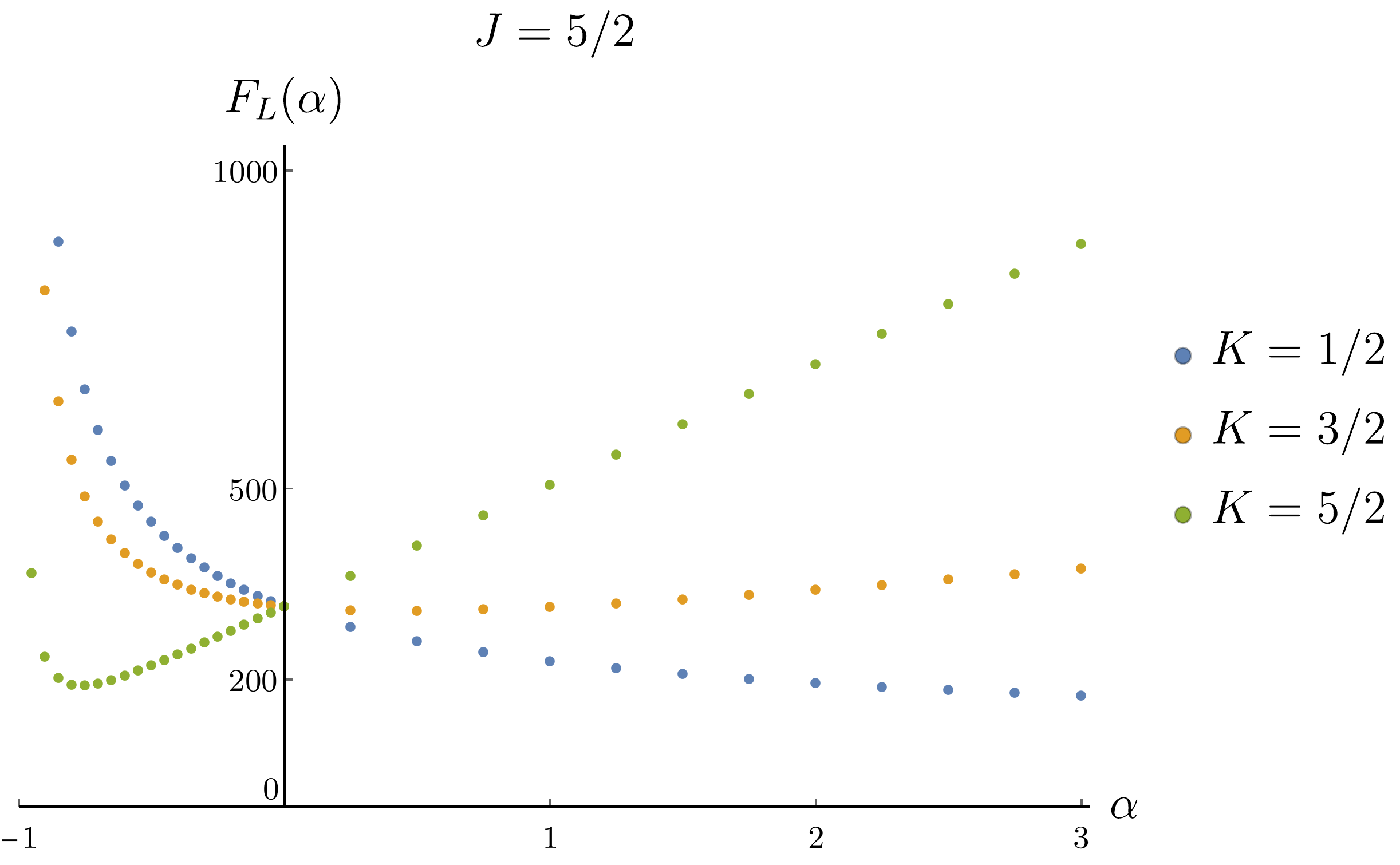}
\caption{\small The coefficient-functions $F_L(\alpha)$, for $L = (5/2,K,M)$, in $\left| \Psi_\text{fluct.}(f_L \, | \, \overline{B^4}) \right| \sim \exp \left[ -F_L(\alpha) f_L^2 / \hbar \Lambda \right]$.}
\label{J=5/2}
\end{figure}

\begin{figure}[b]
\centering
\includegraphics[width=0.70\textwidth]{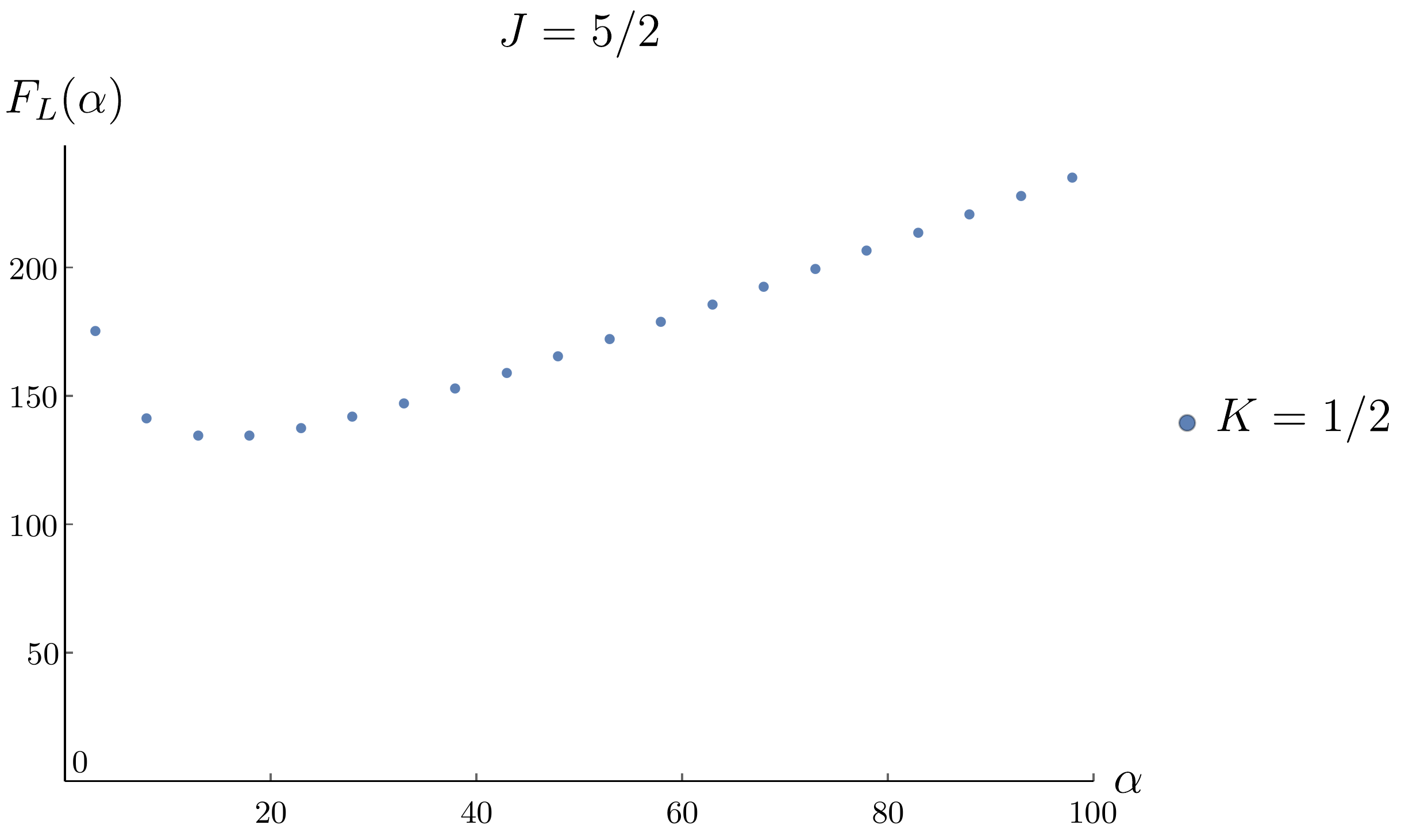}
\caption{\small The coefficient-functions $F_L(\alpha)$ for $L = (5/2,1/2,M)$ in $\left| \Psi_\text{fluct.}(f_L \, | \, \overline{B^4}) \right| \sim$ $\exp \left[ -F_L(\alpha) f_L^2 / \hbar \Lambda \right]$, at larger values of $\alpha$. This is an extension of the same type of $J=5/2,K=1/2$ data points shown in Figure \ref{J=5/2}. From that figure it was not clear what the behavior would be at large $\alpha$, but we see here that the function reaches a minimum at a positive value and increases monotonically from there onwards. All of the $F_L(\alpha)$ with $K \neq 0$ appear to share this qualitative behavior.}
\label{J=5/2bis}
\end{figure}

\begin{figure}[t]
\centering
\includegraphics[width=0.75\textwidth]{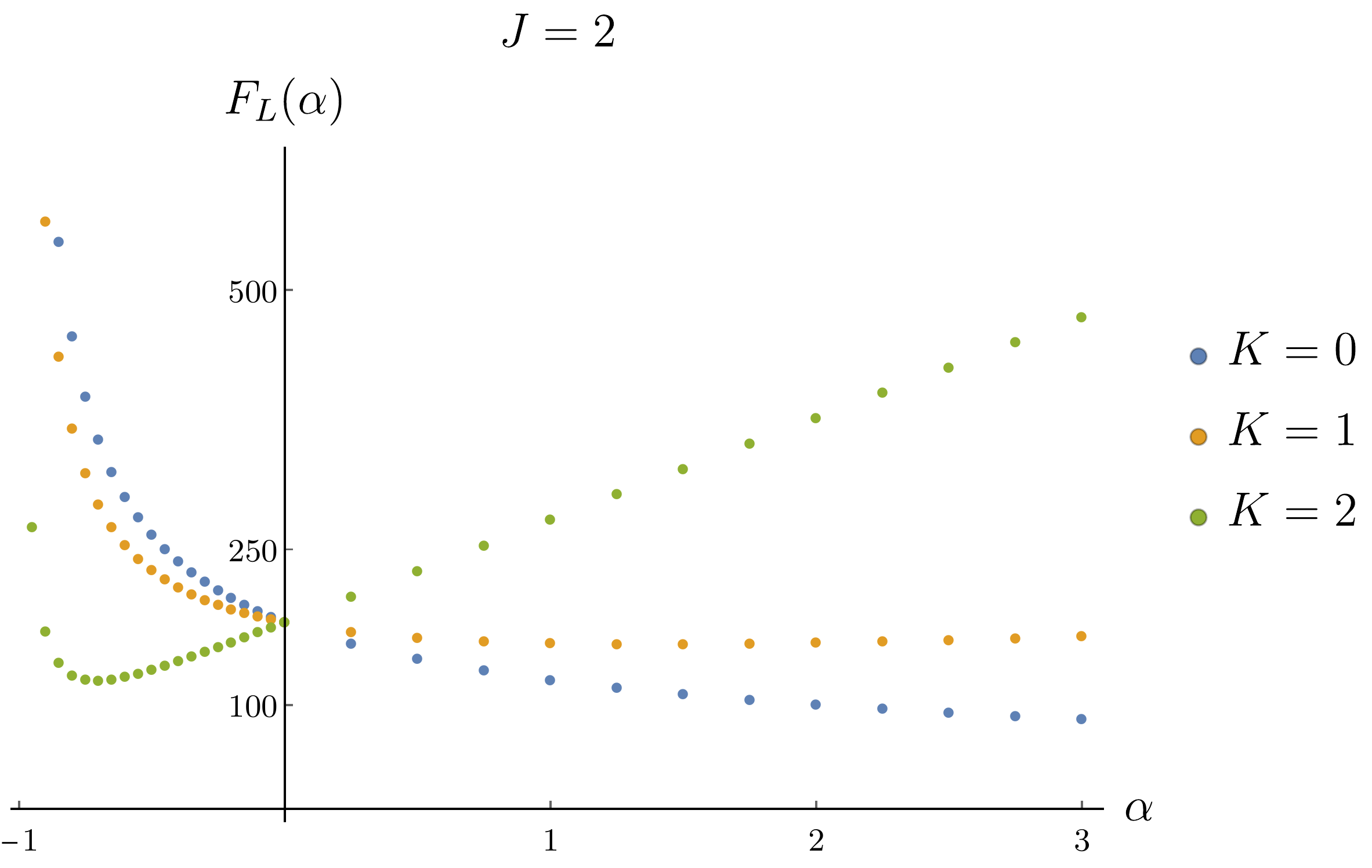}
\caption{\small The coefficient-functions $F_L(\alpha)$, for $L = (2,K,M)$, in $\left| \Psi_\text{fluct.}(f_L \, | \, \overline{B^4}) \right| \sim \exp \left[ -F_L(\alpha) f_L^2 / \hbar \Lambda \right]$.}
\label{J=2}
\end{figure}

\begin{figure}[b]
\centering
\includegraphics[width=0.75\textwidth]{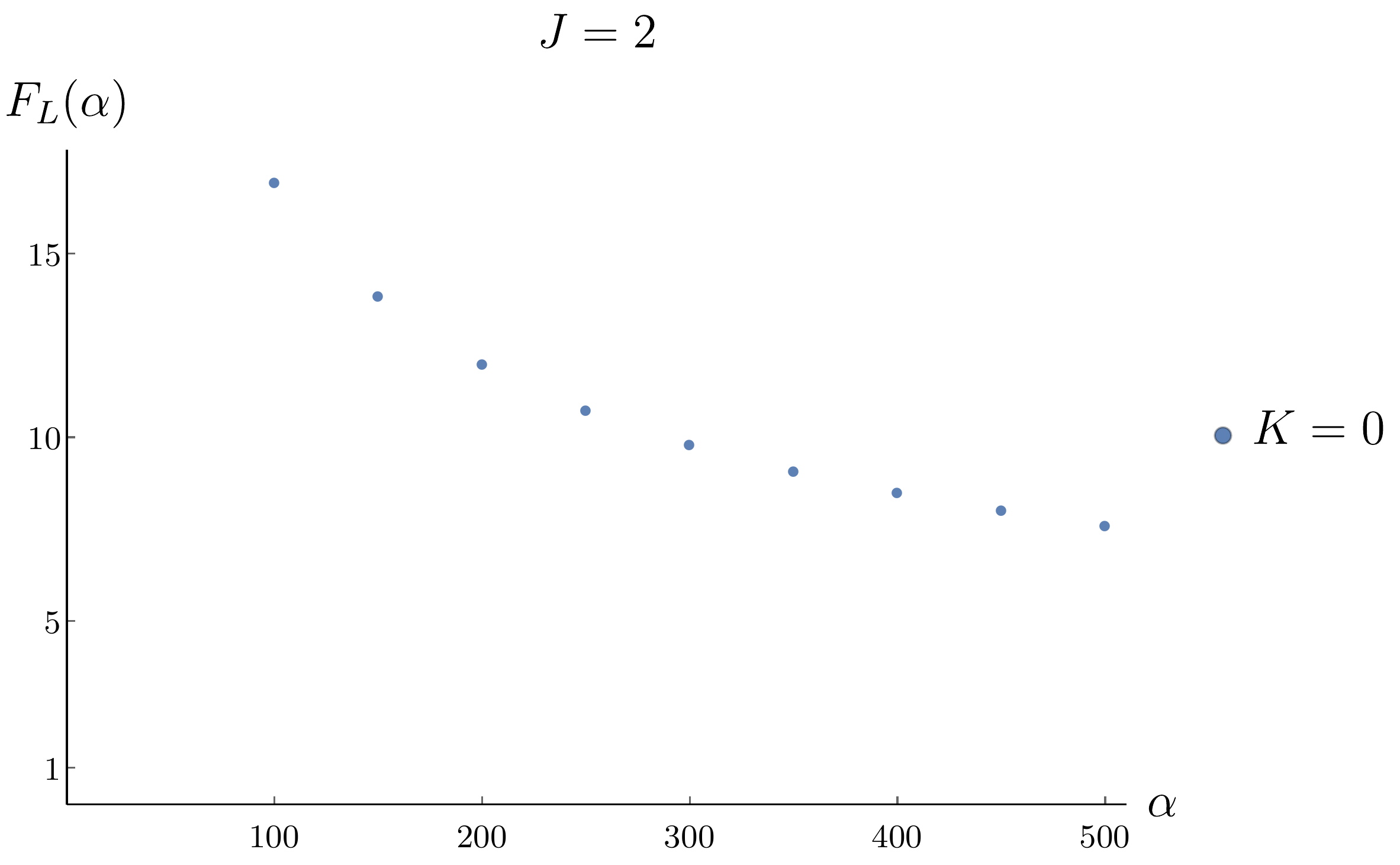}
\caption{\small Behavior of the coefficient-functions $F_L(\alpha)$ for $L = (2,0,M)$ in $\left| \Psi_\text{fluct.}(f_L \, | \, \overline{B^4}) \right| \sim \exp \left[ -F_L(\alpha) f_L^2 / \hbar \Lambda \right]$, at larger values of $\alpha$. This is an extension of the same type of $J=2,K=0$ data points shown in Figure \ref{J=2}. From that figure it was not clear what the behavior would be at large $\alpha$. This figure, and further numerical investigations, suggest that $F_{J,K=0}(\alpha)$ tends to zero as $\alpha$ tends to $+\infty$. Still, every mode function is normalizable.}
\label{J=2bis}
\end{figure}

\noindent As we have mentioned above, while we could not solve Eq. \eqref{fLonshellaction} for a general no-boundary background $(p(\tau),q(\tau),N_s)$, we can solve it explicitly for the special case of the NUT-type solution with $p(1) = q(1)$ (i.e. in the isotropic limit). In this case we have $p(\tau) = q(\tau), \forall \tau \in [0,1]$, and so $\alpha(\tau) \equiv 0$, and one obtains \cite{DiazDorronsoro2017,HarHal1990,FLT2,FLT3} (see also Eq. \eqref{JJeq})
\begin{align} \label{alpha=0action}
	S^\text{(on-shell)}_\phi (\alpha = 0) &= \frac{1}{2} \sum_{L} J (J+1) p \left( \frac{\mp\sqrt{p / 3 - 1} +(2J+1)i}{J(J+1) + p / 12} \right) \, f_{L,1}^2 \\
	&= \sum_{L} \left\{ \mp 2 J(J+1) \sqrt{3 p} \left[ 1 + \mathcal{O}\left(\frac{1}{\sqrt{p}} \right) \right] + 6 i J(J+1)(2J+1) \left[ 1 + \mathcal{O}\left(\frac{1}{p} \right) \right] \right\} f_{L,1}^2 \,. \notag
\end{align}
This result for the isotropic limit suggests that the imaginary part of $S_\phi^\text{(on-shell)}$ -- the part which determines the normalizability of the fluctuation wave function -- tends to an $\alpha$-dependent constant at large volume. Numerically we found this to be the case indeed. Therefore let us define a function $F_L(\alpha)$, the leading term at large volume (or better, in $\mathcal{D}_x$ defined in Eq. \eqref{xdomain}) in the approximation of the imaginary part of $S_\phi^\text{(on-shell)}$:
\begin{equation}
	\text{Im} \left(S_\phi^\text{(on-shell)} \right) \left( p,\alpha ; \overline{B^4} \right) \approx \sum_{L} F_L(\alpha) f_{L,1}^2 \,,
\end{equation}
where $F_L(0) = 6J(J+1)(2J+1)$ from Eq. \eqref{alpha=0action} and we expect corrections to this formula to be small in $\mathcal{D}_x$. The results of our numerical investigations for $F_L(\alpha)$ are summarized in Figures \ref{J=1/2}-\ref{J=2bis}. Most importantly, the numerics support the conclusion that
\begin{equation}
	F_L(\alpha) > 0 \,, ~~~~~ \forall L,\alpha \,.
\end{equation}
This implies that the NBWF for massless, minimally coupled scalar perturbations around any anisotropic NUT-type background are suppressed and normalizable.

\vfill
\vfill

\newpage

\newpage

\clearpage
\newpage
\subsection{Fluctuations around the Bolt solution} \label{Boltinhomogeneoussec}
\begin{figure}[h]
\centering
\includegraphics[width=0.75\textwidth]{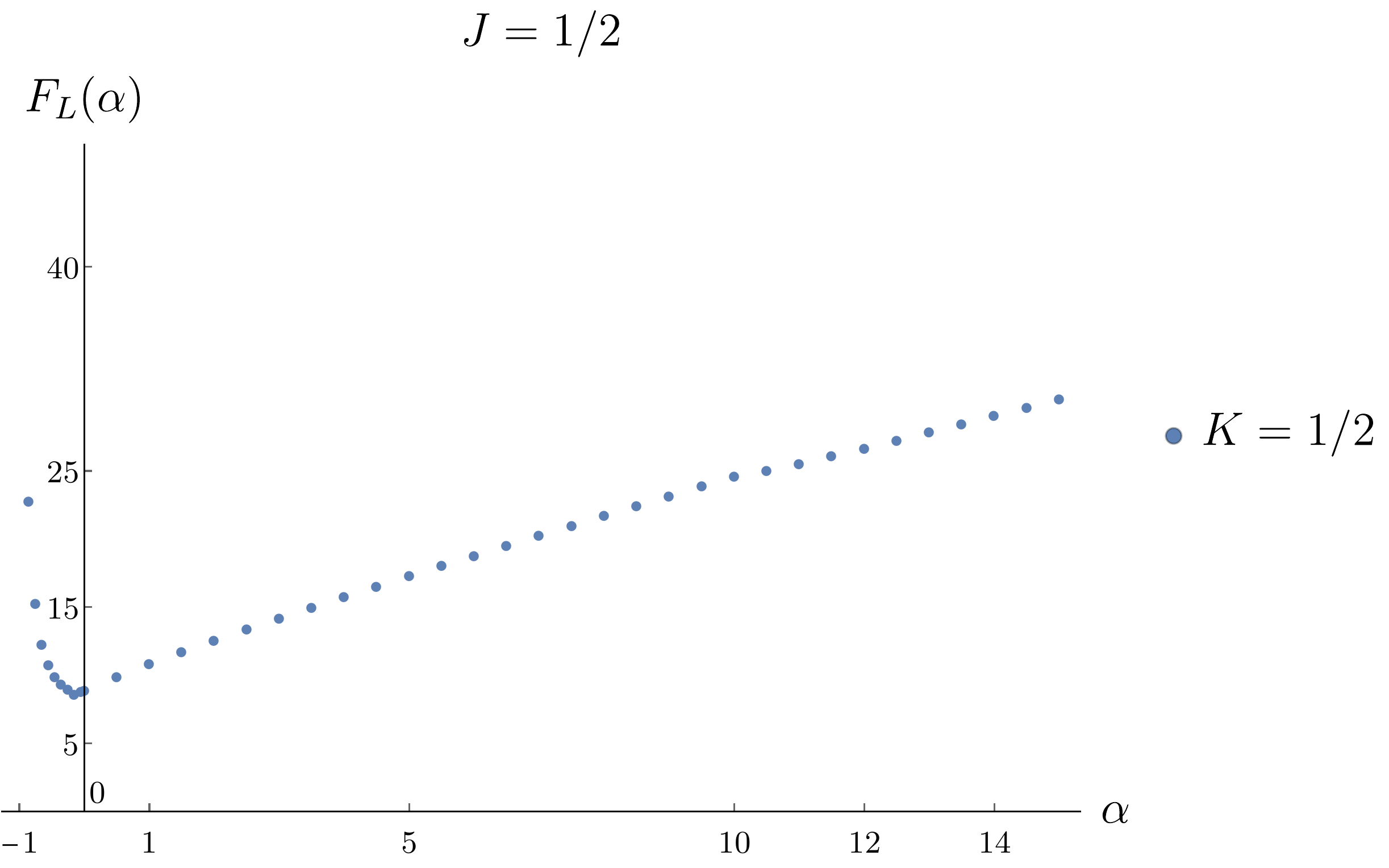}
\caption{\small The coefficient-functions $F_L(\alpha)$, for $L = (1/2,1/2,M)$, in $\left| \Psi_\text{fluct.}(f_L \, | \, \mathbb{C}\text{P}^2 \setminus B^4) \right| \sim \exp \left[ -F_L(\alpha) f_L^2 / \hbar \Lambda \right]$. Here $| \, \mathbb{C}\text{P}^2 \setminus B^4$ means we are considering massless scalar perturbations around a fixed, anisotropic, no-boundary Taub-Bolt-dS instanton at large volume and with squashing $\alpha$.}
\label{J=1/2B}
\end{figure}

\begin{figure}[t]
\centering
\includegraphics[width=0.75\textwidth]{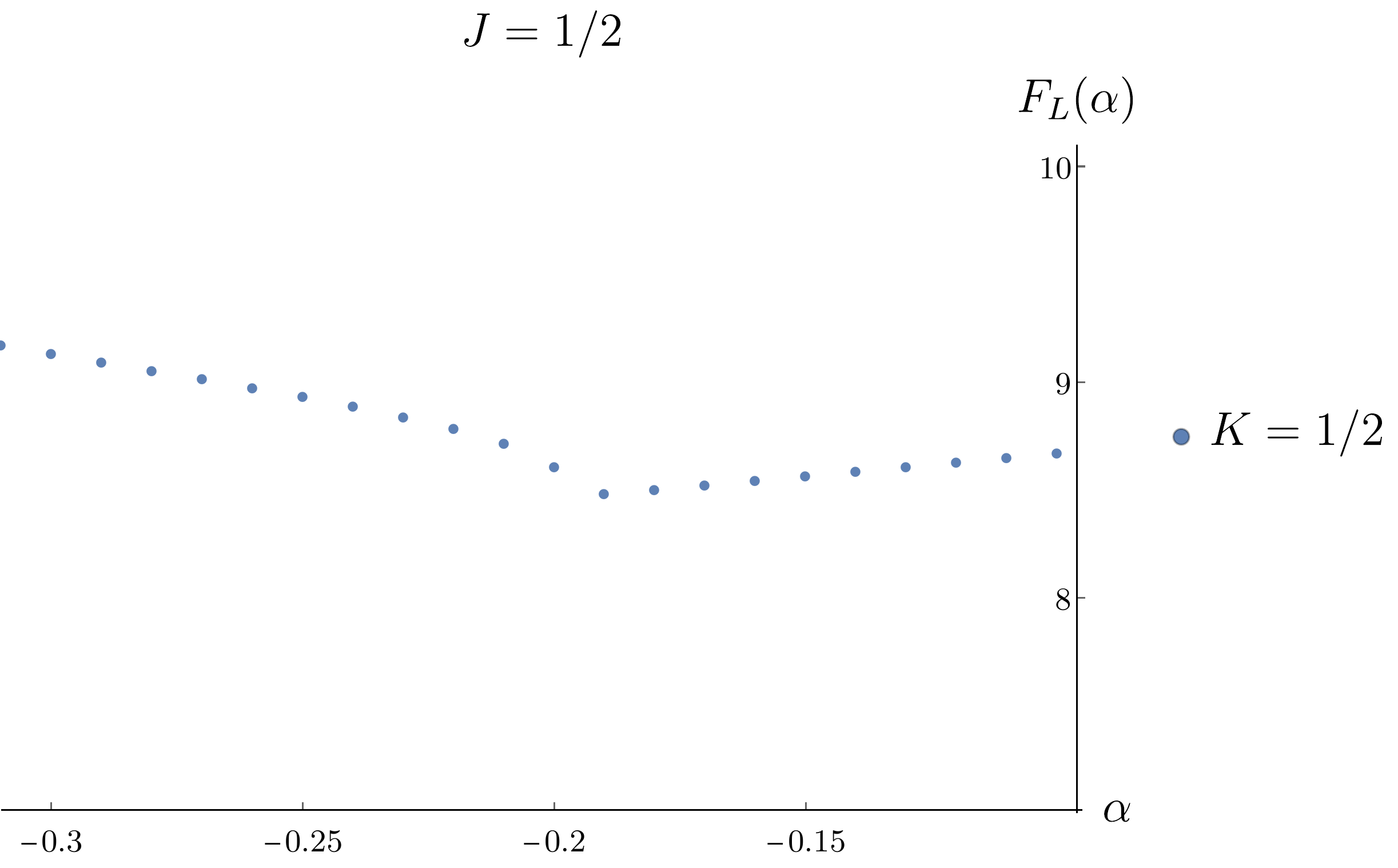}
\caption{\small The same data as in Figure \ref{J=1/2B}, zoomed in on the region near $\alpha_- = 5-3\sqrt{3}$.}
\label{J=1/2bisB}
\end{figure}

\begin{figure}[b]
\centering
\includegraphics[width=0.75\textwidth]{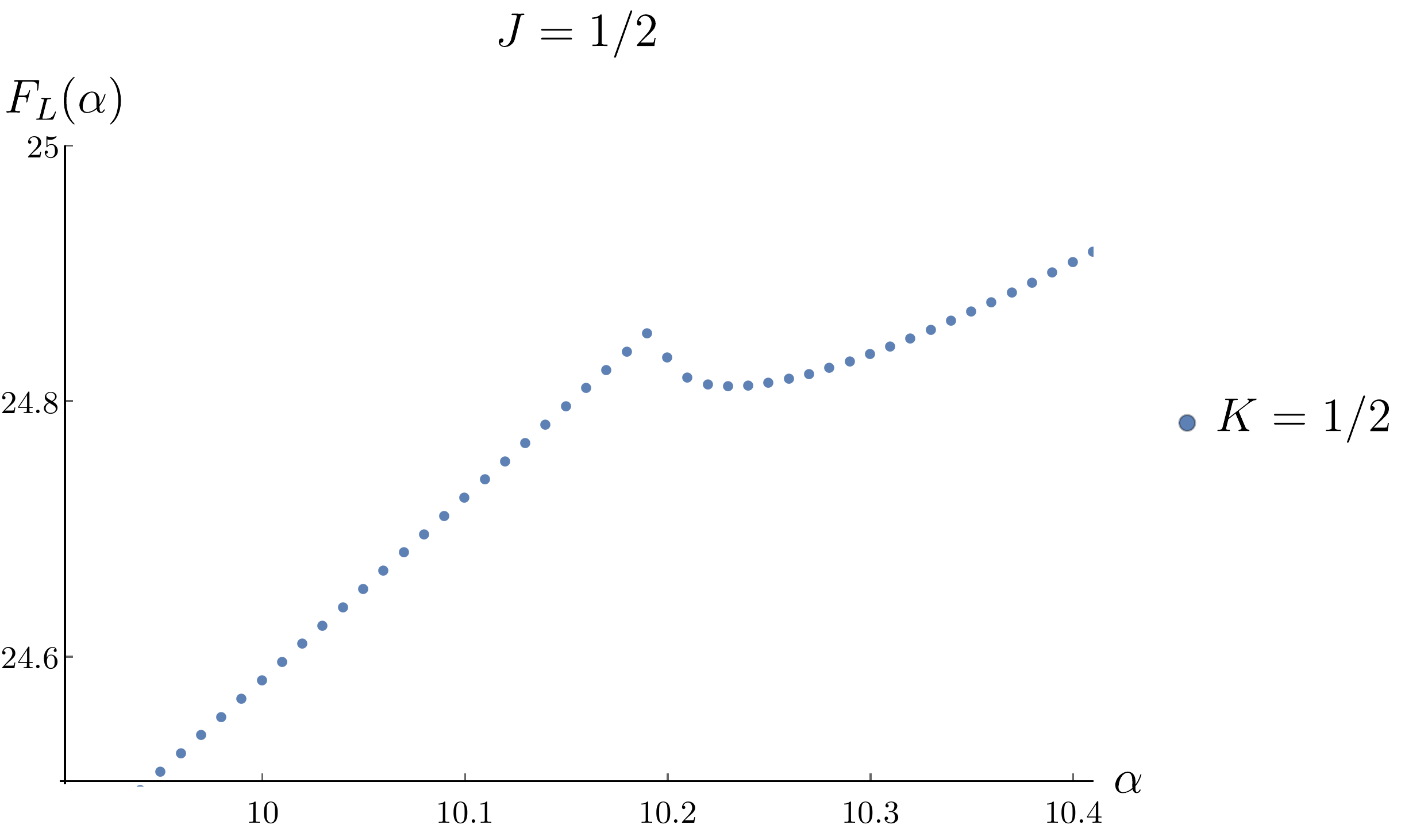}
\caption{\small The same data as in Figure \ref{J=1/2B}, zoomed in on the region near $\alpha_+ = 5+3\sqrt{3}$.}
\label{J=1/2trisB}
\end{figure}

\begin{figure}[h]
\centering
\includegraphics[width=0.75\textwidth]{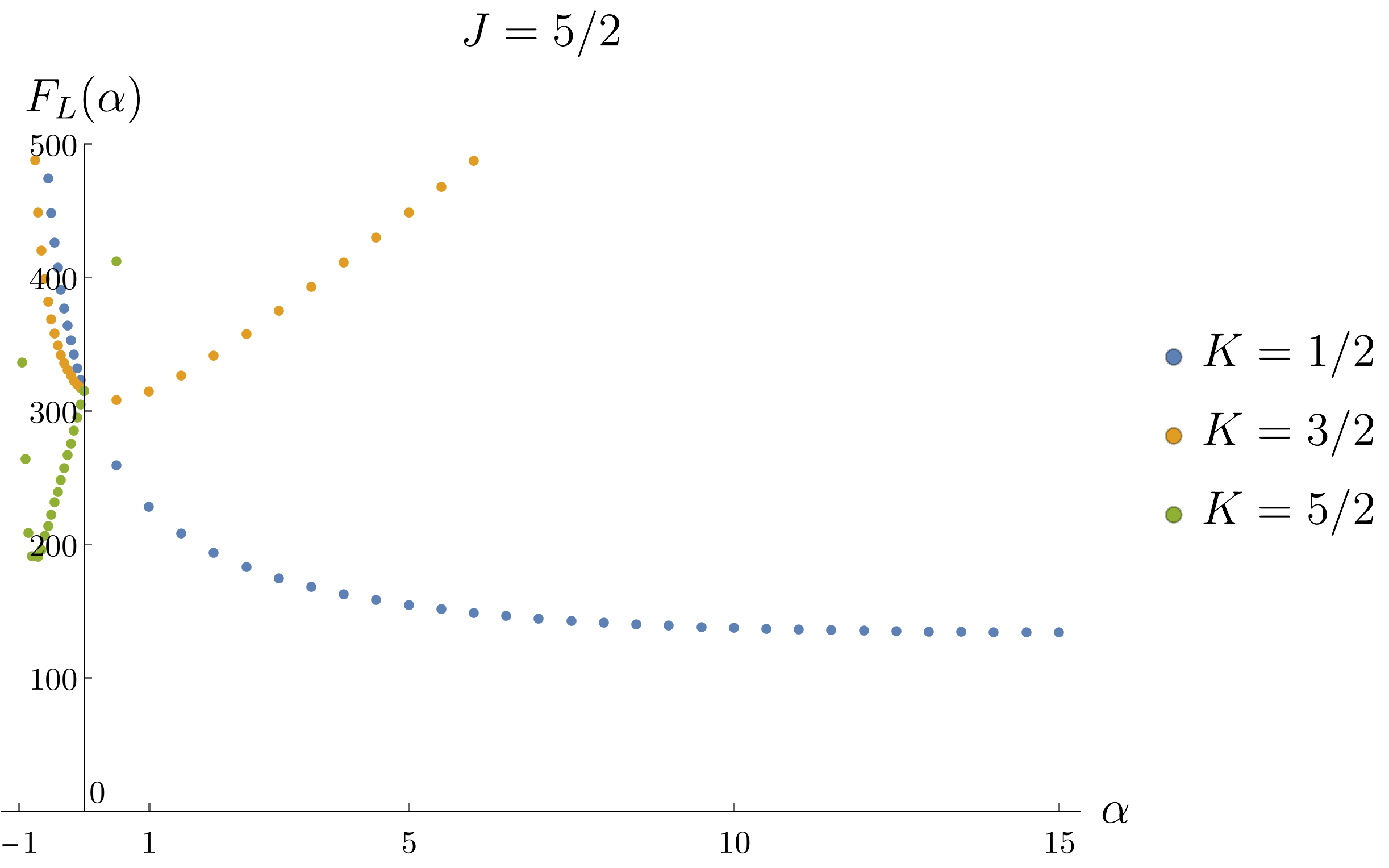}
\caption{\small The coefficients $F_L(\alpha)$, for $L = (5/2,K,M)$, in $\left| \Psi_\text{fluct.}(f_L \, | \, \mathbb{C}\text{P}^2 \setminus B^4) \right| \sim \exp \left[ -F_L(\alpha) f_L^2 / \hbar \Lambda \right]$.}
\label{J=5/2B}
\end{figure}

\noindent In this section we present our results for the NBWF of massless scalar fluctuations around the Bolt-type background discussed in \S\ref{Boltsec}. We use analogous notation as in the previous section. The results are shown in Figures \ref{J=1/2B}-\ref{J=5/2B}.

\vfill
\vfill

\newpage

\clearpage
\newpage

\phantom{XXX}
\vfill
\vfill

\clearpage
\newpage
\section{Tunneling wave function} \label{TWFsec}
\noindent In this section we comment on the tunneling proposal for the wave function of the universe \cite{Vilenkin1982,Vilenkin1984} in the BB9 model \cite{delCampo:1989hy} and its relation to the no-boundary proposal we studied in this paper. Instead of a longspun review of the tunneling wave function (TWF) (see e.g. \cite{PhysRevD.37.898,Vilenkin:2018dch}) we simply note that the TWF result written in \cite{delCampo:1989hy} agrees with ``half'' of the result we have presented in \S\ref{NUTsec} for the contribution to the semiclassical NBWF from the four-disk topology, at least for small $\alpha \approx 0$ and $\alpha \rightarrow -1$ in the regime \eqref{largevolumecondition2} where the TWF is known.\footnote{``Half'' because the NBWF is real and receives contributions from pairs of instantons while the TWF does not per se.} We did not discuss the $\alpha \rightarrow -1$ limit in detail in this paper\footnote{However we did give all the relevant information to do so in Eqns. \eqref{Nplus} and \eqref{on-shellS}.}, because the NBWF does not behave classically there and for that reason is of less immediate interest. However convenient expansions in the small parameter $y \equiv p (1+\alpha) \sim 1/x$ exist in this regime \cite{unpublished} (they are the analogs of Eqns. \eqref{Nplusx} and \eqref{Splus} in the regime $\mathcal{D}_x$). They show that the NBWF behaves as $|\Psi_\text{HH}(p,q)| \sim e^{-q/\hbar} = e^{-p/(1+\alpha)\hbar}$ in this regime\footnote{This behavior can readily be guessed from the on-shell action \eqref{on-shellS}.}, up to small corrections in $y$, which coincides with the result stated in \cite{delCampo:1989hy} for the TWF. In $\mathcal{D}_x$ and for $\alpha \approx 0$ we obtained $|\Psi_\text{HH}(p,\alpha)| \sim e^{-6 \alpha^2 / \hbar \Lambda}$, which again coincides with the result stated in \cite{delCampo:1989hy}. More precisely, one arrives in \cite{delCampo:1989hy} at the following expressions for the TWF $\Psi_\text{T}$ in the BB9 minisuperspace model:
\begin{align}
	\Psi_\text{T}(a>1/H,\text{ lowest non-trivial order in } \beta_+) &\sim \exp \left( -\frac{i}{3 \hbar H^2}(-f_6)^{3/2} \right) \times \exp \left( -\frac{4i a^2 \beta_+^2 / \hbar}{\sqrt{-f_6} + 3i} \right) \,, \label{psiT1} \\
	\Psi_\text{T}(\beta_+ \rightarrow -\infty \,, a \gg e^{2 \beta_+} ) &\sim \exp \left( - x/6 \right) \,, \label{psiT2}
\end{align}
where
\begin{equation}
	f_6 \equiv 1-(aH)^2 \,, \quad \quad x \equiv a^2 e^{-4 \beta_+} \,.
\end{equation}
The connection with our notation in this paper is
\begin{equation}
	6p = a^2 e^{2 \beta_+} \,, \quad \quad 6q = a^2 e^{-4 \beta_+} \,, \quad \quad \Lambda = 18 H^2  \,.
\end{equation}
One may verify that Eqns. \eqref{psiT1} and \eqref{psiT2} coincide with $e^{i \bar{S}_0^+ / \hbar}$ in the appropriate regimes, up to a factor independent of $p$ and $q$, where $\bar{S}_0^+$ is the no-boundary Taub-NUT-dS action that appears in the contribution to the semiclassical NBWF from the four-disk (Eq. \eqref{PsiHHB4}).

\noindent In \cite{delCampo:1989hy} one does not compute the TWF behavior in the intermediate regime $\alpha \in (-1,0)$ or in the regime $\alpha > 0$, so a comparison with our result for the NBWF is not possible there. However the coincidence of the TWF and the NBWF/2 for $\alpha \approx 0$ and $\alpha \rightarrow -1$ in the classical regime makes it tempting to conjecture that the two objects will coincide in large portions of the minisuperspace including $\alpha \in (-1,0)$ (and in the classical regime, i.e., where the saddle points are complex). On the other hand they will certainly not coincide in the entire minisuperspace, e.g. for $p \rightarrow 0$ and $\alpha \approx 0$, let alone in general in other models. Our reasoning behind this conjecture is that the TWF has been claimed to be representable as a (Lorentzian) path integral over geometries which start at zero size \cite{Vilenkin1984,Vilenkin:1994rn,Vilenkin:2018dch,Vilenkin:2018oja}. In the semiclassical limit the TWF should thus be dominated by one or more instantons, and, in a BB9 minisuperspace path integral on the four-disk, this instanton should be the Taub-NUT-dS solution -- the same one which appears in the semiclassical NBWF/2 -- provided it is \textit{regular}. A caveat here is that in recent work on the TWF in perturbative minisuperspace models \cite{Vilenkin:2018dch,Vilenkin:2018oja} this last assumption does not hold -- the relevant instantons are singular there. On the other hand it is unclear how, and indeed whether, the implementation of the TWF as a gravitational path integral discussed in that work can be extended to non-perturbative minisuperspace models like the BB9 model. In any case it is possible to write down a BB9 minisuperspace path integral which yields a Green's function for the BB9 WDW operator and which agrees with the information about the TWF given in \cite{delCampo:1989hy}. (In our recent work \cite{DiazDorronsoro:2018wro}, consider a lapse contour which runs from $N = i \varepsilon, \varepsilon > 0,$ down the negative imaginary axis and which avoids the point $N = 0$.) The path integral we have in mind is not ``Lorentzian'' in the sense of \cite{FLT1}, but as we hope to have made clear with this paper this qualification is of no fundamental importance in minisuperspace path integral constructions of wave/Green's functions. We leave these further investigations into the tunneling wave function for future work.\footnote{Particularly interesting would be to compute the TWF in $\mathcal{D}_x$ for $\alpha > 0$. In particular one would learn if the same phase transitions at $\alpha = 2$ and $\alpha = 2(3+\sqrt{10})$ take place as in the NBWF. This seems implausible to us.}

\clearpage
\newpage
\section{Conclusion} \label{conclusionsec}
\noindent The main purpose of this paper was to continue the development and refinement of the no-boundary proposal in the context of an exactly solvable Bianchi IX minisuperspace model, building on \cite{DiazDorronsoro:2018wro,DiazDorronsoro2017} and staying close to the recently proposed definition of the no-boundary proposal in terms of a collection of saddle points \cite{Halliwell:2018ejl}. Our work was motivated in part by the need to address the challenges to the definition of the no-boundary proposal presented in \cite{FLT2,FLT3}, but it also contributes to the history, by now very long, of the development and applications of the no-boundary proposal.

Our work significantly substantiates and extends our earlier work on the Bianchi IX model \cite{DiazDorronsoro:2018wro} by giving a more detailed analysis of the saddle points, including a second topological contribution, by studying the phase transitions between the various saddle points and by confirming the expected normalization properties. From the latter follows the prediction of suppressed anisotropies, a clear refutation of the claims of Refs. \cite{FLT2,FLT3}, and some more detailed aspects of Refs. \cite{FLT2,FLT3} were addressed in detail. We also showed that this model may be viewed as a non-linear extension of the dS minisuperspace model perturbed by a single mode of either a tensor field or massless minimally coupled field.

We have in addition shown that our model has the expected isotropic limit, and that the no-boundary proposal predicts that massless scalar fluctuations around our BB9 model have the expected decaying Gaussian wave functions. We also carried out a comparison with the tunneling wave function in the BB9 model and found a large regime of parameter space in which the two proposals coincide.

Our work was primarily based on the no-boundary proposal defined as a collection of saddle points \cite{Halliwell:2018ejl}, which we found in practice to be a very useful guiding principle, but we also examined a number of aspects of full minisuperspace quantization using path integrals. We found in particular an unphysical dependency on certain features of the minisuperspace model such as the parameterization of the metric, which only reinforces the approach of Ref. \cite{Halliwell:2018ejl} as the most reliable definition of the no-boundary proposal.

To conclude, we address a recent criticism \cite{Vilenkin:2018oja} of the no-boundary proposal concerning the lack of a general definition as to which compact four-manifolds to include in the sum Eq. \eqref{NBWFsaddledef}. While we generally agree that this is an issue we offer some thoughts on why it may not be such a pressing one. First, a sum over compact manifolds is often held to be ill-defined due to the fact that they cannot be classified. (Although see Ref. \cite{doi:10.1063/1.526571} for an interesting alternative view on this.)
Secondly, in the semiclassical limit (which is the only regime in which we expect the current framework of quantum cosmology to be valid) and for a sufficiently small region of superspace, one expects only a single manifold to be relevant in the sum since the contributions from the various $\mathcal{M}$ are generally exponentially different. Our analysis of the NBWF in the BB9 minisuperspace model confirms this expectation. Finally, the sum-over-manifolds aspect of the definition of $\Psi$ might be circumvented or at least be profoundly redrawn by the holographic definition of the NBWF in terms of the partition function of a Euclidean CFT that is defined directly on the boundary \cite{Maldacena2002,Hertog2011}.

\vfill
\noindent{\bf Acknowledgements:} We are very grateful to Juan Diaz Dorronsoro and Yannick Vreys for their initial collaboration on this work, and to Jim Hartle for many useful conversations. We also thank Suddho Brahma, Reza Javadinezhad, Matthew Kleban, Jorma Louko, Giorgi Tukhashvili, Alex Vilenkin, Masaki Yamada, Dong-han Yeom and Cedric Yu for discussions. TH is supported in part by the C16/16/005 grant of the KULeuven and by the European Research Council grant no. ERC-2013-CoG 616732 HoloQosmos. OJ is supported by a James Arthur Graduate Fellowship, and wishes to thank the Asia Pacific Center for Theoretical Physics and the Belgian-American Chamber of Commerce for their hospitality while part of this work was completed.


\newpage
\appendix
\section{Detailed discussion of saddle points on $\mathbb{C}\text{P}^2 \setminus B^4$} \label{A1}
\noindent The \href{http://www.ma.rhul.ac.uk/~uvah099/Sat/reader.html}{Attentive Reader} will have noticed that in our discussion of the saddle points on the manifold $\mathbb{C}\text{P}^2 \setminus B^4$, in the interval $\alpha \in [5 - 3\sqrt{3} , 5 + 3\sqrt{3}] \equiv [\alpha_-,\alpha_+] \approx [-0.2,10.2]$ and at small $x$, we seem to have found only two saddle points instead of the expected four. The reason is that the (existing) differences between the saddle points appear only at a subleading order in $x$ in this interval, and Eq. \eqref{fpm} is too crude to capture this. More precisely we find the following perturbative expansions in $x$ for the four different saddle points for all values of $\alpha$ except $\alpha_\pm$:
\begin{align} \label{BoltNs}
N_+^\text{Bolt}(+) &= \left\{ \begin{array}{ll}
	\sqrt{3q} + g_0(\alpha) + \mathcal{O}(\sqrt{x}) + i \, \mathfrak{f}_-(\alpha,x) & \mbox{~~if } \alpha \in (\alpha_- , \alpha_+) \,, \\
    \displaystyle\sqrt{3q} + \mathcal{O}(\sqrt{x}) + i \left[ f_+(\alpha) + f_{+,1}(\alpha) x + \mathcal{O}(x^2) \right] & \mbox{~~elsewhere except } \alpha = 5 \pm 3\sqrt{3} \,,
    \end{array} \right. \notag \\
N_-^\text{Bolt}(+) &= N_+^\text{Bolt}(+)^* \,, \notag \\
N_+^\text{Bolt}(-) &= \left\{ \begin{array}{ll}
	\sqrt{3q} - g_0(\alpha) + \mathcal{O}(\sqrt{x}) + i \, \mathfrak{f}_+(\alpha,x) & \mbox{~~if } \alpha \in (\alpha_- , \alpha_+) \,, \\
    \displaystyle\sqrt{3q} + \mathcal{O}(\sqrt{x}) + i \left[ f_-(\alpha) + f_{-,1}(\alpha) x + \mathcal{O}(x^2) \right] & \mbox{~~elsewhere except } \alpha = 5 \pm 3\sqrt{3} \,,
    \end{array} \right. \notag \\
N_-^\text{Bolt}(-) &= N_+^\text{Bolt}(-)^* \,,
\end{align}
where $f_\pm$ were defined in Eq. \eqref{fpm} and
\begin{align}
	\mathfrak{f}_-(\alpha,x) &= -1 + f_1(\alpha)x - f_{3/2}(\alpha) x^{3/2} + \mathcal{O}(x^2) \,, \\
	\mathfrak{f}_+(\alpha,x) &= -1 + f_1(\alpha)x + f_{3/2}(\alpha) x^{3/2} + \mathcal{O}(x^2) \,, \\
	g_0(\alpha) &= \frac{s_-}{1+\alpha} \,, \\
	f_1(\alpha) &= - \frac{(1+\alpha)^2}{18} \,, \\
	f_{3/2}(\alpha) &= \frac{(1+\alpha)^2 (2\alpha-1)}{18 \, s_-} \,, \\
	f_{-,1}(\alpha) &= \frac{-\alpha  \left\{ \alpha  \left[ 3 \left( s_+ - 67 \right) + \alpha \left( \alpha - 2 + s_+ \right)\right] + 3 s_+ + 148\right\} - s_+ - 22}{18 (\alpha +1) s_+ } \,, \\
	f_{+,1}(\alpha) &= \frac{-\alpha  \left\{ \alpha  \left[3 \left(s_+ + 67 \right) - \alpha \left(\alpha -2 - s_+\right)\right] + 3 s_+ - 148 \right\} - s_+ + 22}{18 (\alpha +1) s_+} \,, \\
	s_+(\alpha) &= \sqrt{\alpha^2 - 10\alpha - 2} \,, \label{splus} \\
	s_-(\alpha) &= \sqrt{-\alpha^2 + 10\alpha + 2} \,. \label{sminus}
\end{align}
(In Eqns. \eqref{splus}-\eqref{sminus} we mean the positive square root, and recall that $q \equiv 12/[(1+\alpha)^2 x]$).

There are various useful things note about these formulas. The most basic one is that the series expansions of the saddles around $x = 0$ change form when $\alpha$ crosses $\alpha_-$ or $\alpha_+$. Secondly we must determine the regime in which the approximations implied by the expansions we have presented are accurate. At finite $\alpha$ this is not simply the regime $x \ll 1$. This is because the coefficient functions $f_{3/2}$ and $f_{\pm,1}$ diverge as $\alpha \rightarrow \alpha_\pm$ (since these are the zeroes of $s_\pm$) while the supposedly more leading terms remain finite.\footnote{The higher order terms in the real parts of the saddles, which we have not written, have the same trouble.} So in order to use the approximations implied by Eqns. \eqref{BoltNs} we must stay sufficiently far away from $\alpha = \alpha_\pm$. More precisely, for $\alpha \in (\alpha_- , \alpha_+)$ we must have
\begin{equation}
	x^{3/2} \ll s_- \,,
\end{equation}
and for $\alpha$ outside of this strip we must have
\begin{equation} \label{OUTcondition}
	x \ll s_+ \,.
\end{equation}
Having said this, there is nothing inherently singular about the points $\alpha = \alpha_\pm$, i.e., all $(4+3)$ solutions of Eq. \eqref{BoltHamiltonianconstraint} exist there and when $x \ll 1$ are well-approximated by the estimates we have given in \S\ref{Boltsaddlessec} of the main text (or by the non-singular terms in the expansions given here). We refer the reader to Appendix \ref{A2} for evidence of this. In other words, the expansions we have provided in this section should be viewed as asymptotic series instead of convergent ones, at least for $\alpha$ close to $\alpha_\pm$. This interpretation is further supported by the observation that the singular terms we have not written in equations \eqref{BoltNs} diverge more violently near $\alpha_\pm$ than the ones we have written. E.g. the $\mathcal{O}(x^2)$ term in the imaginary part of the second line in Eq. \eqref{BoltNs} diverges as $x^2/s_+^3$ near $\alpha = \alpha_\pm$, which is a stronger divergence than the previous term in the expansion ($\sim x / s_+$). So if we were to include such a term in an approximation to the saddles, this would further restrict us to the regime $x \ll (s_+)^{3/2}$ (cf. Eq. \eqref{OUTcondition} which less restrictive). So the amount of terms one should ideally retain in the asymptotic series in Eqns. \eqref{BoltNs} depends on the value of $\alpha$ (slightly more precisely, on how close $\alpha$ lies to $\alpha_\pm$).

Finally, to use the expansions for $N_+^\text{Bolt}(-)$ and $N_-^\text{Bolt}(-)$ at large $\alpha$, we see that we must have $\alpha^2 x \ll 1$ (for the $N_+^\text{Bolt}(+)$ and $N_-^\text{Bolt}(+)$ solutions $\alpha x \ll 1$ suffices at large $\alpha$). This should be contrasted with the condition $\alpha x \ll 1$, or $p \gg 1$, which delineates the regime $\mathcal{D}_x$ at large $\alpha$ (recall $\mathcal{D}_x$ was defined in our discussion of the contributions to the NBWF from the manifold $\overline{B^4}$ in \S\ref{B4classicalsec}). In the $(p,\alpha)$-plane this translates to the stronger condition $p \gg \alpha$ at large $\alpha$. In short, we have provided reliable approximations to all the no-boundary instantons on the manifold $\mathbb{C}\text{P}^2 \setminus B^4$ in the parameter regime
\begin{equation}
	p \cdot \frac{\alpha + 1}{(\alpha + 2)^2} \gg 1 ~~~ (\tilde{\mathcal{D}}_x) \,,
\end{equation}
while for the no-boundary instantons on $\overline{B^4}$ we needed only
\begin{equation}
	p \cdot \frac{\alpha + 1}{\alpha + 2} \gg 1 ~~~ (\mathcal{D}_x) \,.
\end{equation}

For the actions of the instantons corresponding to Eqns. \eqref{BoltNs} we obtain
\begin{align} \label{Boltactions}
	S_0 \left( N_+^\text{Bolt}(+) \right) &= \left\{ \begin{array}{ll}
	-\displaystyle\frac{2 \, p^{3/2}}{\sqrt{3(1+\alpha)}} + \mathcal{O}\left( \frac{1}{\sqrt{x}} \right) + i \, \mathcal{F}_+(\alpha,x) & \mbox{~~if } \alpha \in (\alpha_- , \alpha_+) \,, \\
 	-\displaystyle\frac{2 \, p^{3/2}}{\sqrt{3(1+\alpha)}} + \mathcal{O}\left( \frac{1}{\sqrt{x}} \right) + i \Big[ F_+(\alpha) + \mathcal{O}(x) \Big] & \mbox{~~elsewhere except } \alpha = \alpha_\pm \,,
\end{array} \right. \notag \\
	S_0 \left( N_-^\text{Bolt}(+) \right) &= - S_0 \left( N_+^\text{Bolt}(+) \right)^* \,, \notag \\
	S_0 \left( N_+^\text{Bolt}(-) \right) &= \left\{ \begin{array}{ll}
	-\displaystyle\frac{2 \, p^{3/2}}{\sqrt{3(1+\alpha)}} + \mathcal{O}\left( \frac{1}{\sqrt{x}} \right) + i \, \mathcal{F}_-(\alpha,x) & \mbox{~~if } \alpha \in (\alpha_- , \alpha_+) \,, \\
 	-\displaystyle\frac{2 \, p^{3/2}}{\sqrt{3(1+\alpha)}} + \mathcal{O}\left( \frac{1}{\sqrt{x}} \right) + i \Big[ F_-(\alpha) + \mathcal{O}(x) \Big] & \mbox{~~elsewhere except } \alpha = \alpha_\pm \,,
\end{array} \right. \notag \\
	S_0 \left( N_-^\text{Bolt}(-) \right) &= - S_0 \left( N_+^\text{Bolt}(-) \right)^* \,,
\end{align}
where
\begin{align}
	\mathcal{F}_+(\alpha,x) &= F_0(\alpha) + F_1(\alpha) x + F_{3/2}(\alpha) x^{3/2} + \mathcal{O}(x^2) \,, \\
	\mathcal{F}_-(\alpha,x) &= F_0(\alpha) + F_1(\alpha) x - F_{3/2}(\alpha) x^{3/2} + \mathcal{O}(x^2) \,, \\
	F_0(\alpha) &= \frac{2(\alpha-17)}{9} \,, \\
	F_1(\alpha) &= - \frac{2 \alpha}{9}(1+\alpha) \,, \\
	F_{3/2}(\alpha) &= \frac{s_-}{243} (1+\alpha) \left( 11 - 8\alpha - \alpha^2 \right) \,, \\
	F_+(\alpha) &= \frac{2}{9(1+\alpha)^2} \left[ \alpha^3 - 17 + 2 s_+ - \alpha^2 \left( 15 + s_+ \right) - \alpha \left( 33 - 10 s_+ \right) \right] \,, \\
	F_-(\alpha) &= \frac{2}{9(1+\alpha)^2} \left[ \alpha^3 - 17 - 2 s_+ - \alpha^2 \left( 15 - s_+ \right) - \alpha \left( 33 + 10 s_+ \right) \right] \,.
\end{align}
The effects of the divergent terms in Eqns. \eqref{BoltNs} near $\alpha = \alpha_\pm$ only show up at order $x^2$ in Eqns. \eqref{Boltactions} (and even then only in the real parts), giving us a rather reliable approximation of the action in the regime $\tilde{\mathcal{D}}_x$ by using only the non-divergent terms in the asymptotic expansions of Eqns. \eqref{BoltNs}.

\section{Phase transitions on $\mathbb{C}\text{P}^2 \setminus B^4$} \label{A2}
\begin{figure}[h!]
\centering
\includegraphics[width=0.8\textwidth]{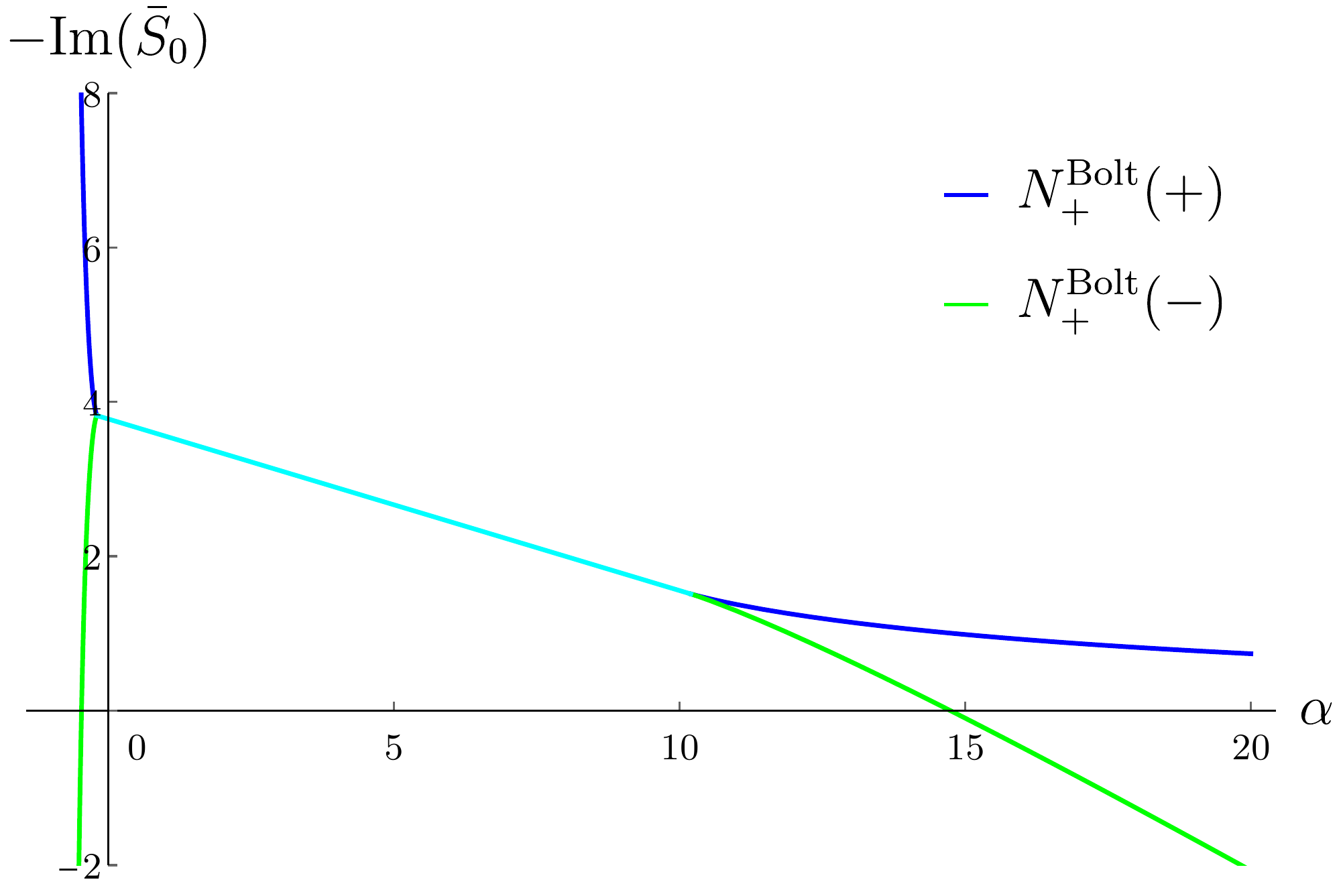}
\caption{\small The opposite of the imaginary part of the on-shell action, $-\text{Im}(S_0(N_s))$, is shown for the two kinds of regular saddle points on the manifold $\mathbb{C}\text{P}^2 \setminus B^4$ (which can be labeled by $N_s$) that have no-boundary initial conditions and match onto the arguments of the wave function $(p,q)$ on some spacelike slice. The potential contributions of these saddle points to the wave function are weighted according to $e^{-\text{Im}(S_0(N_s))/\hbar}$. At infinite $p$ and $q$, $\text{Im}(S_0(N_s))$ is only a function of $\alpha = p/q - 1$, whose shape is shown here.}
\label{BoltactionsFIG}
\end{figure}

\noindent We begin the discussion of phase transitions on the manifold $\mathbb{C}\text{P}^2 \setminus B^4$, or in other words the possible exchange in dominance between the various saddle points on $\mathbb{C}\text{P}^2 \setminus B^4$, by plotting the leading order behavior of the on-shell actions of the various saddle points in $\tilde{\mathcal{D}}_x$ (Eqns. \eqref{Boltactions}). This is shown in Figure \ref{BoltactionsFIG}.

\noindent As we have noted in Appendix \ref{A1}, and as can be seen in Figure \ref{BoltactionsFIG}, the two types of solutions seem to degenerate in the regime $\alpha \in [\alpha_-,\alpha_+] \equiv [5-3\sqrt{3},5+3\sqrt{3}] \approx [-0.2,10.2]$. This only truly happens at infinite volume (infinite $p$ and $q$, or $x = 0$ and finite $\alpha$), however.\footnote{Even there the degeneracy is not complete; while the difference in the imaginary part of the on-shell action tends to zero, there remains a finite difference in the real part of the on-shell actions. This can be traced back to Eqns. \eqref{BoltNs}: in the limit $x \rightarrow 0$ we have $N_+^\text{Bolt}(+) - N_+^\text{Bolt}(-) \rightarrow 2 g_0(\alpha)$. This results in a different real part of the on-shell action in Eqns. \eqref{Boltactions}.} An interesting question is whether a phase transition might happen at finite volume. By ``phase transition'' we mean a ``first order'' one, in which below some temperature $(\alpha)$ there is one metastable phase and one stable phase, where the free energy ($-\text{Im}(S_0(N_s))$) of the former is greater than that of the latter. At some critical temperature the free energy curves intersect at a non-zero angle, so that there is a discontinuity in the first derivative of lowest free energy curve as a function of temperature. (According to this definition the transitions across $\alpha = \alpha_\pm$ that one observes in Figure \ref{BoltactionsFIG} are not phase transitions.)

\begin{figure}[h!]
\centering
\includegraphics[width=0.8\textwidth]{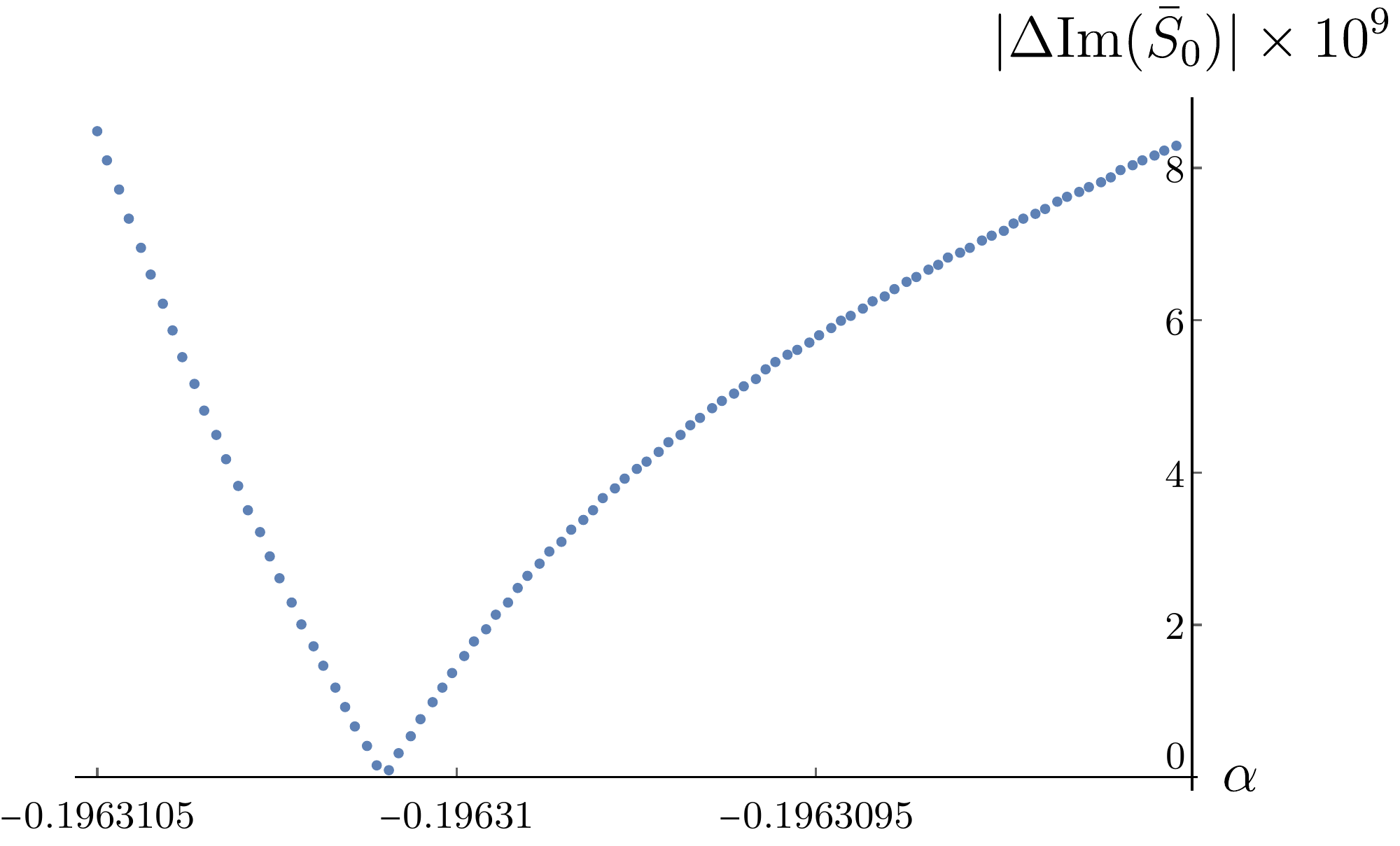}
\caption{\small The absolute difference between the imaginary parts of the on-shell actions of the two types of no-boundary saddle points on the manifold $\mathbb{C}\text{P}^2 \setminus B^4$ is plotted as a function of the squashing parameter $\alpha$, at a finite value of $x$ (chosen to be $10^{-3}$ here) and near $\alpha = \alpha_- = 5-3\sqrt{3} \approx -0.2$. This result confirms that a phase transition around $\alpha = \alpha_-$ is possible. By ``phase transition'' we mean an exchange of dominance amongst saddle points that contribute to a functional integral over geometries on $\mathbb{C}\text{P}^2 \setminus B^4$ that are weighted by $e^{iS/\hbar}$ with $S$ the Einstein-Hilbert action. The approximations given in Appendix \ref{A1} are not valid near $\alpha = \alpha_-$ at finite $x$, so the existence of an intersection between $-\text{Im}(\bar{S}_0)$ for the two types of saddle points near $\alpha = \alpha_-$ and at finite $x$ could not be inferred from the information in Appendix \ref{A1}.}
\label{DeltaImS0_near_alphaminus}
\end{figure}

\begin{figure}[t]
\centering
\includegraphics[width=0.8\textwidth]{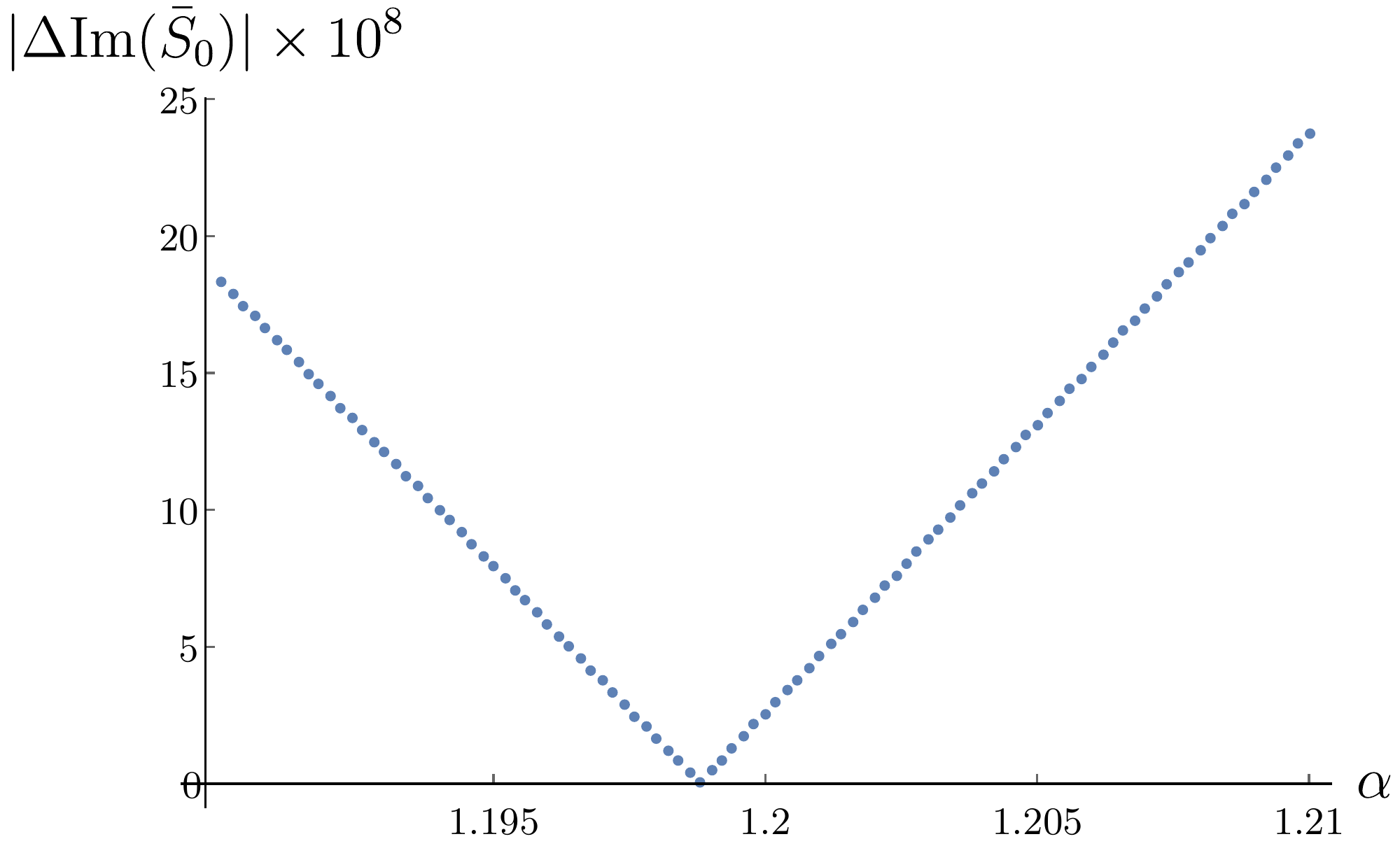}
\caption{\small The same setup as in Figure \ref{DeltaImS0_near_alphaminus}, but now near $\alpha = 3\sqrt{3} - 4 \approx 1.2$. Contrary to the situation in the regime around $\alpha = \alpha_-$, we do expect the approximations given in Appendix \ref{A1} to be valid around $\alpha \approx 1.2$. The approximations suggest an intersection between the curves $-\text{Im}(\bar{S}_0)$ in this regime, and the numerics confirm this expectation.}
\label{DeltaImS0_near_alphac}
\end{figure}

\begin{figure}[b]
\centering
\includegraphics[width=0.8\textwidth]{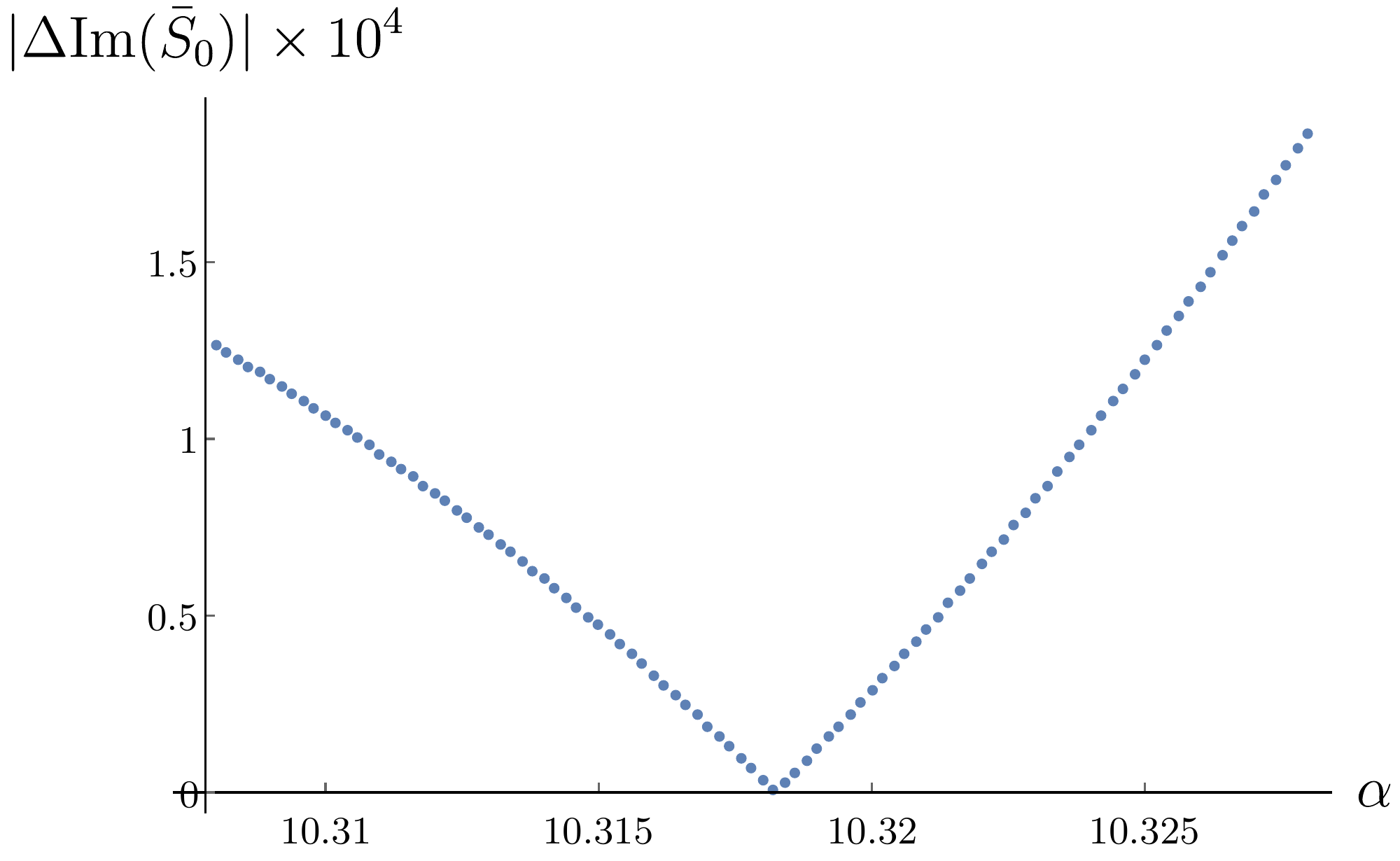}
\caption{\small The same setup and qualitative conclusions as in Figure \ref{DeltaImS0_near_alphaminus}, but now near $\alpha = \alpha_+$.}
\label{DeltaImS0_near_alphaplus}
\end{figure}

In brief we find that phase transitions at finite volume are possible. To definitively establish this we turn to numerics, the results of which are shown in Figures \ref{DeltaImS0_near_alphaminus}-\ref{DeltaImS0_near_alphaplus}. In these numerics we have computed the solutions of Eq. \eqref{BoltHamiltonianconstraint} to machine precision and evaluated the action \eqref{S0bolt} on them.

As we noted in Appendix \ref{A1}, the approximations we have written there are not valid for $\alpha$ close to $\alpha_\pm$ (at fixed $x$). So from the expansions in Appendix \ref{A1} we cannot extract the exact behavior of the actions near $\alpha_\pm$, in particular we cannot confirm or rule out an intersection in this regime at finite $x$. The numerics confirm an intersection near both of these points at finite $x$ (Figures \ref{DeltaImS0_near_alphaminus} and \ref{DeltaImS0_near_alphaplus}). On the other hand, we \textit{can} analytically justify the intersection between the (imaginary parts of the) actions that we observe numerically near the point $\alpha \approx 1.2$. For this we use the equations \eqref{Boltactions}, in particular the expressions for the imaginary part of $\bar{S}_0$ inside the interval $\alpha \in (\alpha_-,\alpha_+)$. From there we infer that at small $x$, there should be an intersection between the curves near the zero of the function $F_{3/2}$ in the interval $(\alpha_-,\alpha_+)$, which is at $\alpha = 3\sqrt{3}-4 \approx 1.2$. This value lies ``far away'' from $\alpha = \alpha_\pm$, so that we are hopeful the calculation is trustworthy. This is indeed confirmed by the numerics (see Figure \ref{DeltaImS0_near_alphac}).

\clearpage
\newpage
\phantom{XXX}
\vfill
\vfill

\newpage

\clearpage
\newpage
\section{A non-linear extension of the de Sitter + ($n=2$) gravitational wave mode minisuperspace} \label{nonlinearsec}
\noindent In this section we explain how the anisotropic BB9 minisuperspace model can be viewed as a non-linear extension of the isotropic dS minisuperspace model perturbed by a particular tensor (or gravitational wave) mode. We begin by reviewing the latter model in the context of the NBWF. The general tensor-perturbed minisuperspace model can be defined through the four-metric Ansatz
\begin{equation} \label{tensorminisuperspace}
	2 \pi^2 \di s^2 = -\frac{N^2}{Q(\tau)} \di \tau^2 + Q(\tau) \left( \Omega_{ij} + \varepsilon_{ij} \right) \di \Omega^i \di \Omega^j \,,
\end{equation}
where $\Omega = (\theta,\phi,\psi)$ are the standard angles on $S^3$ with $\theta,\phi \in [0,\pi]$ and $\psi \in [0,2\pi)$, $\psi \cong \psi + 2\pi$, and $\Omega_{ij} = \text{diag}\left[ 1,\sin^2(\theta), \sin^2(\theta) \sin^2(\phi) \right]_{ij}$ are the components of the round metric on the unit $S^3$ in these coordinates. Here we have assumed that the four-manifold on which the metric \eqref{tensorminisuperspace} lives is the closed four-ball $\overline{B^4}$, with $\tau \in [0,1]$ a radial coordinate, and we will only take into account such geometries in the no-boundary sum.\footnote{Of course one can be more general and sum over other geometries as well. These will require an adjusted metric Ansatz. For example, in \S\ref{Boltinhomogeneoussec} we discussed massless minimally coupled scalar fluctuations around the no-boundary background saddle point which lives on $\mathbb{C}\text{P}^2 \setminus B^4$.} The lapse $N$ has been gauge-fixed to a constant and is integrated over in a minisuperspace path integral representation of the NBWF. The wave function is defined on three-spheres with metrics given by the spatial part of Eq. \eqref{tensorminisuperspace}, and is thus a functional of $Q \in \mathbb{R}^+$ and the collection of functions $\{\varepsilon_{ij}(\Omega)\}$. We will take the perturbation $\varepsilon$ to be transverse and traceless. As such it may be expanded in terms of the (real) transverse traceless tensor harmonics $G_{ij}$ on $S^3$ at any value of $\tau$,
\begin{equation}
	\varepsilon_{ij}(\tau,\Omega) = 2 \sum_{n,l,m} \varphi_{nlm}(\tau) \, (G_{ij})_{nlm}(\Omega) \,,
\end{equation}
where
\begin{equation}
	\nabla^2 (G_{ij})_{nlm} = -(n^2 + 2n - 2) (G_{ij})_{nlm} \,,
\end{equation}
and $n,l,m$ are integers with $n \in \{2,3,\dots \} , l \in \{ 2,3,\dots,n \}$, and $m \in \{ -l, -l+1, \dots, l \}$, and covariant derivatives are with respect to $\Omega_{ij}$. (See Ref. \cite{Halliwell:1984eu} for a more general discussion, and Ref. \cite{PhysRevD.18.1773} for an explicit construction of the harmonics.) The harmonics satisfy
\begin{equation}
	\nabla^i (G_{ij})_{nlm} = 0 = \Omega^{ij} (G_{ij})_{nlm} \,,
\end{equation}
and are normalized according to
\begin{equation}
	\int_{S^3} \di^3 \Omega \, \sqrt{\Omega} \, (G_{ij})_{nlm} (G^{ij})_{n'l'm'} = 2 \pi^2 \delta_{nn'} \delta_{ll'} \delta_{mm'} \,.
\end{equation}
Thus the wave function can equivalently be viewed as a function on the infinite set of real variables $Q$ and $\{ \varphi_{nlm} \}$. From here on we will suppress the $l$ and $m$ indices. The action, expanded to second order in the $\varphi_n$, reads
\begin{align} \label{tensoraction}
	&S_\text{tensor}[Q,\{\varphi_n\};N] = \int_{\overline{B^4}} \di^4 x \sqrt{-g} \left( \frac{R}{2} - 2 \pi^2 \Lambda \right) + \int_{S^3} \di^3 y \sqrt{h} \, K \\ 
	&= S_\text{isotropic}[Q;N] + \sum_n \int_0^1 \di \tau \, N \left[ \frac{Q^2}{2 N^2} \dot{\varphi}_n^2 + \frac{2}{N^2} Q \dot{Q} \varphi_n \dot{\varphi}_n + \left( Q + \frac{3 \dot{Q}^2}{4N^2} - \frac{n(n+2) + 2}{2} \right) \varphi_n^2 \right] \,, \notag
\end{align}
where
\begin{equation} \label{isotropicaction}
	S_\text{isotropic}[Q;N] = \int_0^1 \di \tau \, N \left( - \frac{3}{4N^2} \dot{Q}^2 + 3 - Q \right) \,,
\end{equation}
and we have again absorbed $\Lambda$ into $S$, $N$ and $Q$ in the second line.\footnote{The action \eqref{tensoraction} is appropriate for Dirichlet, or position, boundary conditions at $\tau = 0$ and $\tau = 1$ on all fields. It is also appropriate for Neumann, or momentum, boundary conditions at $\tau = 0$ on the scale factor $Q$ and Dirichlet boundary conditions on $Q$ at $\tau = 1$, and Dirichlet boundary conditions on the $\varphi_n$ at both boundaries. (See also footnote \ref{bconditionsftn}.) Note that our $3+1$ decomposition of the four-manifold has introduced an artificial boundary at $\tau = 0$.}

The correspondence between the BB9 minisuperspace model (in a perturbative limit) and this minisuperspace model only holds \textit{on-shell}. That is, it holds only when the constraint following from the action \eqref{tensoraction} is satisfied and the EOM following from variations of the tensor modes $\varphi_n$ are imposed. (By ``correspondence'' we mean an equality or simple relation between the actions of the two theories after an identification of the degrees of freedom.) To zeroth order in the perturbations $\varphi_n$ the constraint is the constraint following from $S_\text{isotropic}$ alone,
\begin{equation}
	\frac{3 \dot{Q}^2}{4 N^2}+ 3 - Q = \text{terms quadratic in the fluctuations,}
\end{equation}
which, with an appropriate\footnote{For instance $\Pi_Q(0) = -3i$ with $\Pi_Q = -3\dot{Q}/2N$ the momentum conjugate to $Q$.} no-boundary boundary condition at $\tau = 0$ on the field $Q$, fixes $Q$ and $N$ to their background values $\bar{Q}(\tau)$ and $N_s$ up to an inconsequential sign choice \cite{Louko1988,Halliwell1988,DiazDorronsoro2017,DiazDorronsoro:2018wro},
\begin{align}
	N_s &= 3 \left( \pm \sqrt{Q/3 - 1} - i \right) \,, \\
	\bar{Q}(\tau;N_s) &= \left(Q - N_s^2/3 \right) \tau + (N_s^2/3) \tau^2 \,.
\end{align}
(In these equations $Q$ is the real number in the argument of the wave function.) With this, one can show that the EOM for the $\varphi_n$ are, to lowest non-trivial order in $\varphi_n$, given by
\begin{equation}
	\ddot{\varphi}_n + 2\frac{\dot{\bar{Q}}}{\bar{Q}} \dot{\varphi}_n + N_s^2 n(n+2) \frac{\varphi_n}{\bar{Q}^2} = 0 \,.
\end{equation}
With Dirichlet boundary conditions on the $\varphi_n$ at $\tau = 0$ and $\tau = 1$, $\varphi_n(0) = 0, \varphi_n(1) = \varphi_{n,1}$, the action of the solution $(\bar{Q},\bar{\varphi}_n,N_s)$ can be written as
\begin{equation} \label{Sbartensor}
	\bar{S}_\text{tensor} = \bar{S}_\text{isotropic} + \sum_n \frac{Q \, \dot{\bar{Q}}(1)}{N_s} \varphi_{n,1}^2 + \sum_n \int_0^1 \di \tau \, N_s \left( \frac{\bar{Q}^2}{2 N_s^2} \dot{\bar{\varphi}}_n^2  - \frac{n(n+2)}{2} \bar{\varphi}_n^2 \right) \,.
\end{equation}
The solution $\bar{\varphi}_n$ is given by (e.g. \cite{HarHal1990})
\begin{align}
		\bar{\varphi}_n(\tau;N_s) &= \frac{\xi(\tau;N_s)}{\xi(1;N_s)} \varphi_{n,1} \,, \\
	\xi(\tau;N_s) &= \tau^{n+1} \, \bar{Q}^{-\left(n+2\right)/2} \left[ \dot{\bar{Q}} + (n+1) \dot{\bar{Q}}(0) \right] \,.
\end{align}
The integrals in Eq. \eqref{Sbartensor} evaluate to
\begin{equation} \label{JJeq}
	\frac{Q^2}{2 N_s}\varphi_{n,1} \dot{\bar{\varphi}}_n(1) = \frac{1}{2} \sum_n n(n+2) Q \left( \frac{\mp\sqrt{Q / 3 - 1} +(n+1)i}{n(n+2) + Q / 3} \right) \, \varphi_{n,1}^2
\end{equation}
while the middle terms in Eq. \eqref{Sbartensor} evaluate to
\begin{equation}
	\pm 2 Q \sqrt{Q/3-1} ~ \varphi_{n,1}^2 \,.
\end{equation}

The connection with the BB9 model is given via the identifications
\begin{align}
	p &= Q \, e^{\sqrt{2/3} \, \varphi} \,, \label{pQphi2} \\
	q &= Q \, e^{-\sqrt{8/3} \, \varphi} \,, \label{qQphi2} \\
	N(\tau) &\rightarrow e^{-\sqrt{2/3} \, \varphi} N(\tau) \label{tautphi} \,.
\end{align}
With this correspondence, and with the field $\varphi$ treated perturbatively, the anisotropic theory with action given in Eq. \eqref{Sqp} reduces to
\begin{equation} \label{SpertBB9}
	S_{\text{pert. BB9}}[Q,\varphi;N] = S_\text{isotropic}[Q;N] + \int_0^1 \di \tau \, N \left( \frac{Q^2}{2 N^2} \dot{\varphi}^2  - 4 \varphi^2 \right) \,.
\end{equation}
In this expression we recognize part of the action \eqref{Sbartensor} for the $n = 2$ contribution. The difference is that the action \eqref{Sbartensor} is an on-shell expression, while the action \eqref{SpertBB9} is also valid off-shell (neither the constraint nor the EOM have been imposed), and that there is an extra term in \eqref{Sbartensor}. \textit{On-shell}, and to lowest non-trivial order in $\varphi$, we have
\begin{equation} \label{BB9tensorrelation}
	S_{\text{pert. BB9}}[\bar{Q},\bar{\varphi};N_s] = S_\text{tensor}[\bar{Q},\bar{\varphi}_2 = \bar{\varphi};N_s] \mp \sum_n 2 Q \sqrt{Q/3-1} ~ \varphi_{n,1}^2 \,,
\end{equation}
which establishes the correspondence between the (slightly anisotropic) BB9 model and the perturbative $n = 2$ tensor mode model we have alluded to. We say the BB9 model is a non-linear extension of the dS + ($n=2$) tensor mode model because of the relations \eqref{pQphi2}, \eqref{qQphi2} and \eqref{BB9tensorrelation}: the BB9 action reduces, up to a \textit{real} factor which does not affect how perturbations are weighted, to the $n = 2$ tensor action under an appropriate identification of the degrees of freedom and in a perturbative limit.

Since we have computed the NBWF on arbitrarily deformed three-spheres in the main text, we can in particular take the limit of small anisotropy $\alpha \approx 0$ and compare the result\footnote{More precisely it is only the contribution to the NBWF from the $\overline{B^4}$ topology that is relevant here, i.e. Eq. \eqref{PsiHHB4approx}. By ``result'' we mean the leading order in $\hbar$ behavior of the wave function, i.e. the on-shell action. As we have stressed on multiple occasions one should not expect subleading terms to match.} to the result we reviewed above for the perturbative $n = 2$ tensor mode model, thus testing the relation \eqref{BB9tensorrelation} and the consistency of our calculations. (We will perform this check in the regime $Q \gg 1$ since that is the regime we have focussed on in the main text.) According to Eq. \eqref{Splus}, and using the identifications \eqref{pQphi2}-\eqref{qQphi2}, we have
\begin{equation}
	\bar{S}_{\text{pert. BB9}} \approx \mp \frac{2}{\sqrt{3}} Q^{3/2} \pm \sqrt{3} \left( 3 - 4 \varphi^2 \right) \sqrt{Q} + \mathcal{O}(Q^{-1/2}) - 6 \left[ 1 - 6 \varphi^2 + \mathcal{O}(Q^{-1}) \right] i \,.
\end{equation}
According to the discussion in this appendix, this is indeed equal to the RHS of Eq. \eqref{BB9tensorrelation}.

\clearpage
\newpage
\section{A non-linear extension of the de Sitter + ($n=2$) massless scalar \\ minisuperspace} \label{nonlinearsec2}
\noindent Here we discuss how the BB9 minisuperspace model can be viewed as a non-linear extension of the dS minisuperspace model containing a (particular mode of a) massless minimally coupled scalar. Our conventions match those of Appendix \ref{nonlinearsec}, and the discussion is analogous. The metric Ansatz for the dS + massless scalar model is
\begin{equation} \label{scalarminisuperspace}
	2 \pi^2 \di s^2 = -\frac{N^2}{Q(\tau)} \di \tau^2 + Q(\tau) \di \Omega_3^2 \,,
\end{equation}
and the scalar is decomposed into (real) scalar harmonics on $S^3$ at each $\tau$ (see \cite{Halliwell:1984eu,PhysRevD.18.1773}),
\begin{equation}
	\varphi(\tau,\Omega) = \sum_{n,l,m} \varphi_{nlm}(\tau) Y_{nlm}(\Omega) \,,
\end{equation}
where $n \in \{ 0,1,2,\dots \}, l \in \{ 0,1,\dots,n \}, m \in \{ -l, -l+1,\dots,l \},$
\begin{align}
	\int_{S^3} \di^3 \Omega \, \sqrt{\Omega} \, Y_{nlm} Y_{n'l'm'} &= 2 \pi^2 \delta_{nn'} \delta_{ll'} \delta_{mm'} \,, \\
	\nabla^2 Y_{nlm} &= -n(n+2) Y_{nlm} \,.
\end{align}
Suppressing $l$ and $m$, the dS + scalar (off-shell) minisuperspace action reads
\begin{align} \label{scalaraction}
	S_\text{scalar}[Q,\{\varphi_n\};N] &= \int_{\overline{B^4}} \di^4 x \sqrt{-g} \left( \frac{R}{2} - 2 \pi^2 \Lambda - \frac{1}{2} g^{\mu \nu} \partial_\mu \varphi \, \partial_\nu \varphi \right) + \int_{S^3} \di^3 y \sqrt{h} \, K \notag \\ 
	&= S_\text{isotropic}[Q;N] + \sum_n \int_0^1 \di \tau \, N \left( \frac{Q^2}{2 N^2} \dot{\varphi}_n^2  - \frac{n(n+2)}{2} \varphi_n^2 \right) \,.
\end{align}

For the $n=2$ scalar mode, we see that this exactly matches Eq. \eqref{SpertBB9}, which is the action for the BB9 minisuperspace model for slightly anisotropic configurations $\varphi \approx 0$. Thus
\begin{equation}
	S_{\text{pert. BB9}}[Q,\varphi;N] = S_\text{scalar}[Q,\varphi_2 = \varphi;N] \,.
\end{equation}
In contrast to the correspondence between the perturbative BB9 minisuperspace model and the dS + ($n=2$) perturbative tensor model (Appendix \ref{nonlinearsec}), the correspondence between the perturbative BB9 minisuperspace model and the dS + ($n=2$) scalar model holds also \textit{off-shell} (instead of only on-shell). 

\section{The off-shell structure in minisuperspace models depends on the \\ construction: some examples} \label{offshellsec}
\noindent In this appendix we illustrate by means of two simple examples that the off-shell structure of minisuperspace path integrals depends sensitively on the choice of gauge for the lapse function $N(\tau)$ and on the boundary conditions $\mathcal{B}$ imposed on the paths. The examples are well-known from the literature, so we will skip over some details here and focus on the results.

Let us first clarify what we mean by ``off-shell structure of minisuperspace path integrals''. The minisuperspace path integral construction of solutions to the WDW equation, or Green's functions of the WDW operator, takes the general form
\begin{equation}
	\Psi(y) = \int \mathcal{D} N \, \mathcal{D}x^\alpha \, \mathcal{D}\Pi_\alpha ~ \Delta_\text{FP} ~ \delta\left( \dot{N} - \chi(x,\Pi,N) \right) \, e^{iS/\hbar} \,,
\end{equation}
where we singled out the contribution from a single four-manifold (our analysis applies to each contribution separately). Here $\chi$ is an essentially arbitrary gauge-fixing function and $\Delta_\text{FP}$ is the associated determinant which renders the entire expression independent of $\chi$. The only gauge choice that has been used in practical computations to our knowledge is $\chi \equiv 0$. This choice sets the lapse to an undetermined number which labels the gauge orbits, and one can show that $\Delta_\text{FP}$ is constant in this case. Specifying the integration ranges and boundary conditions on the lapse and fields, one arrives at (see Eq. \eqref{NBWFminisuperspace})
\begin{align}
	\Psi(y) &= \int_\mathcal{C}\di N ~ K(y,N;\mathcal{B},0) \,, \label{minisuperspace1} \\
	K(y,N;\mathcal{B},0) &= \hspace{-3mm} \overset{\hspace{3mm}x(1) = y}{\int_\mathcal{B}} \hspace{-3mm} \mathcal{D}x^\alpha \, \mathcal{D}\Pi_\alpha ~ e^{i S[x,\Pi;N]/\hbar} \,, \label{minisuperspace2} \\
	S[x,\Pi;N] &= \int_0^1 \di \tau \left( \Pi_\alpha \dot{x}^\alpha - N H \right) + (\text{boundary term appropriate to } \mathcal{B}) \,. \label{minisuperspace3}
\end{align}
To evaluate Eq. \eqref{minisuperspace1}, perhaps approximately, in the literature one usually employs a two-step semiclassical approximation. One first writes the integrand $K$ of the $N$-integral as an asymptotic series in $\hbar$, keeping only the leading term:
\begin{equation} \label{Kapprox}
	K(y,N;\mathcal{B},0) \approx \mathcal{P} \, e^{i S_0 / \hbar} \,.
\end{equation}
The off-shell structure we mentioned above refers to the properties of the functions $\mathcal{P}$ and $S_0$ in the complex $N$-plane. These properties include poles, essential singularities, branch points and relative homology groups of the Morse function $\text{Re}(i S_0 / \hbar)$. These last groups essentially determine the possible contours $\mathcal{C}$ in the definition \eqref{minisuperspace1} which lead to a convergent integral \cite{Witten2010}. After the step \eqref{Kapprox}, the $N$-integral is in turn approximated in the $\hbar \rightarrow 0$ limit by the method of steepest descent, so that in the end
\begin{equation}
	\Psi \approx \mathcal{A} \, e^{i \bar{S}_0 / \hbar} \,.
\end{equation}
There are several reservations one may have about this two-step procedure. A first point of caution, which has been brought up before \cite{Halliwell1990}, is that the properties of the approximation \eqref{Kapprox} in the complex $N$-plane may not reflect the properties of the exact expression \eqref{minisuperspace2}.\footnote{With ``properties'' we again mean e.g. poles, essential singularities and branch points. Note that in general neither the expression \eqref{minisuperspace2} nor its approximation \eqref{Kapprox} need be analytic functions of $N$ at fixed $y$ and $\mathcal{B}$. This calls for particular caution when evaluating contour integrals such as Eq. \eqref{minisuperspace1}.} Since we know of no way to concretely illustrate this reservation -- it would require a system where the semiclassical approximation to a path integral is not exact and yet known in closed form, a scenario which may not even exist! \cite{Schulman:1981vu} -- we will not take it too seriously. Anyhow this possible issue can be avoided by considering models where the approximation \eqref{Kapprox} is exact. Much of the discussion in the literature (e.g. \cite{Halliwell1988,Halliwell1990,Garay1990,DiazDorronsoro2017,DiazDorronsoro:2018wro}), which includes our recent work and the examples we discuss below, has focussed on such models. A second, not unrelated but more serious point, is that the semiclassical approximation \eqref{Kapprox} may change its functional form as $N$ moves across the complex plane. In other words, something like a ``phase transition'' may occur as $N$ crosses a line in the complex plane.\footnote{These are not the same phase transitions as those that occur for the contributions to the NBWF from the $\mathbb{C}\text{P}^2 \setminus B^4$ topology, which were discussed in \S\ref{Boltsec} of the main text and in Appendix \ref{A2}. Those phase transitions happen when the arguments of the wave function cross a certain real line in the minisuperspace, as opposed to the situation here where the lapse which crosses a generally complex line.} As we demonstrate in Appendix \ref{FLTsec}, this is what happens in the dS + massless minimally coupled scalar minisuperspace model that was reviewed in Appendix \ref{nonlinearsec2} and was recently discussed by Feldbrugge et al. in \cite{FLT2,FLT3}. This phenomenon may complicate the evaluation of the integral \eqref{minisuperspace1} in the $\hbar \rightarrow 0$ limit. \\

Let us now turn to a first example, namely the dS minisuperspace model in $3+1$ dimensions with two different choices for $\mathcal{B}$ in Eq. \eqref{minisuperspace2}. The dS minisuperspace model is defined via the metric Ansatz \eqref{scalarminisuperspace}, which specifies the wave function \eqref{minisuperspace1} up to the choices of $\mathcal{B}$ and $\mathcal{C}$. The possible quantum-mechanical boundary conditions $\mathcal{B}$ that one can impose at the south pole of the geometry in the context of the NBWF ($:=$ the point $\tau = 0$ in a $3+1$ decomposition of the space) have been discussed at length in \cite{Halliwell1988,Halliwell1990}. Their discussion can be summarized as follows: as $\hbar \rightarrow 0$, the no-boundary amplitude should be determined by a regular solution to Einsteins equations on a compact manifold. This condition implies a specific behavior of the metric near the south pole, as we have mentioned in the Introduction to this paper. For the dS minisuperspace model, this is $Q(\tau) \sim \pm 2iN \tau + \mathcal{O}(\tau^2)$ as $\tau \rightarrow 0$, or $Q(0) = 0, \Pi_Q(0) = \mp 3i$. In a path integral over one quantum-mechanical degree of freedom we expect to be able to specify a single boundary condition at $\tau = 0$. In this case the most straightforward options are
\begin{equation}
	\mathcal{B} = \left\{ Q(0) = 0 \right\} \,, \,\,\,\, \text{(D)}
\end{equation}
or
\begin{equation} \label{BdSminisuperspace}
	\mathcal{B} = \left\{ \Pi_Q(0) = -3i \right\} \,, \,\,\,\, \text{(N)}
\end{equation}
where D and N stand for ``Dirichlet'' and ``Neumann'' boundary conditions respectively.\footnote{\label{fluctfootnote}One motivation for the choice of sign of the initial momentum we have made in Eq. \eqref{BdSminisuperspace} is that the Euclidean action of the classical configuration that dominates the path integral in the small volume regime, $Q < 3/\Lambda$, is negative with this choice (following \cite{PhysRevD.28.2960}). More pertinently, the wave function of (massless, scalar) fluctuations around the background that is selected by the choice of sign in Eq. \eqref{BdSminisuperspace} is normalizable when this wave function is defined via a path integral with Dirichlet boundary conditions at the south pole \cite{HarHal1990,DiazDorronsoro2017} (see also \S\ref{inhomogeneoussec}). For the opposite choice of sign in Eq. \eqref{BdSminisuperspace} the fluctuation wave function, defined via Dirichlet boundary conditions, would not have been normalizable. This conclusion can be reversed if other boundary conditions are considered for the fluctuations \cite{Vilenkin:2018dch,DiTucci:2019dji}.} We consider both choices. Straightforward calculation shows
\begin{align}
	\Psi_\text{HH}^{(D)}(Q) &= \int_{\mathcal{C}_\text{D}} \di N \, \frac{1}{\sqrt{N}} \, \exp \left[ \frac{i}{\hbar} \left( \frac{N^3}{36} + \left( 3 - \frac{Q}{2} \right) N - \frac{3 Q^2}{4 N} \right) \right] \,, \label{PsiHHD} \\
	\Psi_\text{HH}^{(N)}(Q) &= e^{3Q/\hbar} \int_{\mathcal{C}_\text{N}} \di N \, \exp \left[ \frac{i}{\hbar} \left( \frac{N^3}{9} - Q N \right) - \frac{N^2}{\hbar} \right] \,, \label{PsiHHN}
\end{align}
where we have absorbed $\Lambda$ into $Q, N$ and $\hbar$, and we have included a label on the contours which are to be determined. (We neglect overall $Q$-independent factors.) The central point we would like to make is that the properties of the integrands in Eqns. \eqref{PsiHHD} and \eqref{PsiHHN} in the complex $N$-plane are different: e.g. while the integrand in Eq. \eqref{PsiHHN} is an entire function, the integrand in Eq. \eqref{PsiHHD} has an essential singularity at $N = 0$ which is also a branch point. Evidently, choosing the same contour for both boundary conditions, $\mathcal{C}_\text{D} = \mathcal{C}_\text{N}$, does not imply that the respective integrals are equal. Instead, to make $\Psi_\text{HH}^{(D)}$ and $\Psi_\text{HH}^{(N)}$ agree, at least in the semiclassical limit, the contours must be chosen such that the same saddle points dominate the respective integrals. If we choose a branch cut for the square root appearing in Eq. \eqref{PsiHHD} that lies along the positive imaginary $N$-axis, appropriate choices are $\mathcal{C}_\text{D} = \mathbb{R}_\downarrow$ -- the contour that lies along the real line except that it avoids the origin by passing into the $\{ \text{Im}(N) < 0 \}$ half-plane \cite{DiazDorronsoro2017} -- and $\mathcal{C}_\text{N} = \mathbb{R}$. Note that changing $\mathcal{C}_\text{N}$ from $\mathbb{R}$ to $\mathbb{R}_\uparrow$ would not alter $\Psi_\text{HH}^{(N)}$, but that changing $\mathcal{C}_\text{D}$ from $\mathbb{R}_\downarrow$ to $\mathbb{R}_\uparrow$ (together with an adjusted choice for the branch cut) would radically alter\footnote{This statement should be qualified: in the pure dS minisuperspace model that we are considering here, the integral in Eq. \eqref{PsiHHD} defined via $\mathcal{C}_\text{D} = \mathbb{R}_\downarrow$ and the same expression defined by the choice $\mathcal{C}_\text{D} = \mathbb{R}_\uparrow$ differ in the semiclassical limit by a relative factor $e^{12/\hbar \Lambda}$ and a phase, which are $Q$-independent. These would be irrelevant to any discussion of normalization or probabilities, so the two wave functions should in fact be identified. More precisely we mean that the choices $\mathcal{C}_\text{D} = \mathbb{R}_{\uparrow,\downarrow}$ imply very different states when fluctuations are included (cf. footnote \ref{fluctfootnote}).} $\Psi_\text{HH}^{(D)}$ \cite{Halliwell1988,FLT1,DiazDorronsoro2017}. Choosing $\mathcal{C}_\text{D} = \mathbb{R}_\downarrow, \mathcal{C}_\text{N} = \mathbb{R}$ we can evaluate \eqref{PsiHHD} and \eqref{PsiHHN} in closed form, $\forall \hbar$:
\begin{equation}
	\Psi_\text{HH}^{(N)}(Q) = \text{Ai}\left[ - \left( \frac{9}{\hbar \Lambda} \right)^{2/3} \left( \frac{\Lambda Q}{3} - 1 \right) \right] = \Psi_\text{HH}^{(D)}(Q) \,,
\end{equation}
where we have reintroduced $\Lambda$. Finally we note that this first example demonstrates that the no-boundary proposal does not have a unique implementation in terms of a minisuperspace path integral. There are various ways to construct it, and these are in agreement with one another. \\

We now turn to our second example, which is meant to illustrate the sensitivity of the off-shell structure of minisuperspace path integrals to the choice of gauge for the histories summed over (see also \cite{PhysRevD.43.2730}). For clarity, we have in mind to remain within the same minisuperspace model (i.e. the wave function is a functional of the same kinds of three-metrics), but to construct the NBWF by summing over four-geometries using a different gauge-fixing for the lapse function. Unfortunately, in almost all the examples in the literature we are familiar with, in a given minisuperspace model there is a single known gauge in which the calculation of the NBWF (formulated as a minisuperspace path integral) can be carried out analytically.\footnote{The model in \cite{PhysRevD.43.2730} is an exception and that paper illustrates our point, perhaps more clearly than via example we give here.} So in fact we will not be able to sharply make the comparison we have in mind (but see [64]). Instead, we will consider two qualitatively similar minisuperspace models, the dS minisuperspace models in 2+1 \cite{PhysRevD.40.4011} and 3+1 \cite{PhysRevD.28.2960,Halliwell1988,DiazDorronsoro2017} dimensions, and compare the computation of the NBWF in two qualitatively different gauges in these models.

\noindent The details of the calculation in 3+1 dimensions have been reviewed above. For the (2+1)-dimensional discussion \cite{PhysRevD.40.4011}, we will sum over metrics of the form
\begin{equation} \label{dSminisuperspacea}
	16 \pi^2 \, \di s^2 = - N^2 \di \tau^2 + a(\tau)^2 \di \Omega_2^2 \,,
\end{equation}
where $\di \Omega_2^2$ is the round metric on the unit two-sphere (compare this to the (3+1)-dimensional gauge Eq. \eqref{scalarminisuperspace}). In this gauge, the Lorentzian action $S[a;N]$ is quadratic in $a$ and so the path integral in Eq. \eqref{minisuperspace2} can be evaluated exactly by the semiclassical ``approximation''. For the analogous gauge choice to Eq. \eqref{scalarminisuperspace} in 2+1 dimensions,
\begin{equation} \label{dSminisuperspaceq}
	16 \pi^2 \, \di s^2 = - \frac{N^2}{q(\tau)} \di \tau^2 + q(\tau) \di \Omega_2^2 \,,
\end{equation}
$S[q;N]$ is not quadratic in $q$ and the calculation of \eqref{minisuperspace2} is more involved (we could not even solve the second order EOM analytically, i.e. we could not write down Eq. \eqref{Kapprox}). The same statement holds for the analogous gauge choice to Eq. \eqref{dSminisuperspacea} in 3+1 dimensions.

 Choosing Dirichlet boundary conditions $\mathcal{B} = \{ a(0) = 0 \}$ in the path integral \eqref{minisuperspace2}, including a three-dimensional cosmological constant $16 \pi^2 \Lambda$ and setting the reduced three-dimensional Planck mass to unity, a straightforward calculation shows \cite{PhysRevD.40.4011}
\begin{equation}
	\Psi_\text{HH}(a) = \int_{\mathcal{C}_{2+1}} \di \tilde{N} \, \frac{1}{\sqrt{\sinh(\tilde{N})}} \, \exp \left[ \frac{i}{\hbar \sqrt{\Lambda}} \left( \tilde{N} - \frac{\Lambda a^2}{\tanh(\tilde{N})} \right) \right] \,, \label{PsiHH2d}
\end{equation}
where $\tilde{N} \equiv \sqrt{\Lambda} N$. Our point is that the integrand in Eq. \eqref{PsiHH2d} bears little resemblance to the integrands in Eqns. \eqref{PsiHHD}-\eqref{PsiHHN}; for example the former has infinitely many singularities and saddle points in the complex $\tilde{N}$-plane\footnote{These saddles correspond to a series of complete three-spheres (with (real) round metrics on them of radius $1/\sqrt{\Lambda}$) which eventually match on to Hartle \& Hawking's (generally complex) solution on part of a three-sphere \cite{PhysRevD.40.4011}.} while the latter expressions have a finite amount. In order for $\Psi_\text{HH}(a)$ to be controlled by the same saddle point as the $\Psi_\text{HH}^{(D,N)}(Q)$ discussed earlier in this appendix, the contour $\mathcal{C}_{2+1}$ could be chosen as any contour that runs from the upper-left imaginary $\tilde{N}$-plane to the upper-right imaginary $\tilde{N}$-plane by passing an infinitesimal distance below the origin $\tilde{N} = 0$. All such contours yield a convergent integral in Eq. \eqref{PsiHH2d}. This statement is not true for say the integral in Eq. \eqref{PsiHHD}, which would require the contour to run to infinity in the wedges $0 < \theta < \pi/3$ or $2\pi/3 < \theta < \pi$ of the upper-half $N$-plane (where $\theta \equiv \arg(N), \theta = 0$ representing the positive real line). Contours running to infinity in the $\pi/3 < \theta < 2\pi/3$ wedge would not yield a convergent integral for $\Psi_\text{HH}^{(D)}(Q)$, but they would for $\Psi_\text{HH}(a)$. This illustrates that the relative homology groups of the Morse functions which feature in the semiclassical approximation of the NBWF in various parameterizations of the same minisuperspace model, are inequivalent \bibnote{\label{caveatnote}As we mentioned at the beginning of this example, we have not truly compared the calculation of the NBWF in different parameterizations in the same minisuperspace model. However, while we mentioned that we could not perform the ``off-shell approximation'' \eqref{minisuperspace2} analytically in the gauge $\di s^2 = - N^2 \di \tau^2 + a(\tau)^2 \di \Omega_3^2$, we could solve the constraint equation and thus determine the (on-shell) saddle points $\bar{a}(\tau;N_s)$ analytically (of course this is well-known). With an appropriate definition of $\Lambda$ the on-shell lapse points $N_s$ are in fact equal to the on-shell lapse values following from the expression \eqref{PsiHH2d} in the $(2+1)$-dimensional model, in particular they are equally infinite in number. This hints at a strong similarity between the off-shell discussion of the $(2+1)$-dimensional minisuperspace model in the gauge \eqref{dSminisuperspacea} and the $(3+1)$-dimensional model in the gauge quoted above in this remark, which strengthens our argument. Furthermore we have numerically computed $S_0(N)$ in Eq. \eqref{Kapprox} in part of the complex $N$-plane for the $(3+1)$-dimensional minisuperspace model in the gauge written above, and the numerics support the conclusion that the off-shell properties of $S_0(N)$ are qualitatively similar in the $(2+1)$- and $(3+1)$-dimensional models.}.

The takeaway message from this appendix is that the contour $\mathcal{C}$ for the gauge-fixed lapse, which appears in a minisuperspace path integral expression of solutions to the WDW equation or Green's functions of the WDW operator (see Eqns. \eqref{minisuperspace1}-\eqref{minisuperspace3}), is not an invariant in all constructions of either of these objects. In one construction, which involves specifying a class of metrics summed over and a choice of boundary conditions $\mathcal{B}$ on the path integral, a particular $\mathcal{C}$ may lead to a well-defined object (:= solution or Green's function) while the same $\mathcal{C}$ may lead to an ill-defined object in another attempted construction of the same object. The reason is that the integrands of the lapse contour integral in different constructions may have a different singularity structure in the complex plane. This conclusion has been arrived at before in \cite{PhysRevD.43.2730} (see also references therein), and is further supported by all the minisuperspace models we know of that have been studied in quantum cosmology via a path integral approach since that work.

\clearpage
\newpage
\section{Comments on the calculation of Feldbrugge et al.} \label{FLTsec}
\noindent In this appendix we comment on the recent calculation performed by Feldbrugge et al. in \cite{FLT3}, which lead the authors to conclude that any quantum state constructed via a minisuperspace path integral is inevitably ill-defined. We will point out a technical error in their computation which invalidates their conclusion. An explicit counter-example to their claim is provided by the no-boundary quantum state in the BB9 minisuperspace model that has been constructed via a minisuperspace path integral in \cite{DiazDorronsoro:2018wro} and has been further elaborated upon in the main body of this paper. In \S\ref{isotropicsec} we have argued why this counter-example is in no way inconsistent with previous work (on the contrary), addressing further criticism that the same authors made in Ref. \cite{FLT4}. In Ref. \cite{DiazDorronsoro:2018wro} it was claimed that the computation in \cite{FLT3} is plagued by the breakdown of perturbation theory -- in this appendix we give the details that substantiate this claim.

In \cite{FLT3} the authors consider the dS + massless scalar minisuperspace that was reviewed in Appendix \ref{nonlinearsec2}.\footnote{The authors incorrectly state that they are studying the \textit{tensor}-perturbed dS minisuperspace model that we have reviewed in Appendix \ref{nonlinearsec}. As we have explained, the dS + scalar and dS + perturbative tensor theories have a different off-shell structure (compare Eqns. \eqref{tensoraction} and \eqref{scalaraction}). In \cite{FLT3} one assumes the off-shell action \eqref{scalaraction} and is thus studying the dS + massless scalar minisuperspace model.} The action is written in Eq. \eqref{scalaraction}, and the second order EOM are
 \begin{align}
 	\ddot{Q} - \frac{2 N^2}{3} &= - \frac{2}{3} \sum_n Q \dot{\varphi}_n^2 \,, \label{FLTeq1} \\
 	\forall n: ~~~ \ddot{\varphi}_n + 2\frac{\dot{Q}}{Q} \dot{\varphi}_n + N^2 n(n+2) \frac{\varphi_n}{Q^2} &= 0 \,. \label{FLTeq2}
 \end{align}
In order for the effective field theory description of gravity that we are using to be valid, the $\varphi_{nlm}$ must remain small (compared to $M_\text{Pl} = 1$) at all $\tau$. At this point one proceeds in \cite{FLT3} with the general algorithm to compute solutions of the WDW equation or Green's functions of the WDW operator in this model, reviewed in Appendix \ref{offshellsec}. The first step is the approximation \eqref{Kapprox}, for which one requires a solution to the classical EOM \eqref{FLTeq1}-\eqref{FLTeq2} for arbitrary $N \in \mathbb{C}$. Initially, one attempts to proceed analytically with this programme -- let us discuss this aspect of their calculation first.

\begin{figure}[h!]
\centering
\includegraphics[width=0.9\textwidth]{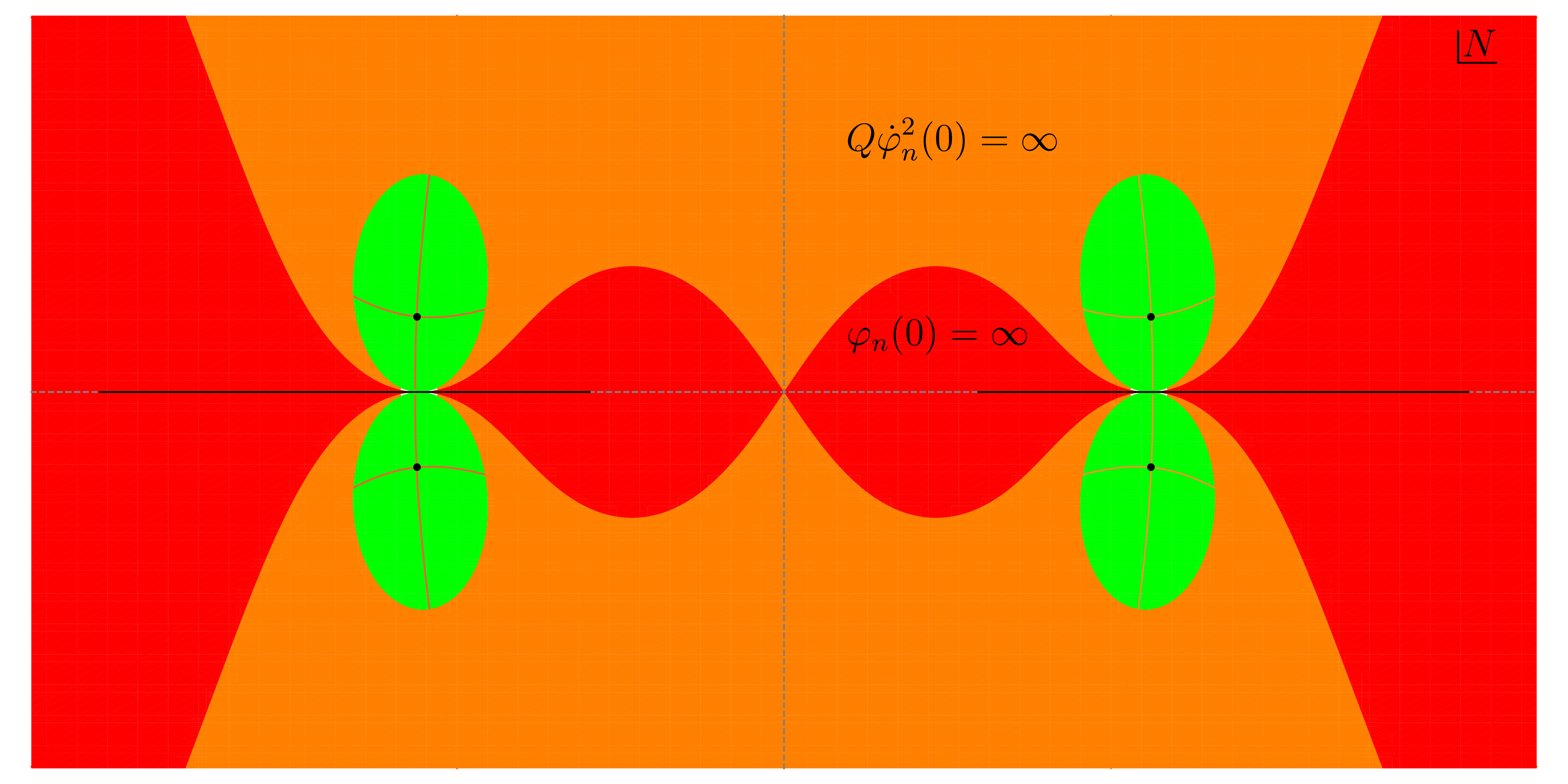}
\caption{\small The complex $N$-plane, illustrating the deficiencies of the calculation in \cite{FLT3}. To make this plot we have chosen the numerical values $n=3, Q=100$ and $\Lambda = 1$, but our qualitative conclusions apply generally to the parameter space. First, the green, orange and red regions are respectively the regions defined in Eqns. \eqref{greenregion}-\eqref{redregion}. The green regions are the subset of the complex plane where the analytic computation done in \cite{FLT3} is consistent, while in the orange and red regions the computation is inconsistent because those points correspond either to a singular solution which violates the effective field theory approximation and/or does not correspond to a solution to the EOM because at those locations a term in the EOM was neglected while it was inconsistent to do so. In the green regime we have used the off-shell action $S_0(N)$ found in \cite{FLT3} to compute its critical points (the four black dots) and their corresponding steepest descent and ascent curves (the descent curves run in the vertical direction). The black line segments that lie along the real $N$-axis represent the branch cuts for the function $S_0(N)$, analytically continued from the green regime where its form is known to as much of the rest of the complex plane as possible, which were found in \cite{FLT3}.}
\label{FLTfigure}
\end{figure}

At least initially one does not discuss in detail the boundary conditions $\mathcal{B}$ that are required to specify this classical solution. At a later stage it becomes apparent that the authors have in mind the Dirichlet boundary conditions $\mathcal{B} = \{ Q(0) = 0, \varphi_n(0) = 0 \}$ at the south pole of the geometry (the manifold on which their solutions live is implicitly assumed to be $\overline{B^4}$), and the usual Dirichlet boundary conditions that match the argument of the wave function / Green's function at $\tau = 1$, $\{ Q(1) = Q, \varphi_n(1) = \varphi_{n,1} \}$. Whatever the case may be, one proceeds in their analytic computation by disregarding the term on the RHS of the equality in Eq. \eqref{FLTeq1}. It turns out that this assumption is inconsistent for some values $N \in \mathbb{C}$. These values are easily determined: under the assumption, for all $N \in \mathbb{C} \setminus \{ \pm \sqrt{3Q} \}$, $Q(\tau;N) \propto \tau$ as $\tau \rightarrow 0$ according to Eq. \eqref{FLTeq1} and
\begin{equation}
	\varphi_n(\tau;N) \propto \tau^{[\gamma_n(N)-1]/2} ~~ \text{as } \tau \rightarrow 0
\end{equation}
according to Eq. \eqref{FLTeq2}, with
\begin{equation} \label{gamman}
	\gamma_n = \sqrt{1 - \frac{36 n(n+2) N^2}{(3Q-N^2)^2}} \,.
\end{equation}
The sign ambiguity for the square root in Eq. \eqref{gamman} is resolved in \cite{FLT3} by choosing, for each $N$, the branch which has $\text{Re}[\gamma_n(N)] \geq 0$. This is certainly possible, but 1) it does not ensure that $\varphi_n(\tau;N)$ is bounded on $\tau \in [0,1]$ for all $N$, and 2) it does not ensure that
\begin{equation}
	Q \dot{\varphi}_n^2 \propto \tau^{\gamma_n(N)-2} ~~ \text{as } \tau \rightarrow 0
\end{equation}
is small compared to the terms on the LHS of Eq. \eqref{FLTeq1} -- conditions that are required for the consistency of the calculation. Let us define three regimes in the complex $N$-plane
\begin{align}
\text{Re}[\gamma_n(N)] &\geq 2 ~~~ \text{(green)} \,, \label{greenregion} \\
\text{Re}[\gamma_n(N)] &< 2 ~~~ \text{(orange)} \,, \label{orangeregion} \\
\text{Re}[\gamma_n(N)] &< 1 ~~~ \text{(red)} \,, \label{redregion}
\end{align}
where we have assigned a color to each regime (note red $\subset$ orange). In the green regime the analytic computation in \cite{FLT3} is consistent, while in the orange and red regimes the computation is inconsistent. We pictorially represent these regions in Figure \ref{FLTfigure}, together with some other elements that are relevant to the discussion of the calculation done in \cite{FLT3}.

Because the assumption to neglect the term on the RHS of Eq. \eqref{FLTeq1} is only consistent in the green regime, the function $K(Q,\varphi_{n,1},N;0,0,0)$ which features in the integrand in Eq. \eqref{minisuperspace1} is only known analytically in that regime. From Figure \ref{FLTfigure} however it is clear that to evaluate the integral in Eq. \eqref{minisuperspace1} semiclassically, one will generally require knowledge of the integrand outside of the green regime. (This holds if the contour $\mathcal{C}$ is chosen to be any infinite or half-infinite curve, or a closed curve around the origin.) This knowledge includes the location and type of eventual singularities in the orange regime and the behavior of the integrand at infinity.

\noindent The conclusions the authors make in \cite{FLT3} are based upon the extrapolation of the consistent analytic result in the green regime to the orange and red regimes where their computation is inconsistent. At a later stage of the paper one attempts to justify this extrapolation with a numerical investigation of the EOM and off-shell action $S_0(N)$ in Eq. \eqref{minisuperspace1} in the orange and red regimes, claiming that despite the inconsistency in their analytic calculation their extrapolated analytic result is nevertheless a ``good approximation'' to the integrand $K$. For instance, one finds numerical evidence for the existence of a branch cut in the numerical integrand which is exhibited by the analytic result, even though the branch cut lies outside of the green regime except for one point (see Figure \ref{FLTfigure}). For the scenario at hand -- the semiclassical evaluation of an integral -- however, numerical hints do not suffice. The reason is that to evaluate the integral semiclassically in a controlled manner, defined say by a contour that encircles the origin, one must deform the contour onto a sum of steepest descent lines. For the sake of the argument let us assume that there are no singularities in the orange region which prohibit us from deforming the contour at will (except for a singularity at the origin), and let us take the branch cut structure of the integrand seriously. As one correctly mentions in \cite{FLT3}, the branch cut prohibits us from choosing a deformation of the contour which always remains on a steepest descent line. To evaluate the integral we are forced to leave the steepest descent lines, going around the branch cut in some way. However once we leave a steepest descent line, we immediately lose the control that the steepest descent approximation usually gives us. Generally, as $\hbar \rightarrow 0$, both the real and the imaginary part of the integrand will oscillate heavily and the standard Gaussian approximation scheme is lost. Of course this does not mean that it is impossible in general to evaluate a contour integral of some function $e^{f(z) / \hbar}$ which has a finite branch cut, around that branch cut. However, as $\hbar \rightarrow 0$, the result of the integral will depend sensitively on the details of the function $f(z)$. If $f$ and $g$ have the same finite branch cut, and $f = g$ to the accuracy ``$\varepsilon$'', this implies nothing about the contour integrals of $e^{f(z) / \hbar}$ and $e^{g(z) / \hbar}$ around the branch cut as $\hbar \rightarrow 0$.

We conclude that the computation performed in \cite{FLT3} does not provide evidence for the statement that all wave functions (or Green's functions for that matter) defined via minisuperspace path integrals are necessarily ill-defined. On the other hand, we could not complete the calculation one had in mind in \cite{FLT3} either. This would require solving the EOM \eqref{FLTeq1}-\eqref{FLTeq2} analytically in the entire complex $N$-plane, or at least in a larger part than the green region in Figure \ref{FLTfigure}, which is very challenging.

There are less ambitious variants of this calculation that one can complete, however. A first variant is the computation of the NBWF of massless scalar fluctuations under the assumption of vanishing backreaction on the geometry. This is the calculation that was done in \cite{HarHal1990} for perturbations around a fixed homogeneous and isotropic background (also reviewed in \cite{DiazDorronsoro2017}), and was extended in this paper in \S\ref{inhomogeneoussec} for perturbations around specific (BB9) fixed homogeneous but anisotropic backgrounds. We stress that in this calculation the background is completely fixed, meaning a solution to the Einstein equations is fixed which includes, in the terminology of this paper, a specification of the lapse $N = N_s$. This is how one generally proceeds when claiming to do quantum field theory in curved background spacetimes. We also stress that this is not the computation \cite{FLT3} had in mind and that was reviewed above: in their computation the authors wished to include the effects of backreaction. This can be seen from the EOM \eqref{FLTeq1}-\eqref{FLTeq2}, where the scale factor $Q$ and the fields $\varphi_n$ are treated at the same level.\footnote{Note also that neglecting the term on the RHS of Eq. \eqref{FLTeq1} does not mean one is neglecting backreaction. The lapse is not fixed either way.} A second variant is the computation of the NBWF in the BB9 minisuperspace model we have illustrated in \cite{DiazDorronsoro:2018wro} and in this paper. As we have explained in Appendices \ref{nonlinearsec} and \ref{nonlinearsec2} the BB9 model can be viewed as an extension of either the dS + $(n=2)$ perturbative tensor mode model or the dS + $(n=2)$ massless scalar model. Similar statements hold for the calculation of the tunneling proposal in these simple models (e.g. \cite{delCampo:1989hy,Vilenkin:2018dch,Vilenkin:2018oja}).

\bibliographystyle{klebphys2}
\bibliography{references}

\end{document}